\documentclass[12pt,preprint]{aastex}
\usepackage{amsmath}
\usepackage{natbib}
\usepackage{lscape}
\usepackage{color}
\usepackage{subfigure}
\usepackage{amssymb}
\usepackage{amsmath}
\usepackage{url}
\usepackage{amsfonts}
\usepackage{amsbsy}
\usepackage{graphicx}
\usepackage{subfigure}
\usepackage{verbatim}
\usepackage{multicol}
\usepackage[normalem]{ulem}

\shorttitle{OH megamaser galaxy IRAS16399-0937}
\shortauthors{Dinalva. A. Sales et al.}

\begin{document}


\title{An embedded active nucleus in the OH megamaser galaxy IRAS16399-0937\footnote{Based partly on observations made with the NASA/ESA Hubble Space Telescope, obtained [from the Data Archive] at the Space Telescope Science Institute, which is operated by the Association of Universities for Research in Astronomy, Inc., under NASA contract NAS 5-26555.}}

\author{Dinalva A. Sales\altaffilmark{1,2}, A. Robinson\altaffilmark{2}, D. J. Axon\altaffilmark{2,3,\dagger}, J. Gallimore\altaffilmark{4}, P. Kharb\altaffilmark{5}, R. L. Curran\altaffilmark{2}, C. O'Dea\altaffilmark{2}, S. Baum\altaffilmark{6}, M. Elitzur\altaffilmark{7}, R. Mittal\altaffilmark{2}} 
\altaffiltext{1}{Departamento de Astronomia, Universidade Federal do Rio Grande do Sul. 9500 Bento Gon\c calves, Porto Alegre, 91501-970, Brazil}
\altaffiltext{2}{School of Physics and Astronomy, Rochester Institute of Technology, 84 Lomb Memorial Drive, Rochester, NY 14623, USA}
\altaffiltext{3}{School of Mathematical \& Physical Sciences, University of Sussex, Falmer, Brighton, BN2 9BH, UK}
\altaffiltext{$\dagger$}{1951-2012}
\altaffiltext{4}{Department of Physics, Bucknell University, Lewisburg, PA 17837, USA}
\altaffiltext{5}{Indian Institute of Astrophysics, II Block, Koramangala, Bangalore 560034, India}
%\altaffiltext{5}{School of Physics and Astronomy, Rochester Institute of Technology, 84 Lomb Memorial Drive, Rochester, NY 14623, USA}
\altaffiltext{6}{Chester F. Carlson Center for Imaging Science, Rochester Institute of Technology, 54 Lomb Memorial Drive, Rochester, NY 14623, USA}
\altaffiltext{7}{Physics \& Astronomy Department, University of Kentucky, Lexington, KY 40506-0055 }

\begin{abstract}

We present a multiwavelength study of the OH Megamaser galaxy (OHMG) IRAS16399-0937, based 
on new HST/ACS F814W and H$\alpha$+[N{\sc\,ii}] images and archive data from HST, 2MASS, Spitzer, 
Herschel and the VLA. This system has a double nucleus, whose northern (IRAS16399N) and 
southern (IRAS16399S) components have a projected separation of $\sim$\,6\arcsec\,(3.4\,kpc) 
and have previously been identified based on optical spectra as a Low Ionization Nuclear Emission Line Region (LINER) and starburst
nucleus, respectively. The nuclei are embedded in a tidally distorted common envelope, 
in which star formation is mostly heavily obscured. The infrared spectrum is dominated by 
strong polycyclic aromatic hydrocarbon (PAH), but deep silicate and molecular absorption features
are also present, and are strongest in the IRAS16399N nucleus. The 0.435-500\micron\,SED was 
fitted with a model including stellar, ISM and AGN torus components using our new MCMC code, 
{\sc clumpyDREAM}. The results indicate that the IRAS16399N contains an AGN (L$_{bol}\,\sim10^{44}$\,ergs/s) 
deeply embedded in a quasi-spherical distribution of optically-thick clumps with a covering 
fraction $\approx1$. We suggest that these clumps are the source of the OHM emission in 
IRAS16399-0937. The high torus covering fraction precludes AGN-photoionization as the origin 
of the LINER spectrum, however, the spectrum is consistent with shocks (v\,$\sim100-200$\,km\,s$^{-1}$).
We infer that the $\sim10^8$\,M$_{\odot}$ black-hole in IRAS16399N is accreting 
at a small fraction ($\sim1$\%) of its Eddington rate. The low accretion-rate and modest 
nuclear SFRs suggest that while the gas-rich major merger forming the IRAS\,16399-0937 
system has triggered widespread star formation, the massive gas inflows expected from merger 
simulations have not yet fully developed.

\end{abstract}


\section{Introduction}\label{sec:intro}

%Ultra-Luminous Infrared Galaxies (ULIRGs) are the most luminous galaxies in the Local Universe, 
%with most of their quasar-like power emitted as infrared dust emission ($L\,\sim\,10^{12}$\,L$_{\odot}$). 

Luminous infrared galaxies (LIRGs) dominate the bright end of the galaxy luminosity function in the nearby universe 
\citep[$z\lesssim 0.3;$][]{Soifer1987} and are known to host both starbursts and active galactic nuclei (AGN), frequently in the same 
system. A variety of observational evidence -- such as double nuclei, starburst activity and disturbed 
morphologies -- suggests that most LIRGs ($L\,\gtrsim \,10^{11}$\,L$_{\odot}$) are gas-rich disk galaxies that are
undergoing strong interactions or mergers. The most luminous systems, ultra-luminous infrared galaxies (ULIRGs), have
quasar-like luminosities ($L\,\gtrsim\,10^{12}$\,L$_{\odot}$) and all appear to be advanced mergers \citep[see][for a review]{Sanders1996}.
Much observational and theoretical work has established that mergers of gas-rich galaxies play a key role in galaxy evolution 
\citep[e.g.,][]{Sanders1988a,Barnes1992,Hopkins2006,Haan2011}: tidal torques 
generated by the merger drive gas into the galaxy core(s), triggering starbursts and fueling embedded AGN. 
It has been proposed that in this scenario, ``cool'', starburst 
dominated (U)LIRGs evolve into ``warm'', AGN-dominated ULIRGs as the circum-nuclear dust is 
dispersed by starburst and AGN-induced outflows \citep{Sanders1988b}.

However, the role of galaxy mergers in triggering AGN remains uncertain. There is 
increasing evidence that major mergers only trigger the most luminous AGN, while less luminous 
AGNs seem to be driven by secular processes \citep[e.g.][]{Treister2012}. Furthermore, in their 
study of 62 ULIRGs, \citet{Rigopoulou1999} found no correlation between merger stage and 
evidence of AGN activity, suggesting that an AGN-dominated system is not an inevitable outcome 
of a gas rich merger and that ULIRG evolution is strongly affected by  the available gas mass 
and the individual structures of the progenitor galaxies.

%(U)LIRGs are therefore of great interest as local snapshots of the processes that are believed 
%to govern both the assembly of a massive spheroid and the growth of a central supermassive 
%black hole, ultimately leading to the black hole mass -- bulge velocity dispersion relationship \citep{Ferrarese2000,Gebhardt2000}.
%{\bf There are also increasing evidence that major mergers only trigger the most luminous AGN,
%while less luminous AGNs seem to be driven by secular processes \citep{Treister2012}.
%A complementary study using a sample of 62 targets also suggests that available gas from the 
%progenitor galaxies, and the morphology of the interacting objects are thought to play a key role
%to the ULIRG evolution as well \citep{Rigopoulou1999}.

%developed by 
%\citet{Rigopoulou1999} led an observational test of the ULIRG evolutionary scenario 
%using a sample of 62 targets. They found no trend for advanced mergers to be dominated by 
%AGN-like types. In addition, \citeauthor{Rigopoulou1999} suggest that available gas from 
%the progenitor galaxies, as well as the morphology of the interacting objects maybe the 
%principal component to the ULIRG evolution.
%rigo led a complementary study which suggest that 
%available gas from the progenitor galaxies, and the morphology of the interacting 
%objects may play a important role to the ULIRG evolution as well.

Approximately 20\% of (U)LIRGs contain extremely luminous OH masers, emitting primarily in the 1667 and 1665\,MHz 
lines with luminosities $\sim 10^{2-4}$\,L$_{\odot}$ \citep{Darling2002,Lo2005}. In the most recent 
and well-developed models,
this OH megamaser (OHM) emission is produced by amplification of a background radio continuum 
source by clumps of molecular gas, with inverted level populations arising from pumping by 
far infra-red (FIR; $\sim 50\mu$m) radiation from dust heated by a starburst and/or an embedded 
AGN \citep[e.g.,][]{Baan1985,Parra2005,Lockett2008}. 
As they are found in gas-rich mergers, it has been suggested that OHMs can 
be used to trace galaxy merger rates and associated processes (dust obscured star formation and 
black hole growth) over a wide redshift range \citep[e.g.,][and references therein]{Lo2005}. 
Moreover, studies of local OH megamaser galaxies (OHMG) seem likely to provide 
important insights into processes occurring
in gas-rich mergers. Thus, it has been argued that OHM emission requires 
exceptionally high concentrations of dense molecular gas, perhaps associated with a temporal spike in tidally  
driven gas inflow \citep{Darling2007}. In some individual OHMGs, detailed interferometric  mapping 
studies have found that the OH masers arise in dense, edge-on rotating molecular 
gas disks or rings on scales $\le 100$\,pc, which have been identified as compact starburst 
rings in some objects \citep{Rovilos2002,Parra2005,Momjian2006}, or in others as the circum-nuclear torii \citep{Yates2000,
Klockner2004,Pihlstrom2005,Richards2005} hypothesized by the unified scheme for AGN 
\citep{Antonucci1985,Antonucci1993,Urry1995}. In addition, the OH lines often show broad asymmetric 
profiles and velocity shifts suggestive of outflows, 
particularly in systems containing optically identified AGN \citep{Darling2006}, 
perhaps indicating that starburst or AGN-driven outflows are clearing away enshrouding, dense molecular material.

In this paper we present a detailed multiwavelength study of IRAS16399-0937, a spectacular but
relatively little studied interacting galaxy pair at a redshift $z=0.027$. It is both a
LIRG ($L_{FIR} \approx 10^{11.2}$\,L$_{\odot}$) and an OHMG ($L_{OH}\approx10^{1.7}$\,L$_{\odot}$). 
The OHM emission has a flux $\sim1.3\times10^{-22}$ W\,m$^{-2}$,
with a peak flux density at 1.6~GHz of $\sim25$mJy
\citep{Staveley-Smith1986,Norris1989,Staveley-Smith1992}. Morphologically, IRAS16399-0937
is a mid-to-late  stage merger, with a double nucleus situated in a common envelope
\citep{Haan2011}. \citet{Baan1998} classified the northern nucleus (denoted IRAS16399N, hereafter)
as a LINER \citep[see][]{Heckman1980} and the southern nucleus (denoted IRAS16399S, hereafter)
as a starburst, based on the optical emission line ratios \citep{Baldwin1981,Veilleux1987}. 
Therefore, IRAS16399-0937 appears to contain
both an AGN and a starburst in a merging system, making it a good system in which to study
the relative importance of star formation and black hole accretion in powering the IR
emission.

Our study is based on new HST imaging observations and an analysis of archival Spitzer, Herschel 
and VLA data. The paper is organized as follows. In Section \ref{sec:data} and \ref{sec:results} 
we describe the data reduction procedures and the measurements and results obtained, including 
inferred star formation rates (SFR). In Section \ref{sec:modeling} we present a decomposition of 
the optical -- IR spectral energy distribution (SED), using a new, robust fitting technique. 
We discuss the implications of our results in Section \ref{sec:discussion}. The main results 
and conclusions are summarized in Section \ref{sec:summary}. Throughout this paper, we adopt the
Hubble constant as H$_0$ = 67.3\,$\pm$\,1.2 km s$^{-1}$ Mpc$^{-1}$, $\Omega_{\Lambda}$=0.73, and $\Omega_m$=0.27 \citep{Ade2013,Lahav2014}.
%, in agreement with the most recent WMAP 9-year
%\citep{Hinshaw2013,Bennett2013}, as well as consistent with the value adopted by \emph{NASA/IPAC EXTRAGALACTIC DATABASE}.}

\section{Observation and Data Reduction}\label{sec:data}

\subsection{Hubble Space Telescope Images}\label{sec:halpha_obser}
%\subsection{F814W, FR914M and H$\alpha+$[N{\sc ii}] Hubble Image}\label{sec:halpha_obser}

We observed IRAS16399-0937 using the Hubble Space Telescope (HST) Advanced Camera for Surveys (ACS) 
as part of a snapshot program to obtain continuum and emission line imaging of a large sample 
of OHMGs (Program id 11604; PI: D.J. Axon). Images of IRAS16399-0937  were acquired in the 
wide-field channel (WFC)  using the broad-band F814W filter and the narrow and medium band
ramp filters, FR656N and FR914M, respectively. The broad-band image in F814W was 
obtained to map the continuum morphology of the host galaxy. The ramp filter images were 
obtained to study the ionized gas distribution,  with the central wavelengths set to cover 
H$\alpha$ in the narrow-band filter and the nearby continuum, for continuum subtraction, in 
the medium-band filter. The band pass of the FR656N ramp filter includes H$\alpha$ and the 
[N{\sc ii}]$\lambda 6548, 83$ lines. The total integration times were 600 s in the broad-band (I) F814W 
filter, 200 s in the medium-band FR914M filter and 600 s in the narrow band H$\alpha$ FR656N 
filter. 

In addition to the new ACS images, we also retrieved B and H broad-band images of 
IRAS16399-0937 from the Hubble Legacy Archive (HLA)\footnote{Based on observations 
made with the NASA/ESA Hubble Space Telescope, and obtained from the Hubble Legacy Archive, 
which is a collaboration between the Space Telescope Science Institute (STScI/NASA), the 
Space Telescope European Coordinating Facility (ST-ECF/ESA) and the Canadian Astronomy Data 
Centre (CADC/NRC/CSA).}. These images were obtained with the ACS/WFC and NICMOS2 cameras 
using the F435W and F160W filters as part of Programs 10592 (PI: Aaron Evans) and 11235 
(PI: Jason Surace), respectively. Both images were enhanced data products generated from 
the standard HST pipeline.

The pipeline image products were used for further processing with IRAF\footnote{IRAF is 
distributed by the National Optical Astronomy Observatory, which is operated by the Association 
of Universities for Research in Astronomy (AURA), Inc., under cooperative agreement with 
the National Science Foundation.} packages. Cosmic rays were removed from individual images 
using the IRAF task $lacos_{im}$ \citep{Vandokkum2001}. Finally, the standard IRAF tasks 
were used to yield one final reduced image for each filter (see Figure~\ref{fig:multi_wave}).

In order to construct a continuum free H$\alpha+$[N{\sc ii}] image of IRAS16399-0937 we 
first measured count rates for several foreground stars in both the medium (FR914M) band and narrow band (FR656N) 
ramp filter images. A mean scaling factor was determined from the count rate ratios
and applied to the medium band FR914M image, which was then subtracted from the narrow band FR656N image. 
Finally, the continuum subtracted image H$\alpha+$[N{\sc ii}] was visually inspected to confirm that the residuals
at the positions of the foreground stars are negligible. This is a well established procedure that
turns out in practice to give typical uncertainties of 5-10\% 
\citep[see][]{Hoopes1999,Rossa2000,Rossa2003}

\subsection{Spitzer Mid-Infrared data}\label{sec:spitzer_obser}

Infrared spectroscopy and image data of IRAS16399-0937 were extracted from the Spitzer Space 
Telescope archive. The IRAC and MIPS images were obtained in Spitzer program 3672 (PI J. Mazzarella). 
Spitzer's IRAC camera is composed of four channels covering a area of 5.2\arcmin x 5.2\arcmin\,
at 3.6, 4.5, 5.8, and 8.0$\mu$m. All four detectors have a 256 x 256 array of pixels, with each 
pixel having a size of 1.2\arcsec x 1.2\arcsec. The FWHM of the IRAC point response function 
varies between 1\farcs7 and 2.0\arcsec\,\,from channels 1 (3.6$\mu$m) to 4 (8.0$\mu$m), so the 
two nuclei of IRAS16399-0937 are well resolved in all the IRAC bands.

%The Spitzer MIPS is an imaging photometry
%at 24, 70, and 160$\mu$m with different field of view and pixels resolution. The 24$\mu$m,
%70$\mu$m, and 160$\mu$m filters have a angular sizes of 2.5\arcsec/pixel, 5.3\arcsec/pixel, 
%and 16\arcsec/pixel x \arcsec/pixel, respectively. The two nuclei of IRAS16399-0937 could 
%not been spatially resolved at MIPS's wavelength ($\geq 24\mu$m).}

The Spitzer MIPS instrument employs 3 detector arrays which provide imaging and photometry 
at 24, 70, and 160$\mu$m, respectively.  The 24$\mu$m, 70$\mu$m, and 160$\mu$m arrays have 
pixel sizes of 2.5\arcsec, 5.2\arcsec\,\, (in narrow field of view mode) and 16$\times$18\arcsec/pixel, 
respectively. As the PSF at these wavelengths has FWHMs of 6\arcsec, 18\arcsec and 40\arcsec, 
respectively, the two nuclei are not spatially resolved in the MIPS images.

To avoid saturation of the bright source, IRAS16399-0937 
was imaged with IRAC at 3.6, 4.5, 5.8, and 8.0$\mu$m and MIPS at 24, 70 and 160$\mu$m
using short integration times \citep[see][]{Armus2009,Petric2011}. For the data reduction and 
mosaicking of IRAC and MIPS images, we use Corrected BCD (cbcd) and Filtered BCD image (fbcd)
data from versions 18.18 and 18.12 of the pipeline, respectively. These data were used as input to the software 
package MOPEX\footnote{http://irsa.ipac.caltech.edu/data/SPITZER/docs/dataanalysistools/} 
\citep{Makovoz2005a,Makovoz2005b}. In order to extract and estimate the galaxy photometric 
fluxes in each IRAC and MPIS filter we used the APEX package of 
the MOPEX code \citep{Makovoz2005c}. The flux error for each passband includes the combined
effect of uncertainty caused by post processing of the images, as well as that of the flux 
density calibration of the instruments \citep{Fazio2004,Rieke2004}. Hence, we computed the 
flux uncertainty by measuring the residual flux density from the background subtracted images 
\citep[typically $\sim 2$\%,][]{Reach2005} and adding an absolute flux calibration uncertainty 
of 10\% for the IRAC and MIPS 24$\mu$m\ measurements, and of 20\% for MIPS 70 and 160$\mu$m\ 
fluxes \footnote{http://ssc.spitzer.caltech.edu/irac/calib/extcal/}. Our measured flux 
densities and uncertainties are given in Table ~\ref{tab:fluxes}. A composite image of IRAS16399-0937
using H$\alpha$, 1.49GHz and 8.0$\mu$m is presented in Figure \ref{fig:rgb}.

The mid-IR spectrum of  IRAS16399-0937 was observed as part of program  30323 \citep{Armus2009}.
The data were obtained using the Short-Low (SL), and Long-Low (LL) Infrared Spectrograph \citep[IRS;][]{Houck2004}
modules, which cover the interval between 5$\mu$m and 37$\mu$m with a resolving power of 64-128. 
The widths of the SL, and LL slits correspond to, respectively, $\approx$ 1.9\,kpc (3.7\arcsec), and 5.5\,kpc 
(10.7\arcsec) for IRAS16399-0937. The raw data were reduced using the IRS 
pipeline version 18.18 at the Spitzer Science Center (SSC)\footnote{http://ssc.spitzer.
caltech.edu/irs/features/}. The spectra are background-corrected and bad/hot-pixel corrected. 
Rogue pixels were removed and replaced using IRSCLEAN software available from the SSC. 

A one-dimensional spectrum of the entire IRAS16399-0937 system was extracted using the 
default parameters of the Gaussian extraction method of the {\sc smart} code \citep{Higdon2004}.
However, we used the optimal extraction method \citep{Lebouteiller2010}\footnote{\citep{Lebouteiller2010}'s optimal 
extraction algorithm reduces the impact of noise from pixels containing little flux from 
the source and substantially improves the S/N compared to spectra extracted with more conventional 
algorithms.} within {\sc smart} to extract spectra at 7 locations along the slit, with two apertures 
centered on the IRAS16399N and IRAS16399S nuclei, respectively, and the others sampling the extended envelope. 
Figure \ref{fig:slit_aperture} shows the position of each 
extracted spectrum superposed on the 8.0$\mu$m IRAC image. To extract spectra using the
optimal method we selected an aperture of radius 2.02\arcsec~while the spectrum of the entire system 
was obtained with an aperture of radius 7\arcsec. The spectra were combined and order matched and are 
shown in Figures \ref{fig:spec_system} and \ref{fig:spec_agn_sb_ism}. 

%\subsubsection{The Mid-Infrared Emission Line Measurements}

The IRS spectra were decomposed using the  {\sc pahfit}\footnote{Source and 
documentation of {\sc pahfit} are available in http://tir.astro.utoledo.edu/jdsmith/research/pahfit.php} 
code \citep{Smith2007}. This code assumes that the mid-IR spectrum is composed of dust continuum, 
starlight, prominent emission lines, individual and blended PAH emission bands, and that the 
light is attenuated by extinction due to silicate grains. The fluxes and equivalent widths of spectral features  
properties derived from {\sc pahfit} code are given in Tables \ref{tab:mirlines_sns} and
\ref{tab:mirlines_ism}. The {\sc pahfit} decomposition of the IRAS16399-0937 system aperture
is presented in Figure \ref{fig:spectra_decompostion}.

\subsection{Herschel Data Reduction}\label{sec:herchel_obser}

The SPIRE photometric observations were made in the small-map mode (appropriate for fields 
smaller than 4 arcmin by 4 arcmin) on 2011-09-22 (PI: D.~Sanders, ObsID: 1342229188). The 
SPIRE photometer \citep{Griffin2010}, with a field of view of 4~arcmin by 8~arcmin, is
capable of carrying out simultaneous observations in three spectral bands, PSW (250$\mu$m), 
PMW (350$\mu$m) and PLW (500$\mu$m), providing resolutions of about 18\arcsec, 25\arcsec 
and 36\arcsec, respectively.

We reduced the data using the Herschel Interactive Processing Environment \citep[{\sc HIPE},][]{Ott2010} 
version 9.0~CIB~3071 (equivalent of HIPE~9.1.0). We used the simplified version of the
small-map mode POF10 pipeline script and the na{\'i}ve mapping algorithm. We did not include 
the turnarounds at the end of the scan lines in the processing and mapmaking procedures. Of 
the additional options available, we made maps with and without applying the relative
bolometer gain corrections (useful for extended sources), and using both the baseline 
subtraction method (which removes baseline from the scans individually) and the destriper 
method (which subtracts a median baseline from all the scans), separately. The resulting
flux-densities were consistent within uncertainties.

%The data processing was commenced from level 0.5 products -- the SPIRE pointing product was 
%created, the wavelet deglitcher was applied on the timeline data, the temperature drift correction 
%was performed, and maps were created, the units of which were Jy/beam. The flux-densities
%were calculated using our self-developed aperture photometry code. %({\bf Rupal, do you have
%some article that describe how you measure the sky contribution etc?}).
%The measured flux densities are listed in Table \ref{tab:fluxes}.

The data processing was commenced from level~0.5 products -- the SPIRE
pointing product was created, the wavelet deglitcher was applied on
the timeline data, the temperature drift correction was performed, and
maps were created, the units of which were Jy/beam. The flux-densities 
were measured by integrating the source counts 
within a circular aperture of radius 36\arcsec. The sky 
back-ground was calculated by placing a circular aperture at a region 
far-removed from the source and subtracting the mean background flux 
from each pixel of the source aperture. Due to the non-uniformity of 
the sky background, we determined an average using three different 
locations.  The flux measurements are given in Table \ref{tab:fluxes}.

%The flux-densities
%were calculated using an aperture photometry code written in
%C-programming language. The source counts were integrated within an
%circular aperture of radius 36\arcsec. The sky background was
%calculated by placing a circular aperture at a region far-removed from
%the source and subtracting the mean background flux from every pixel
%of the source aperture. Due to the non-uniformity in the background,
%we determined the sky background at three different locations. Given
%in Table \ref{tab:fluxes} are the average flux estimates.

\subsection{VLA Radio Data}\label{sec:radio_obser}
%We reduced VLA A-array radio data at 1.4 GHz for IRAS16399-0937 galaxy using the NRAO data 
%archive (Project IDs: AD0483, AD0462, AD0461, AN0104, AS0402). The data reduction was carried 
%out following standard calibration and reduction procedures in the Astronomical Image Processing 
%System (AIPS). Due to the weakness of the radio emission, no self-calibration was possible 
%(there were too many failed solutions when self-calibration was attempted). Figure~\ref{fig:radio_contour} 
%display the radio contour images, while Tables~\ref{tab:radio} list the observed image parameters: 
%$S_{peak}$, $S_{tot}$ and $rms$ are the peak and total flux densities and the {\it r.m.s.} 
%noise estimates. The contour levels and frequency of observation are noted in the figure 
%panel. The beam-sizes for the 1.4 GHz images is typically $1.5\arcsec\times1.4\arcsec$.

%IR16399-0937 shows multiple structure embedded within extended disk at L-band (radio band) 
%with P.A.(triple) = 27$^{\circ}$. C-band (radio band) structure composed of
%multiple SNR-like distribution at same PA as L-band structure. NC = 000.

IRAS16399$-$0937 was observed at 1.49 GHz in the VLA A-array configuration on March 7, 1990 (Project code: AM293).
These are unreduced data that were obtained from the NRAO archive\footnote{https://archive.nrao.edu/}
and processed with the Astronomical 
Image Processing System (AIPS) software using the standard calibration technique. The final 
radio image (Figure \ref{fig:multi_wave}) was made after several iterations of phase and 
amplitude self-calibrations using the AIPS tasks CALIB and IMAGR. The final resultant $rms$ 
noise in the image was $\sim7\times10^{-5}$~Jy~beam$^{-1}$. The radio image shows the presence 
of an unresolved core centered on the IRAS16399N nucleus. There is additional radio emission 
associated with the merging galaxy to the South. Table \ref{tab:fluxes} shows the 1.4 GHz fluxes
for the entire IRAS16399-0937 system, as well as for apertures centered on the IRAS16399N and IRAS16399S nuclei.

We combined the 1.49 GHz image with another dataset extracted from the NRAO  
archive, a pipeline-reduced 4.9 GHz VLA A-array image (Project code: AB660)\footnote{https://archive.nrao.edu/archive/archiveimage.html}, 
to create a 1.49$-$4.9 GHz spectral index image (Figure ~\ref{fig:radio_contour}). For the 
4.9~GHz image, UVRANGE was restricted to 0-180 k$\lambda$ in the task IMAGR and a UVTAPER 
of 150 k$\lambda$,0 was used. Both images were convolved with an 
intermediate-sized circular beam of $1.7\arcsec\times1.7\arcsec$, and aligned with the task 
OGEOM before creating the spectral index map using the task COMB. 
%We find that the Northern component has a steep spectral index of $\alpha = -0.7$\pm$0.1, 
The spectral index,  $\alpha$, is defined 
such that flux density S$_{\nu}$ at frequency $\nu$ is S$_{\nu} \propto \nu^{\alpha}$.

\subsection{Chandra X-ray Image}\label{sec:x-ray_obser}

The X-ray image of IRAS16399-0937 was extracted from the Chandra public 
data archive. The image was observed with Chandra on 2013 June 30 (Obs ID 15055) using the 
focal plane detector ACIS-S operated in VFAINT imaging mode. This data set is part of the
Chandra component of the Great Observatories All-sky LIRG Survey \citep[GOALS,][]{Armus2009}
program 14700854 (PI: D. Sanders). The exposure time of the image is 14.87 ks. The data were 
reduced with the standard Chandra data analysis package CIAO 4.6 with the latest calibration 
files in CALDB 4.6.1, and the script ``srcflux'' in CIAO to estimate the flux. The flux
and luminosity derived for the whole system IRAS16399-0937, as well as for the two nuclei, 
IRAS16399N and IRAS16399S, are presented in Tables \ref{tab:fluxes} and \ref{tab:luminosity}. 
Figure \ref{fig:radio_xray} shows a composite of the 1.49\,GHz radio and X-ray images.

\section{Results}\label{sec:results}

\subsection{Optical to Near-IR Features}\label{sec:result_opt_nir}

\subsubsection{Morphology}\label{sec:result_onir_morph}

\begin{landscape}
\begin{figure*}
\centering
\begin{tabular}{ccc}
\includegraphics[angle=0,scale=0.46]{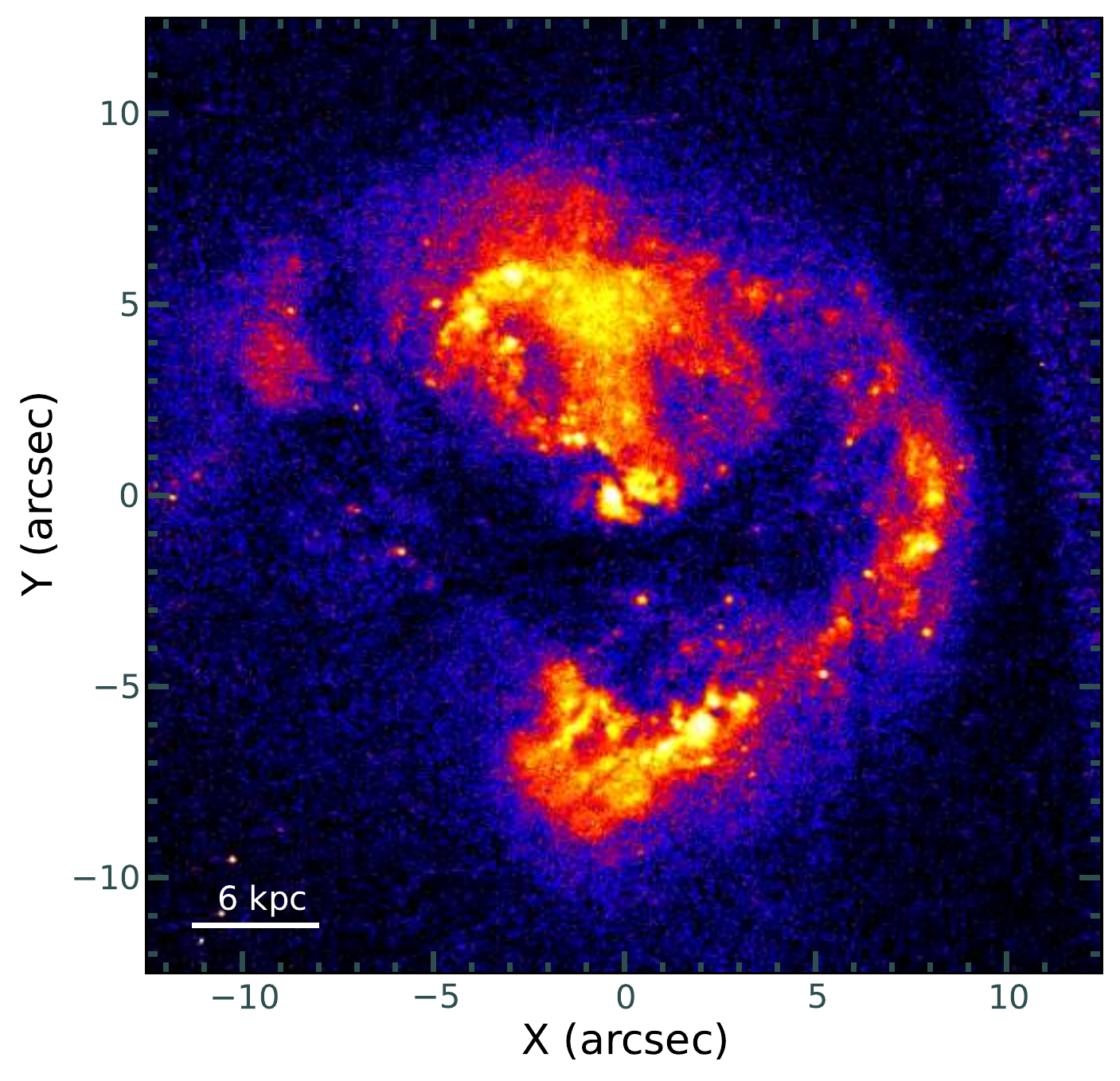}&
\includegraphics[angle=0,scale=0.46]{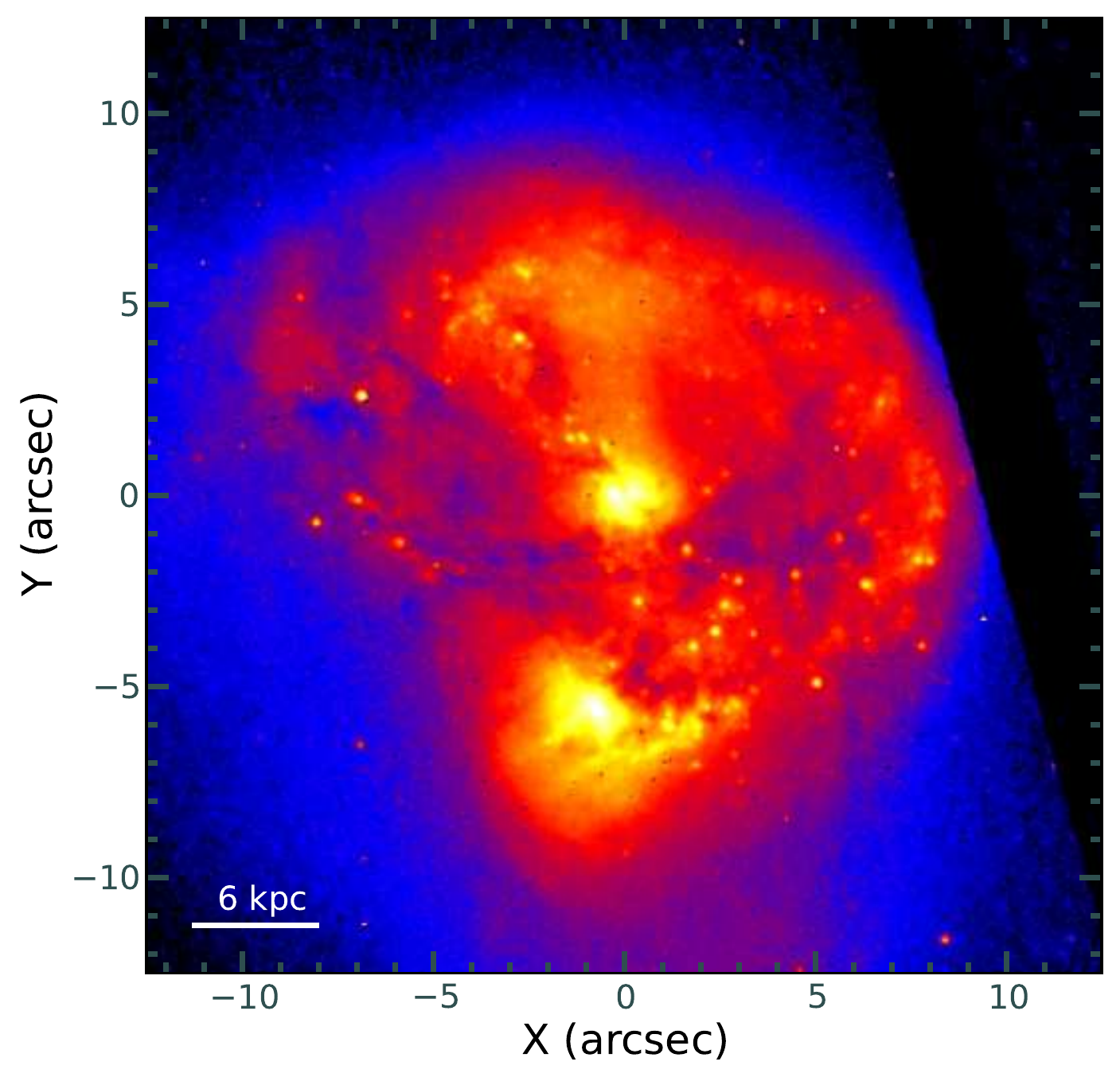}&
\includegraphics[angle=0,scale=0.46]{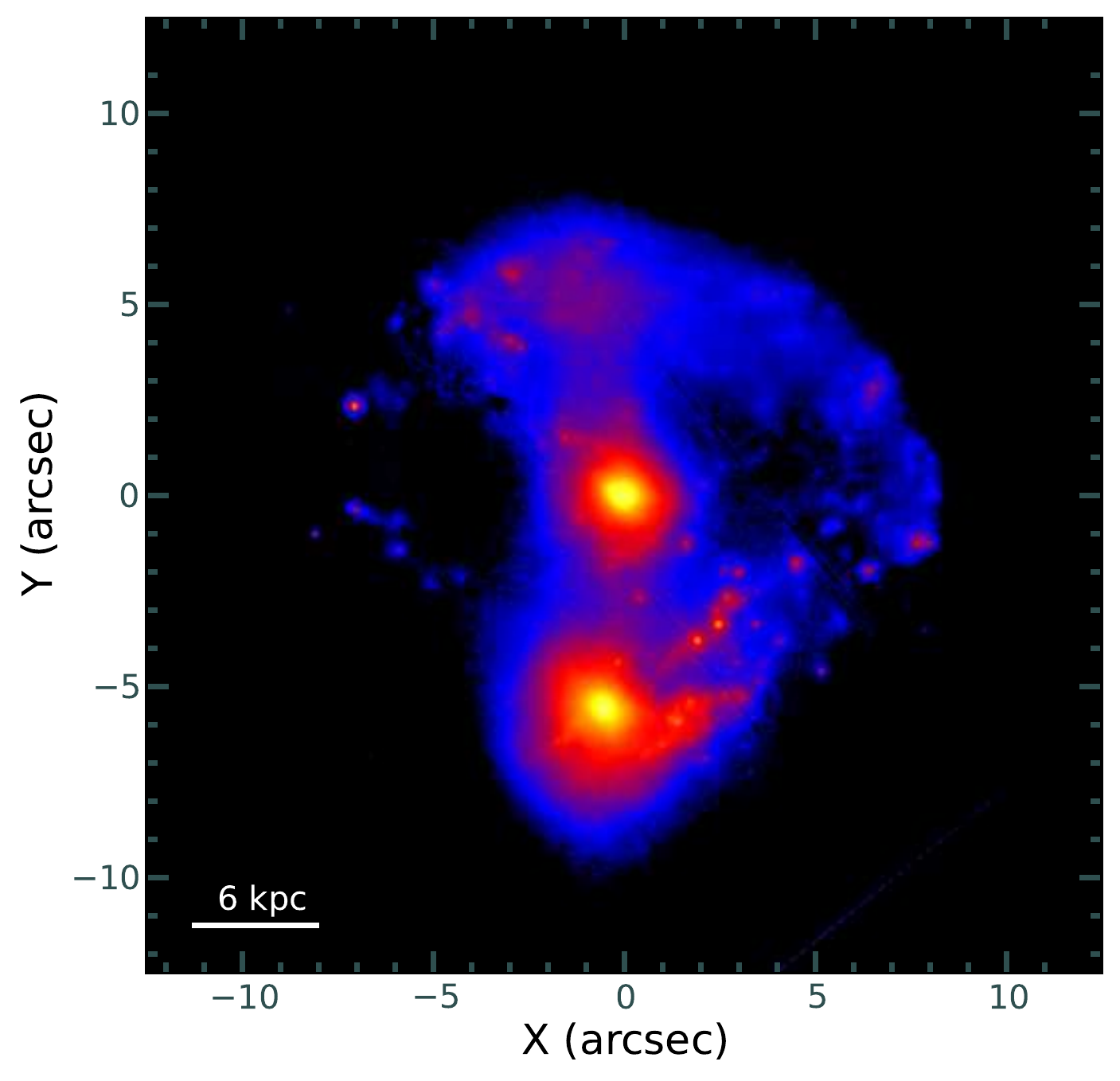}\\
\includegraphics[angle=0,scale=0.46]{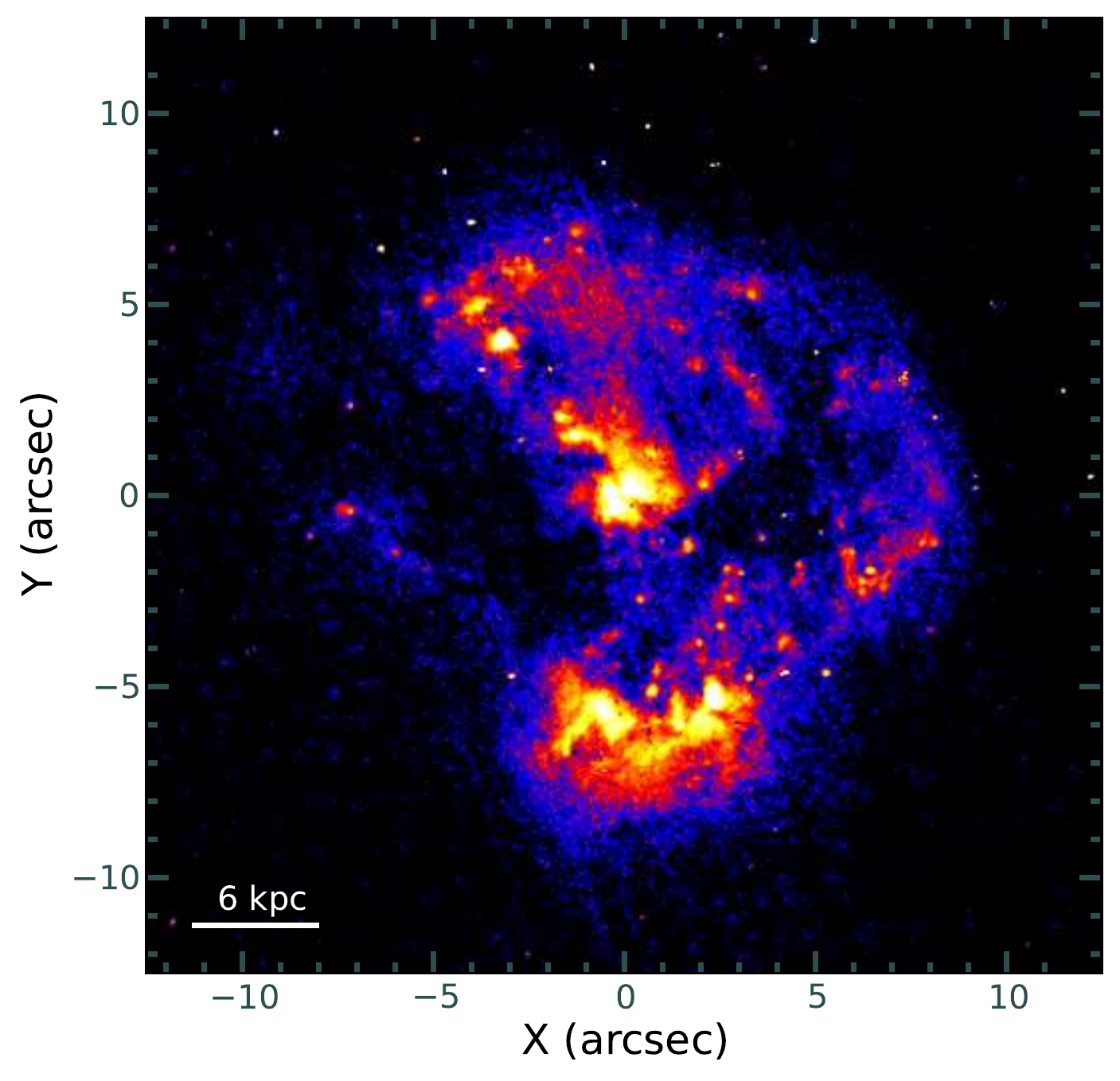}&
\includegraphics[angle=0,scale=0.46]{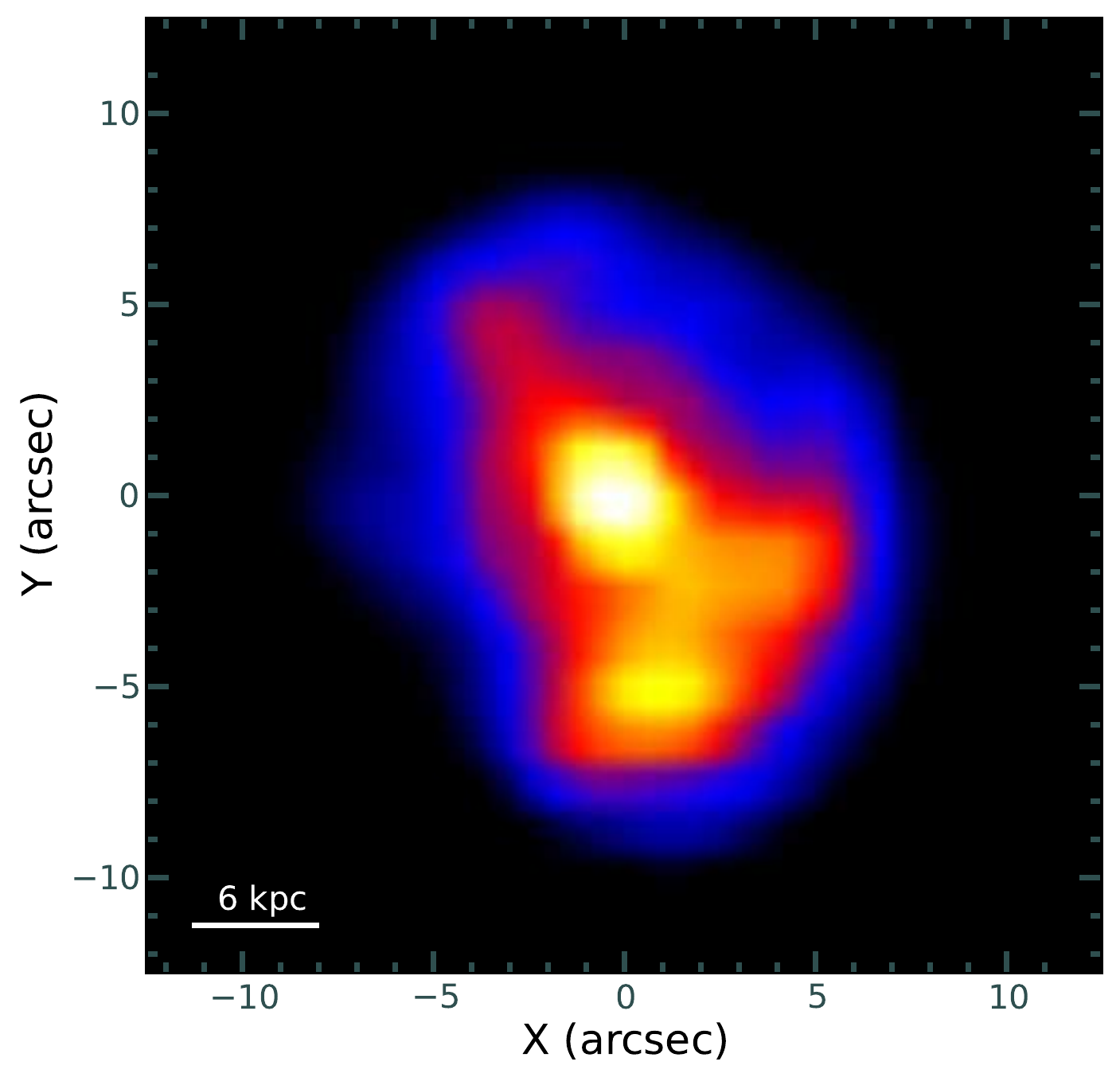}&
\includegraphics[angle=0,scale=0.46]{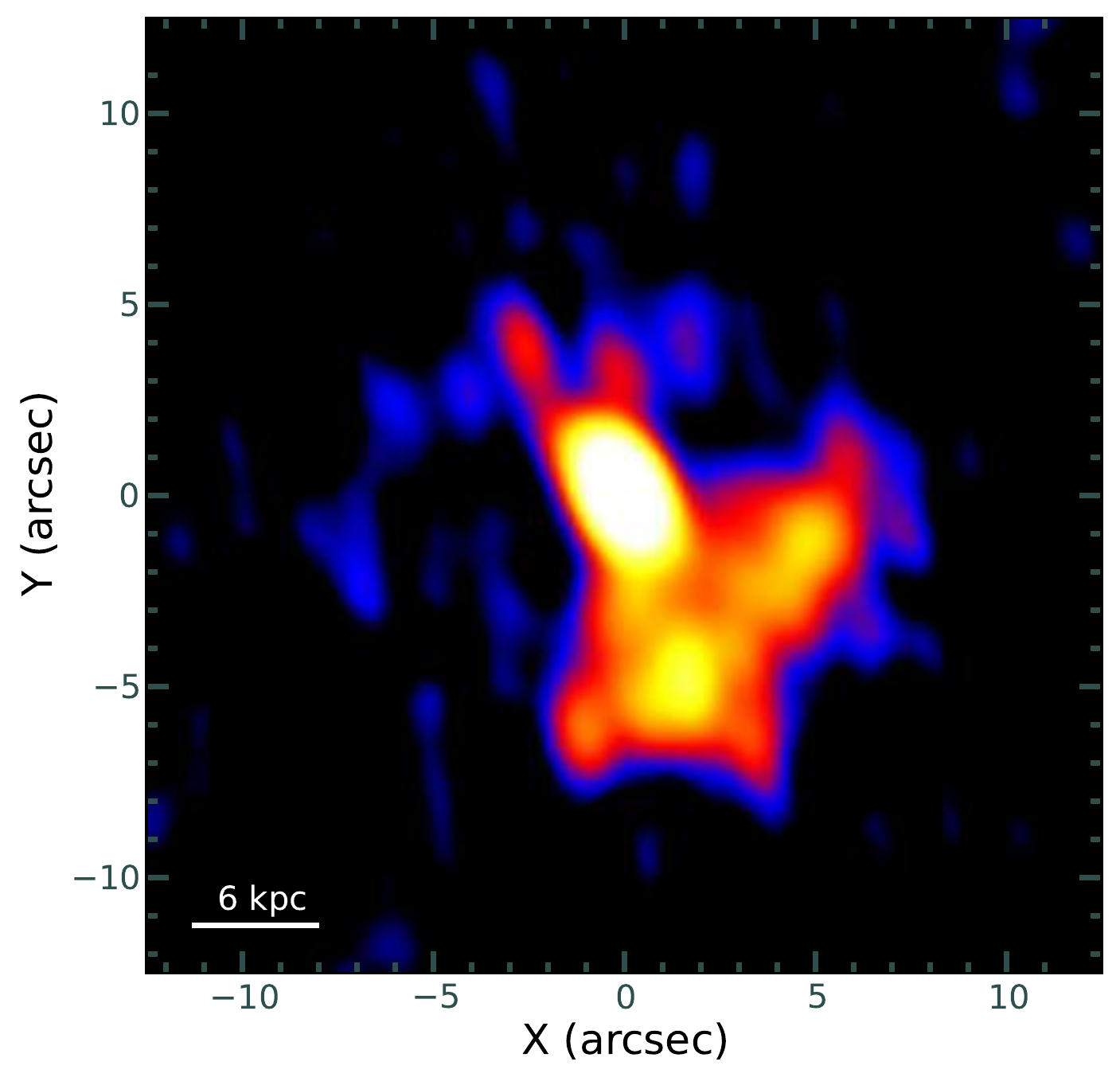}\\
\end{tabular}
\caption{Multiwavelength images of IRAS16399-0937. Top panels, from the left to the right: 
HST images obtained with ACS F435W (left), ACS F814W (middle),  NICMOS F160W (1.6$\mu$m, right). Bottom panels:
continuum-subtracted H$\alpha+$[N{\sc ii}], from HST ACS (left), ISM PAH-dust 8.0 $\mu$m emission, from Spitzer IRAC  (middle), 
VLA 1.49\,GHz (right). North is up and East is to the left.}
\label{fig:multi_wave}
\end{figure*}
\end{landscape}

The ACS medium and broad band images are presented in Figure \ref{fig:multi_wave}, along 
with the continuum subtracted H$\alpha+$[N{\sc ii}] emission line image. These show IRAS16399-0937 
to be a merger system in which two galaxy cores (the IRAS16399N and IRAS16399S nuclei) reside in 
a highly disturbed common envelope. As already noted, the IRAS16399N (centered at RA: 16h42m40.18s
and DEC: -09d43m13.26s) and IRAS16399S (centered at RA: 16h42m40.15s and DEC: -09d43m19.02s) 
nuclei have been classified as a LINER and a starburst, respectively, based on their optical 
spectra \citep{Baan1998}. The two nuclei are separated by a projected distance of $\sim$6\arcsec\, (3.4\,kpc). 
Extensive dust lanes interpenetrate the region between the two nuclei and extend several 
arcseconds across the system to the east and west.

The IRAS16399N nucleus is the more compact of the two but is crossed by a dust lane which is 
clearly visible in the broad-band F435W image (Figure \ref{fig:rgb}). The IRAS16399S nucleus has 
a more extended and disturbed morphology. In addition to the nuclei, two other main morphological 
features are prominent in the blue (F435W) continuum emission (Figure \ref{fig:multi_wave}). 
An arc-like structure, loops around the system to the West, connecting the IRAS16399S nucleus to a 
large amorphous region of emission that extends $\sim 3$\,kpc to the North and East of the 
IRAS16399N nucleus (Figure \ref{fig:multi_wave}). Hereafter, we refer to these features as the ``western 
arc'' and the ``north-eastern blob'', respectively.

Line emission in H$\alpha+$[N{\sc ii}] is widely distributed around the system, with bright knots embedded in more 
diffuse patches associated with the main morphological features apparent in the continuum 
images, including both nuclei, the north-eastern blob and the western arc. In addition, a 
smaller arc of ionized gas emerges from the north west side of the IRAS16399N nucleus and curves back 
in a north-easterly direction. The line emission associated with the IRAS16399S nucleus is extended 
and patchy, while that associated the IRAS16399N nucleus is more concentrated. Strings of knots are visible in 
both the H$\alpha+$[N{\sc ii}]  and the continuum images (e.g., NW of the IRAS16399S nucleus and NE 
of the IRAS16399N nucleus). Given that the B-band (F435W) and H$\alpha$+[N{\sc ii}] emission 
trace, respectively, young stars and H{\sc ii} regions, it is clear that recent or ongoing (and 
relatively unobscured) star formation, while concentrated in the structures mentioned above, 
is widespread across the envelope over scales of several kpc. 

The 1.6$\mu$m emission, mapped in the NICMOS F160W image, is less sensitive to dust extinction 
and predominantly traces the old stellar population, providing a clearer picture of the stellar 
mass distribution, which can be compared with the gas and dust morphology. Although the overall 
morphology of the system at 1.6$\mu$m is similar to that seen in the optical images, the two 
nuclei dominate and are much more clearly defined, appearing as approximately elliptical 
structures with compact central cores. The IRAS16399S nucleus, however, also has a prominent spur 
pointing NW, which is clearly associated with the base of the western arc. Several of the 
brighter knots (H{\sc ii} regions) are also prominent at 1.6$\mu$m. The same features 
appear in the Near-IR J and K band images obtained from the 2MASS data archive.

IRAS16399-0937 was included in the sample of (U)LIRGs studied by \citet{Haan2011}, 
who classify it as a mid-stage merger (stage 3: two nuclei % merger scheme - Veilleux, S., Kim, D. C., & Sanders, D. B. 2002, ApJS, 143, 315
in a common envelope, according to their classification scheme) based on HST/ACS B- and I-band 
images. Haan et al. also modeled the surface brightness distribution in an HST/NICMOS H-band 
image, using a combination of two S\'ersic  profiles. Notably, a point source was also required 
to fit the IRAS16399N nucleus. From the decomposition, \citeauthor{Haan2011} find that the two nuclei 
have very similar H-band luminosities $\log (L_{bulge}/L_{\odot})\approx 10.6$, but the IRAS16399S 
nucleus is more compact ($R_{bulge}\approx 0.5$\,kpc, compared to $\approx 1$ for the IRAS16399N nucleus) 
and therefore has a higher luminosity density ($L_{bulge}/R_{bulge}^2$). The S\'ersic indices 
also differ, indicating radial brightness profiles consistent with a disk-like pseudo-bulge 
in the IRAS16399N nucleus ($n_{s}\approx 1.3$) and a classical bulge ($n_{s}\approx 2.9$) in the IRAS16399S 
nucleus, respectively \citep{Fisher2008}, suggesting that the progenitor galaxies followed 
different evolutionary paths \citep[e.g.,][]{Kormendy2004}. \citeauthor{Haan2011} also derived 
black hole masses from the H-band luminosities using \citet{Marconi2003}'s relation, finding 
$M_{BH}\approx  10^8$\,M$_{\odot}$ for both nuclei\footnote{It should be noted, however, 
that recent work suggests that pseudo-bulges may not follow the same scaling relationships 
with black hole mass as classical bulges and elliptical galaxies \citep[e.g.,][]{Kormendy2011}. 
\citet{Marconi2003} do not 
distinguish between bulge types, but their sample is dominated by ellipticals and lenticulars.}.

%As the compact nuclei 
%are clearly defined in the 1.6$\mu$m NICMOS image, \citet{Haan2011} performed a two dimensional 
%decomposition using the {\sc galfit} code \citep{Peng2002,Weinzirl2009}, finding 
%that the Sersic power-law index of the bulge radial profile of the North nucleus was 
%$1.26\pm0.07$, while the South nucleus exhibited a Sersic index of $2.93\pm0.15$. 

%While the two nuclei are the dominant features in the H-band image, they appear more diffuse 
%and are less prominent relative to the extended envelope in other wavebands. The S nucleus, 
%in particular, resides within a patchy complex of knots visible in both the ACS B-band image 
%and in the while the two nuclei are the dominant features in the H-band image, they appear 
%more diffuse and are less prominent relative to the extended envelope in other wavebands. 
%The S nucleus, in particular, resides within a patchy complex of knots visible in both the 
%ACS  B-band image and in the $H\alpha + [NII]$ line emission. The N nucleus is located at 
%the south-western edge of an even larger, though more diffuse region of B-band and line 
%emission that dominates the northern side of the system. These two regions are connected by 
%an arc-like structure reminiscent of a tidal tail. 

%IRAS16399-0937 The same conclusion was noted by \citet{Haan2011}
%who also claim that only roughly 35\% of the LIRG merger systems have their double
%nuclei detectable at optical and near-IR wavelength, while the remaining are nuclei
%very obscured by dust. 

\subsubsection{H$\alpha$ photometry}\label{sec:result_Ha_phot}

Emission line fluxes were measured from the continuum subtracted FR656N ramp filter image 
using circular apertures of radius 2.02\arcsec\, centered on the IRAS16399N and IRAS16399S nuclei, 
respectively. A circular aperture of radius 7.0\arcsec\ was used to measure the total 
emission line flux of the whole system. 

As the FR656N ramp filter includes both H$\alpha$ and the [N{\sc ii}]$\lambda\lambda 6548, 83$ 
lines in its passband, it is necessary to correct for both dust extinction and the contribution of the [N{\sc ii}] lines
in order to estimate the intrinsic H$\alpha$ fluxes.
Approximate corrections can be derived from the optical line intensities published by \citet{Baan1998} 
and \citet{Kewley2001}, both of whom obtained long slit spectra crossing both nuclei. However, 
while most of the line intensity ratios given in these studies agree to within 20--30\%, 
there is serious disagreement as to the reddening corrections derived from the H$\alpha$/H$\beta$ 
ratio. The reddening estimated by \citeauthor{Baan1998} implies extinctions at H$\alpha$ 
of $A_{H\alpha}\approx  5.9$ ($A_V\approx 6.8$) and 3.0 ($A_V\approx 3.4$) magnitudes for 
the IRAS16399N and IRAS16399S nuclei, respectively, whereas \citeauthor{Kewley2001}'s results imply $A_{H\alpha}\approx  3.2$ 
($A_V\approx 3.8$)  and 2.0 ($A_V\approx 2.4$) magnitudes. For the IRAS16399N nucleus, in particular, 
the difference in $A_{H\alpha}$ of $\approx 3$ magnitudes corresponds to an order of magnitude in 
the intrinsic flux after correcting for extinction. While there is no compelling reason to 
favor either extinction estimate, the extraction aperture used by \citeauthor{Kewley2001} 
(1\,kpc, corresponding to $\approx 2$\arcsec\ at the redshift of IRAS 16399-0937) is 
approximately the same size as our own, whereas that used by \citeauthor{Baan1998} 
was $\approx 3\times$ larger and must therefore include emission from the regions surrounding 
each nucleus. In the following, therefore, we adopt the extinction and line ratios given by 
\citeauthor{Kewley2001}, on the grounds that they are more likely to represent reasonable 
average values for the two nuclei. Nevertheless, we note that if \citeauthor{Baan1998}'s 
results are used, the H$\alpha$ luminosities of the IRAS16399N and IRAS16399S nuclei increase by factors of 
11 and 2.4, respectively.

We first correct the measured emission line fluxes for extinction using the selective extinction 
given by \citeauthor{Kewley2001} for each nucleus and the standard Milky Way reddening curve 
($R_V=3.1$). The reddening corrected [N{\sc ii}$\lambda 6583$]/H$\alpha$ ratios from 
\citeauthor{Kewley2001} were then used to estimate the H$\alpha$ contribution to the total 
H$\alpha$ + [N{\sc ii}]$\lambda\lambda 6548, 83$ flux in each nuclear aperture. We find H$\alpha$ 
contributions of $\sim$\,41\% and $\sim$\,58\% for the IRAS16399N and IRAS16399S nuclei, respectively, 
yielding dereddened H$\alpha$ fluxes of 1.8\,$\times\,10^{-13}$ ergs\,s$^{-1}$\,cm$^{-2}$ 
and 1.0\,$\times\,10^{-13}$ ergs\,s$^{-1}$\,cm$^{-2}$  (Table \ref{tab:fluxes}).

As optical spectroscopy is available only for the nuclei, we have no direct measurement of 
the average extinction or [N{\sc ii}$\lambda 6583$]/H$\alpha$ ratio for the extended emission 
line regions pervading the envelope. However, as the extended line emission is probably 
dominated by star formation we adopt the value of [N{\sc ii}]/H$\alpha$ for the IRAS16399S nucleus (which has 
line ratios characteristic of H{\sc ii} regions, \citet{Baan1998,Yuan2010}; see Sec.~\ref{sec:discussion}) 
and estimate the total H$\alpha$ flux using the values of $A_{H\alpha}$ derived for the 
IRAS16399N and IRAS16399S nuclei as upper and lower limits, respectively, for the extinction. This yields, a total H$\alpha$ 
flux for the whole system in the range F$_{H\alpha}\,\approx$\,5.8\,-\,17.4\,$\times\,10^{-13}$ 
ergs\,s$^{-1}$\,cm$^{-2}$ (Table \ref{tab:fluxes}).

As the  H$\alpha$ luminosity produced by a star forming region is proportional to the global rate 
of production of ionizing photons, it can be related to the total SFR over all stellar masses. 
We adopt the following relation, from \citet{Calzetti2007}:
\begin{equation}
SFR_{H\alpha}(M_{\odot}\,yr^{-1})\,=\,5.3\,\times\,10^{-42}\,L_{(H\alpha,corr)},
\end{equation}
where $L_{(H\alpha,corr)}$ is the H$\alpha$ luminosity in ergs\,s$^{-1}$.
For the IRAS16399N and IRAS16399S nuclei, the reddening corrected H$\alpha$ fluxes imply 
SFR$_{H\alpha}$ of $\sim$\,1.73\,$M_{\odot}\,yr^{-1}$ and $\sim$\,0.96\,$M_{\odot}\,yr^{-1}$,
respectively. It should be noted, however, that as the IRAS16399N nucleus has a LINER spectrum 
(\citet{Baan1998, Yuan2010}; see Sec.~\ref{sec:discussion}), its H$\alpha$ emission is likely to include
a large contribution due predominantly to processes other than stellar photoionization. The 
derived SFR should therefore be regarded as an extreme upper limit. 

The global star formation rate implied by the total $H\alpha$ luminosity of the whole 
system (7\arcsec) is SFR$_{H\alpha}$ $\sim$\,5.58\,-\,16.76\,$M_{\odot}\,yr^{-1}$ (Table \ref{tab:fluxes}).

\subsection{The mid- to far-IR Features}\label{sec:result_mir}

\subsubsection{Mid-IR Spectroscopy}\label{sec:mir_spectra}

The mid-IR spectral range is very rich in emission features of molecular gas, such as H$_2$, 
PAH and silicate bands \citep{Puget1989,Genzel1998,Draine2003}. Prominent ionic emission 
lines of neon, oxygen, sulfur and silicon are also present \citep[e.g.][]{Genzel1998,Sturm2002,Gallimore2010,Sales2010}. 
The observed spectrum of the entire IRAS16399-0937 merger system (Figure \ref{fig:spec_system}) 
clearly shows prominent emission from the well-known PAH bands at 6.2, 7.7, 8.6, 11.3, and 
12.7$\mu$m, as well as molecular hydrogen lines due to vibrational modes at 5.5, 6.9, 8.0, 
9.6, 12.2, and 17.0$\mu$m. In addition, the mid-IR spectrum exhibits emission lines due to 
ions of moderate ionization potential (IP\,$<$\,50eV), such as [Ne{\sc\,ii]\,}12.8$\mu$m, 
[Ne{\sc\,iii]\,}15.5$\mu$m, [S{\sc\,iii]\,}18.7$\mu$m and 33.48$\mu$m, and [Si{\sc\,ii]\,}34.8$\mu$m. 
However, IRAS16399-0937 does not show the higher ionization lines ([Ne{\sc\,v]\,} at 14.3$\mu$m and 
24.2$\mu$m or [Ne{\sc\,vi]\,} at 7.6$\mu$m) that are often taken as indicators of the presence 
of the hard radiation field (IP\,$>$\,90eV) associated with an AGN \citep{Genzel1998,Sturm2000,Armus2007}. 
The PAH and emission line fluxes are listed in Table \ref{tab:mirlines_sns}.

Absorption bands arising from the water ice stretching mode at 6.0$\mu$m and  
from hydrogenated amorphous carbon grains (HACs) at 6.85 and 7.25$\mu$m \citep[see also][]{Chiar2000,Dartois2007} 
are also clearly present. While the water ice absorption band is an indicator of a heavily 
extincted and enshrouded nucleus, AGN or starburst \citep{Spoon2001}, the HAC features are 
known to be associated with dust in the diffuse ISM. Such features are commonly observed 
in LIRGs and ULIRGs \citep{Spoon2001,Spoon2002,Willett2011a}. According to \citet{Willett2011a,Willett2011b},
both water ice and HAC features are common in OHMGs, having been detected in 50\% percent 
of their sample of 51 OHMGs. The ice and HAC optical depths were measured by adjusting a 
spline continuum to the average fluxes within 1$\mu$m windows with pivots at 5.2, 5.6, 7.8, and
13.5$\mu$m \citep{Spoon2007,Willett2011a} and taking the ratio between observed and continuum 
fluxes of these absorption features (see Table \ref{tab:mirlines_sns}).

% Compared with 0 and 3 of non-OHMG control sample of 15 ULRIGs

In order to investigate the spatial variation of the mid-IR emission 
features in IRAS16399-0937, we extracted SL ($5.2\,-\,14\mu$m) spectra from seven regions 
along the slit (Figure \ref{fig:slit_aperture}), including the two nuclei and 5 contiguous 
locations in the extended envelope, one between the two nuclei (ISM 1), three north of 
nucleus N (ISM 2--4) and one south of nucleus IRAS16399S (ISM 5). The extracted spectra are 
shown in Figure \ref{fig:spec_agn_sb_ism}. The spectrum of the IRAS16399N nucleus exhibits 
the same ionic and molecular emission lines as those present in the integrated spectrum of 
the entire system. It shows deeper ice and HAC absorption bands than the integrated spectrum, 
or indeed those of the IRAS16399S nucleus and ISM regions. These features may indicate the 
presence of dense and cold material in the IRAS16399N nucleus. However, the most striking
difference between the IRAS16399N nucleus spectrum and the spectra obtained from the other 
apertures is the steep rise in the continuum to wavelengths $\gtrsim 12\mu$m, which implies 
that an additional component of warm dust  ( $\sim 100$\,K) contributes strongly to the IR 
emission from this region. This rise in the spectrum at longer wavelengths is consistent 
with the composite NICMOS F160W and MIPS 24$\mu$m image presented by \citet{Haan2011}, which 
shows that the IRAS16399N nucleus dominates the 24$\mu$m emission from the merger system.

The mid-IR spectrum (5\,-\,14$\mu$m) of the IRAS16399S nucleus shows strong PAH emission bands
similar to those of the IRAS16399N nucleus. However, the emission lines of [Ne{\sc\,ii]\,} at 12.8$\mu$m,
and H$_2$ at 9.6$\mu$m, as well as the water ice and HAC absorption bands are fainter than  
in the IRAS16399N nucleus. While the ISM 1 spectrum shows prominent PAH bands, but fainter 
ice and HAC absorption bands compared to the IRAS16399N and IRAS16399S nuclei, the ISM 2, 3, 
and 4 extractions show relatively weak PAH emission and no molecular absorption features.

\citet{Draine2001} pointed out that the strengths of the PAH bands at 6.2$\mu$m, 7.7$\mu$m, 8.6$\mu$m 
and 11.3$\mu$m depend on the size and on the charge state of the molecules \citep[see also,][]{Draine2003,Tielens2008,Peeters2002}. 
and proposed a diagnostic diagram based on the PAH emission ratios 6.2$\mu$m/7.7$\mu$m
and 11.3$\mu$m/7.7$\mu$m, which are sensitive to the size and ionization state, respectively,  
of these molecules \citep{Draine2001,Li2001}. \citet{Sales2010} used that diagram to argue 
that AGNs exhibit higher ionization fractions and larger PAH molecules than starburst galaxies
and proposed that PAH ratios can therefore distinguish star-forming galaxies from AGNs. 
It has also been suggested by several authors that the ratio of the [Ne{\sc\,ii]\,}12.8$\mu$m 
and [Ne{\sc\,iii]\,}15.5$\mu$m lines, compared to the relative strengths of PAH bands, can 
be used to classify the mid-IR spectra of galaxies according to the nature of the ionizing 
source \citep[e.g.][]{Smith2007,Sales2010,Baum2010}.

In Figure \ref{fig:pah}, we plot the measured values of the PAH 6.2$\mu$m/7.7$\mu$m and 
11.3$\mu$m/7.7$\mu$m ratios for the extracted spectra as well as the spectrum of the 
IRAS16399-0937 system. We also include data from the sample compiled by \citet{Sales2010} 
representing H{\sc\,ii\,} galaxies, starburst galaxies, LINERs and Seyfert 1 and 2 galaxies. 
The point representing the system spectrum of IRAS16399-0937 falls in the region populated 
with a mix of Seyfert and starburst galaxies \citep{Sales2010}, while the IRAS16399N and 
IRAS16399S nuclei are located in regions occupied by majority of the Seyfert and starburst 
galaxies, respectively. The ISM extractions are distributed widely over the diagram, such 
that the ISM 1, 3, and 4 are present in the region dominated by the Seyfert galaxies and 
ISM 2 and 5 are present in the region of starburst galaxies. 

The ISM extractions that lie in the Seyfert region of the diagram are those from positions 
closest to the IRAS16399N nucleus, either between the two nuclei (ISM1) or to the north
(ISM 3 and 4; Figure \ref{fig:slit_aperture}). It is also worth noting that these ISM points 
have their 6.2$\mu$m PAH emission suppressed (Table \ref{tab:mirlines_ism}) relative to those 
that fall well within the starburst region. As possible explanation is that the small ($\le$50 
carbon atoms) PAHs are destroyed by shocks, with velocities greater than 100\,km\,s$^{-1}$ 
\citep{Micelotta2010a,Micelotta2010b}, caused by the merger interaction. In contrast, the 
ISM positions furthest away (ISM 2 and 5) from the IRAS16399N nucleus -- those located near 
the north and south edges of the common envelope -- fall in the starburst region of the 
diagram \citep{Sales2010,ODowd2009}.

%From extensive study of the PAH ratios measured for a large sample of Seyfert galaxies 
%\citet{Sales2010} claim that the 7.7$\mu$m/11.3$\mu$m $\times$ [Ne{\sc\,ii]\,}12.8$\mu$m/[Ne{\sc\,iii]\,}15.5$\mu$m 
%diagnostic diagram can separate the star-forming galaxies from the AGNs, although no tendency 
%is found using only 7.7$\mu$m/11.3$\mu$m ratio. %\citep[see also discussion in][]{Diamond-Stanic2010,Baum2010,Smith2007}  

In Figure \ref{fig:pah} we plot the 7.7$\mu$m/11.3$\mu$m PAH ratio versus [Ne{\sc\,ii]\,}12.8$\mu$m/[Ne{\sc\,iii]\,}15.5$\mu$m,
again using data from the samples described by \citet{Sales2010}. As the [Ne{\sc\,iii]\,}15.5$\mu$m 
line falls in the IRS LL spectrum, which does not resolve the two nuclei, we plot only one 
point for IRAS16399-0937, that for the integrated spectrum of the system. It can be seen that 
the IRAS16399-0937 merger system falls in the ``transition'' region of the diagram where 
there is most overlap between the  populations of starburst galaxies and AGN.

%Also, we present, in Figure \ref{fig:fork_diagram}, the well-known ``fork diagram'' (EW 
%6.2$\mu$m PAH x S$_{sil}$) proposed by \citet{Spoon2007} with the IRAS16399-0937 data points 
%as well as those OHM and non-OHM galaxies from the sample of \citet{Willett2011a}. To be 
%physically consistent with the methodology applied by these authors we used a spline-fit to 
%measure the EW of 6.2$\mu$m PAH of IRAS16399-0937. Therefore, we defined a local continuum 
%pivots at 5.15, 5.55, 5.95, 6.55, and 7.1$\mu$m afterwards the integrated PAH fluxes were
%divided by the flux densities below the peak of the 6.2$\mu$m PAH band (see Table \ref{tab:mirlines_sns} 
%- \ref{tab:mirlines_sns}).

The well-known ``fork diagram'', which relates the optical depth of the silicate dust feature 
(S$_{sil}$) to the equivalent width of the 6.2$\mu$m PAH feature, provides another diagnostic 
of the relative strengths of AGN and starburst activity in (U)LIRGs \citep{Spoon2007}. 
In this diagram galaxies appear to be systematically distributed along two distinct branches.
The majority of ULIRGs are found along the upper branch which appears to trace obscuration, 
from optically identified starburst nuclei (strong PAH EW; weak silicate absorption) to deeply 
embedded nuclei (weak PAH EW, deep silicate absorption). The lower branch is characterized 
by weak silicate absorption and a range in PAH EW, and is occupied by AGN (low EW) and 
starburst nuclei (high EW).

%{\bf In this diagram galaxies appear to be systematically distributed along two distinct branches, in which 
%one of AGN and starburst-dominated spectra, the other branch extending diagonally from deeply 
%obscured nucleus to an unobscured nuclear starburst. In the light of the ULIRGs evolution
%it is not clear how a target hosting an extremely embedded dust nucleus evolves to AGN and/or
%starburst dominated classes. However, the large majority of the ULIRGs are distributed along 
%the branch of the galaxies containing strong silicate absorption with weak 6.2$\mu$m PAH strengths
%\citep{Spoon2007}.}

In Figure \ref{fig:fork_diagram} we plot these quantities for the IRAS16399-0937 system, 
the two nuclei and the ISM regions, together with data for samples of OHM and non-OHM galaxies 
taken from \citet{Willett2011a}. To be consistent with the measurement methodology used by 
the latter authors we used a spline-fit to measure the equivalent widths of the 6.2$\mu$m 
PAH feature in IRAS16399-0937. Local continuum pivots were defined at 5.15, 5.55, 5.95, 6.55, 
and 7.1$\mu$m and the integrated 6.2$\mu$m PAH flux  was divided by the continuum flux density 
obtained from the spline fit at the peak wavelength of  the PAH band (see Table \ref{tab:mirlines_sns} 
and \ref{tab:mirlines_ism}).

As noted by \citet{Willett2011b}, most OHMGs occupy the upper branch, only a handful being 
found along the lower branch (defined by a wide range of PAH EW but relatively weak silicate 
absorption) which is where most non-OHMG ULIRGs and optically identified AGN are also found.

IRAS16399N is located in the upper branch, among the OHMGs, and consistent with a strong 
obscuration of the nuclear source. However, the nature of the nuclear source is unclear 
from the ``fork diagram''. On the other hand, IRAS16399S 
is located near the ``knee'' (high PAH EW, but low silicate absorption strength), where the 
OHMG and non-OHMG samples overlap. This is also where starburst galaxies tend to be found \citep{Spoon2007}.

\subsubsection{Mid-IR  Photometry}\label{sec:mir_data_points}

The flux in 8$\mu$m Spitzer passband is normally attributed to PAH molecules 
\citep{Leger1984,Allamandola1985,Puget1989}, which are heated by single UV and optical photons 
in the interstellar radiation field (ISRF) of galaxies or near B stars \citep[e.g.][]{Li2002,Haas2002,Peeters2004,Wu2005}. 
The 8$\mu$m emission has been used as an indicator of photodissociation regions (PDRs) 
associated with H{\sc ii} regions \citep[][]{Helou2004,Wu2005} and correlates very well with 
optical and radio tracers of ionizing photons \citep{Wu2005,Calzetti2005}. As can be seen 
in Figure \ref{fig:multi_wave}, the 8$\mu$m image of IRAS16399-0937 shows both nuclei as 
compact sources embedded in an extended envelope. The two nuclei are linked by a ridge of 
emission which crosses the dust lane complex that is prominent at optical wavelengths.
Spurs extend to the west and from the IRAS16399N nucleus to the north-east, the latter 
corresponding to the north-eastern blob observed in shorter-wavelength images. The 8$\mu$m 
emission arising between and to the east of the nuclei is brightest within the region bounded 
(at shorter wavelengths) by the IRAS16399S nucleus and the western arc; it is presumably 
due to dust heated by embedded star formation.

%These large molecules can also be destroyed, 
%fragmented, or ionized in an environment harboring hard radiation field \citep[e.g.][]{Genzel1998,Smith2007,Sales2010}. 

While the 24$\mu$m emission is known to be dominated by interstellar dust, there is an increasing contribution
from stellar photospheric emission at shorter wavelengths. In the IRAC channels, stellar emission dominates
at 3.6$\mu$m,  whereas the strong 7.7$\mu$m PAH feature falls into the 8.0$\mu$m channel, which therefore
predominantly maps ISM emission. The stellar contribution to the 8$\mu$m channel can be approximately
removed by extrapolating the 3.6$\mu$m emission, following the prescription of \citet{Helou2000},
\begin{equation}
F_{(8\mu m,dust)} = F_{(8\mu m)} - 0.232F_{(3.6\mu m)}.
\end{equation}

\begin{figure*}
\centering
\includegraphics[scale=0.45]{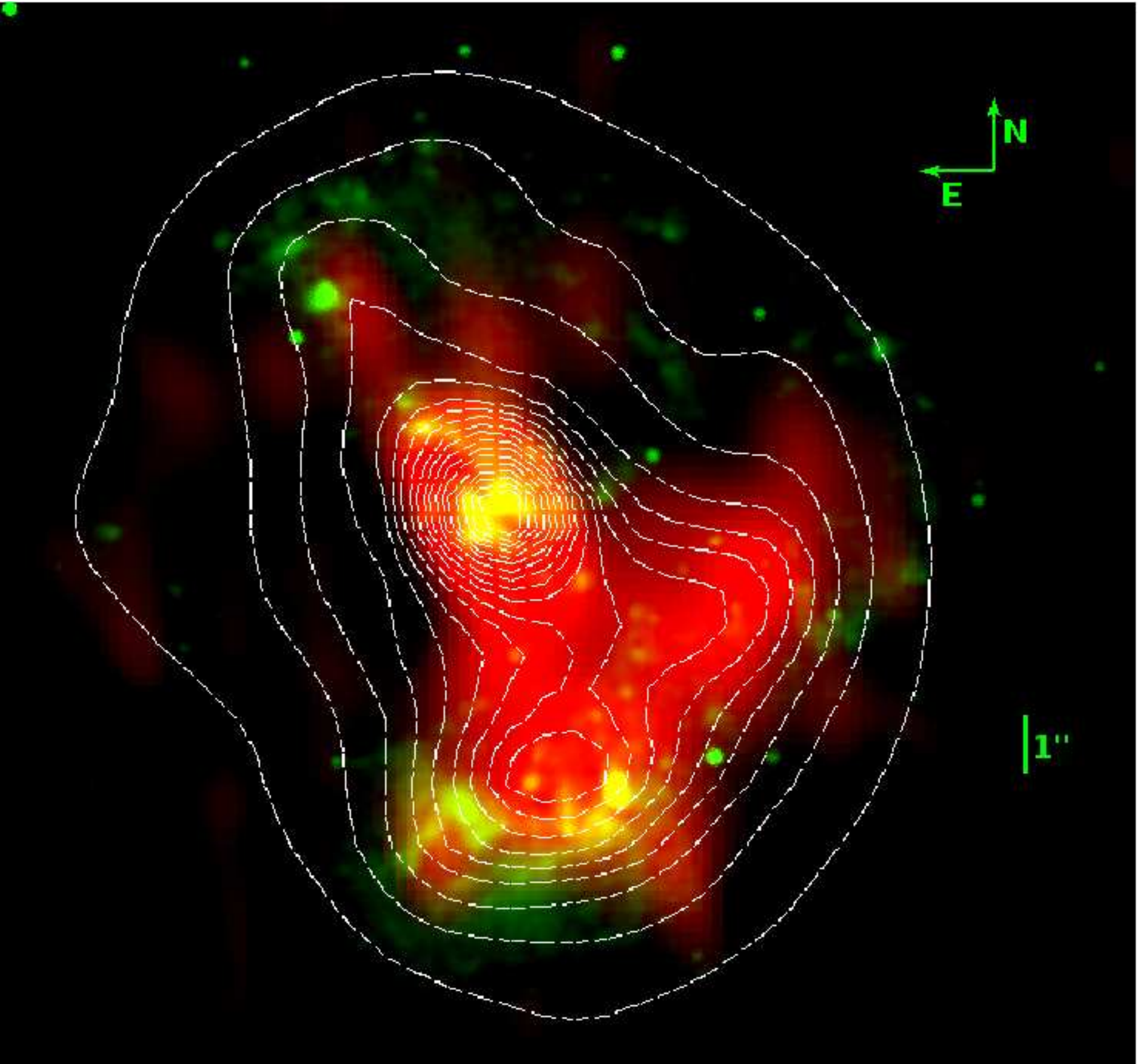}
\caption{Panel shows a composite image of IRAS16399-0937. The green and red channels 
represent the H$\alpha+$[N{\sc ii}] and 1.49\,GHz VLA images, respectively. The contours (dotted lines) show the 
ISM PAH-dust 8$\mu$m emission (contours are linear and stepped in 10\% increments of the peak).}
%The bottom panel shows an expanded view of the HST/ACS FR435W image, centered on the IRAS16399N 
%nucleus and clearly showing the dust lane bisecting it.}
\label{fig:rgb}
\end{figure*}

\begin{figure*}
\centering
\includegraphics[scale=1]{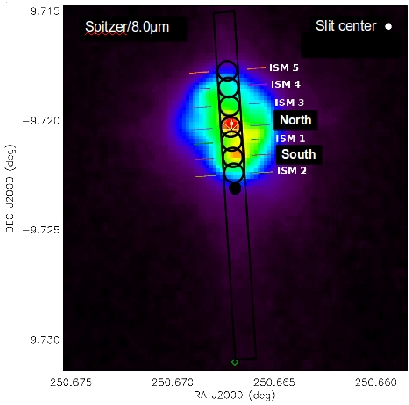}
\caption{Spitzer IRAC 8$\mu$m image showing the apertures used within {\sc smart} to extract 
IRS SL module spectra. The slit position is indicated by solid parallel lines. The {\sc smart} 
apertures are circular, each having a radius of 2.02\arcsec. Two apertures are centered on 
the IRAS16399N (LINER) and IRAS16399S (SB) nuclei, respectively, the other 5  (ISM 1, 2, 3, 4, and 5) 
sample the extended envelope. The corresponding spectra are shown in Fig.~\ref{fig:spec_agn_sb_ism}.}
\label{fig:slit_aperture}
\end{figure*}

\begin{figure*}
\centering
\includegraphics[scale=0.6]{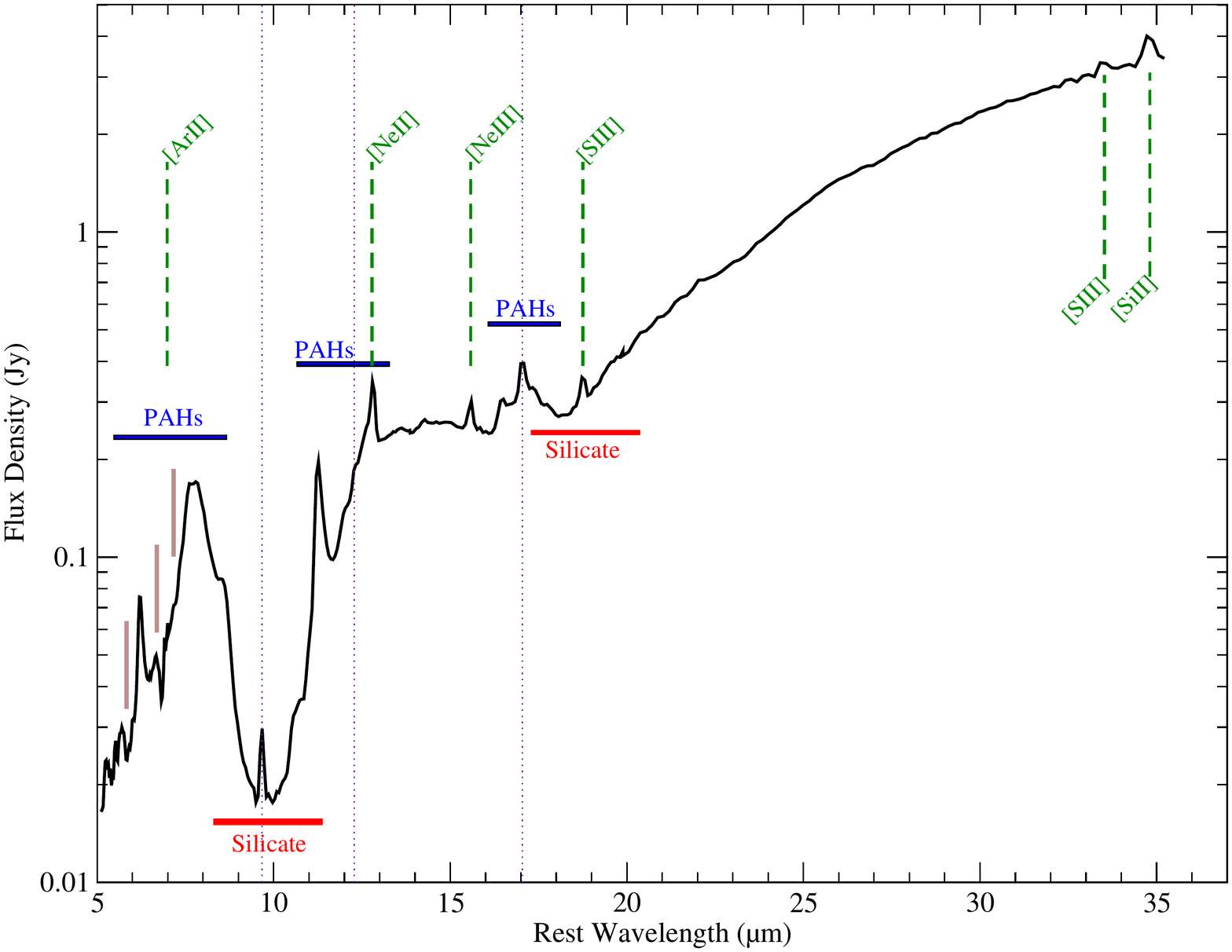}
\caption{Low-resolution Spitzer IRS spectrum of the IRAS16399-0937 system, extracted from a circular aperture of radius 7\arcsec. 
The dotted lines indicate the positions of the H$_2\,S(3)\,9.66\mu$m, H$_2\,S(2)\,12.27\mu$m,
and H$_2\,S(1)\,17.03\mu$m emission lines. The brown solid lines indicate absorption bands
of water ice ($\sim6.0\mu$m) and HACs ($\sim 6.85\mu$m and 7.25$\mu$m). The PAH and silicate
features, as well as various ionic emission lines, are labeled.}
\label{fig:spec_system}
\end{figure*}

\begin{figure*}
\centering
\includegraphics[scale=0.55]{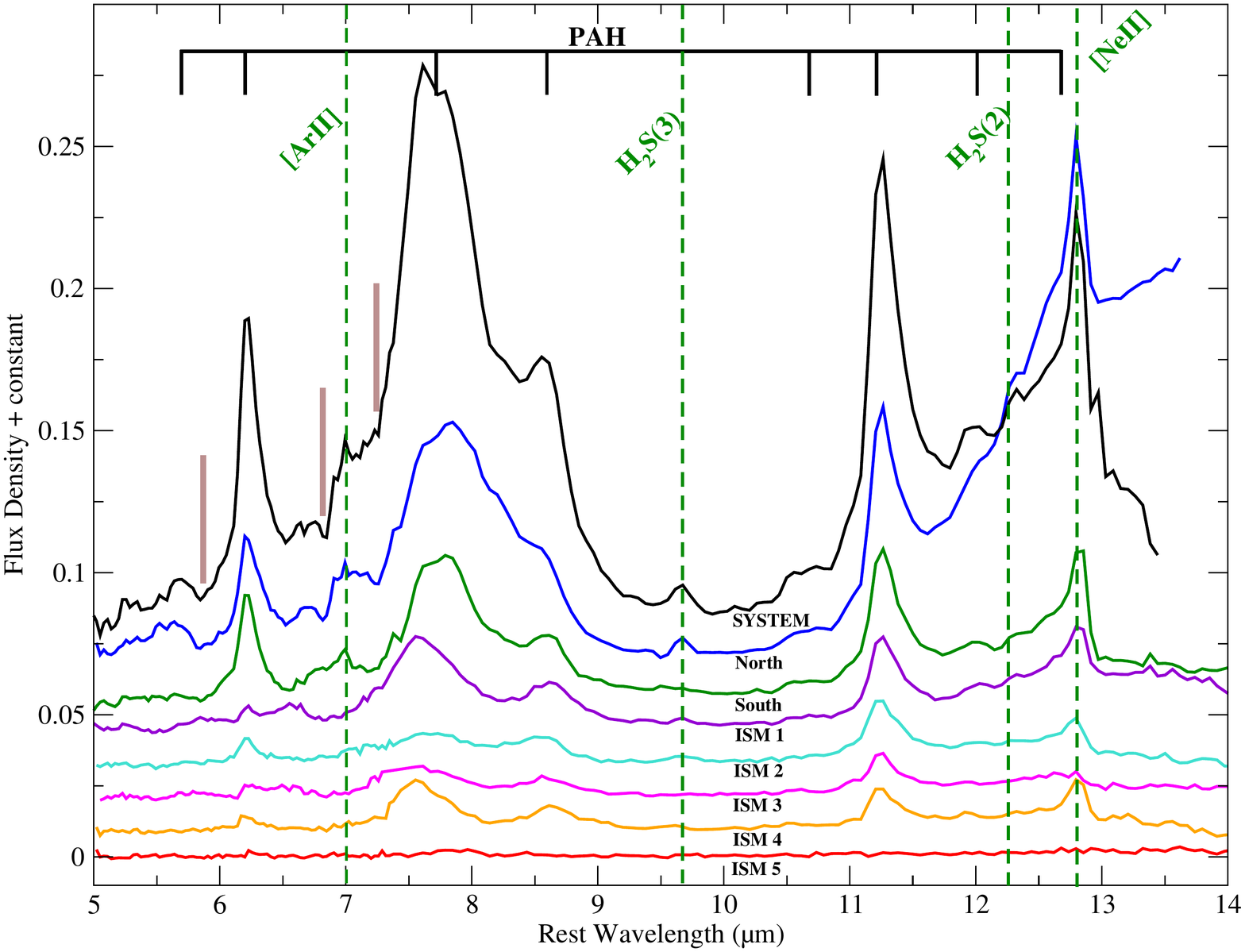}
\caption{Low-resolution (SL module) IRS spectra of IRAS16399-0937. The integrated spectrum 
of the system (labeled {\sc system}), extracted from an aperture of radius 7\arcsec\ is shown, 
along with spectra extracted from the apertures shown in Fig. ~\ref{fig:slit_aperture} 
centered on the IRAS16399N and IRAS16399S nuclei, as well as several locations in the extended 
envelope (ISM 1--5). The PAH bands and emission lines are labeled. The vertical brown solid 
lines show the positions of the water ice ($\sim6.0\mu$m) and HAC ($\sim 6.85\mu$m and 7.25$\mu$m) 
absorption bands.}
\label{fig:spec_agn_sb_ism}
\end{figure*}

\begin{figure*}
\centering
\includegraphics[scale=0.8]{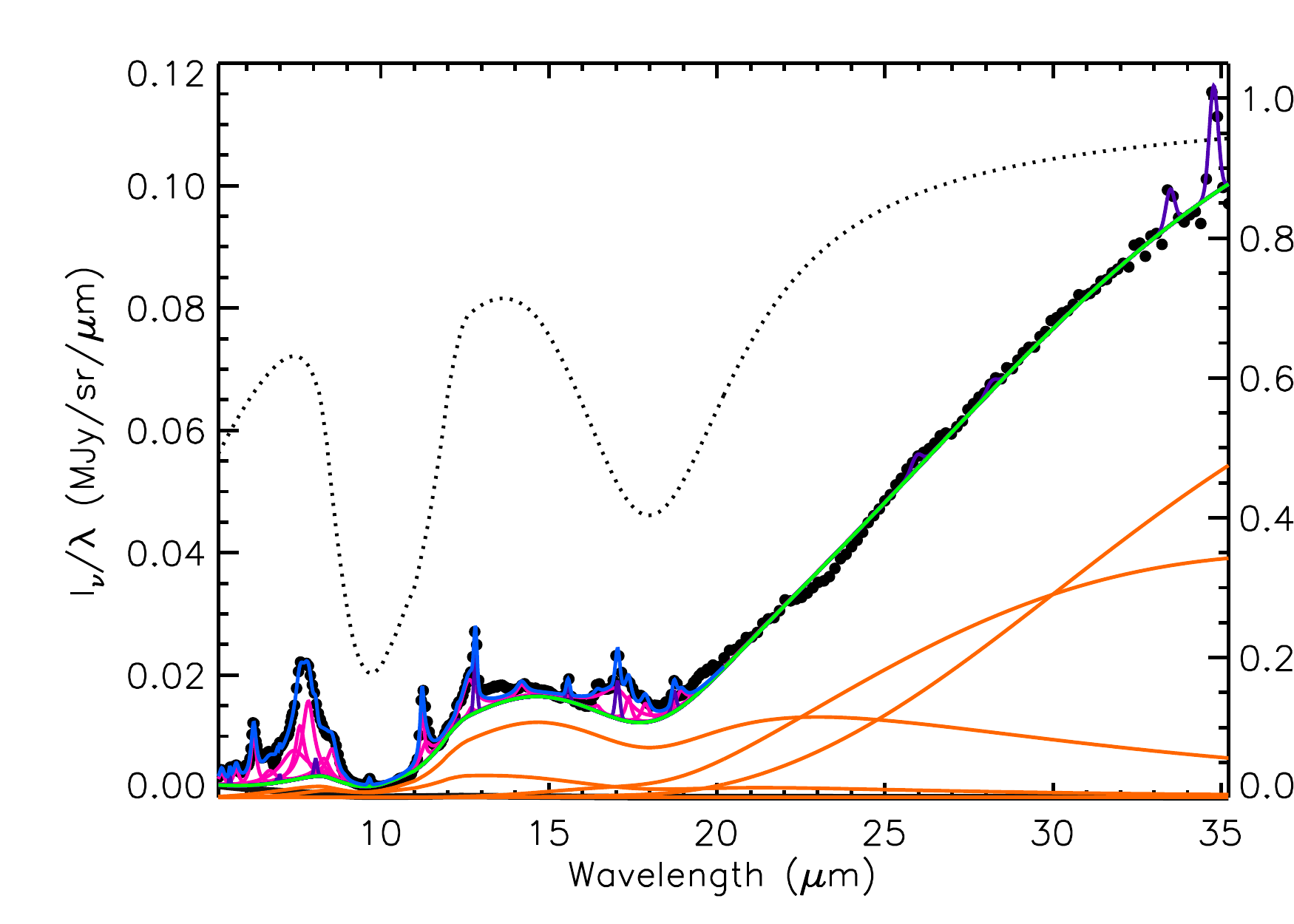}
\caption{Result of the {\sc pahfit} decomposition of the combined (SL+LL) IRS spectrum 
of the whole IRAS16399-0937 system. The data points are represented by filled circles, with 
uncertainties plotted as vertical error-bars that are smaller than the symbol sizes. The best 
fit model is represented by the blue line. The dotted black line indicates the mixed 
extinction components, while solid green and orange lines represent total and individual 
thermal dust continuum components, respectively. The violet lines represent the ionic and
hydrogen molecules, while the magenta lines represent PAH features.}
\label{fig:spectra_decompostion}
\end{figure*}

\begin{figure*}
\centering
\includegraphics[scale=0.4]{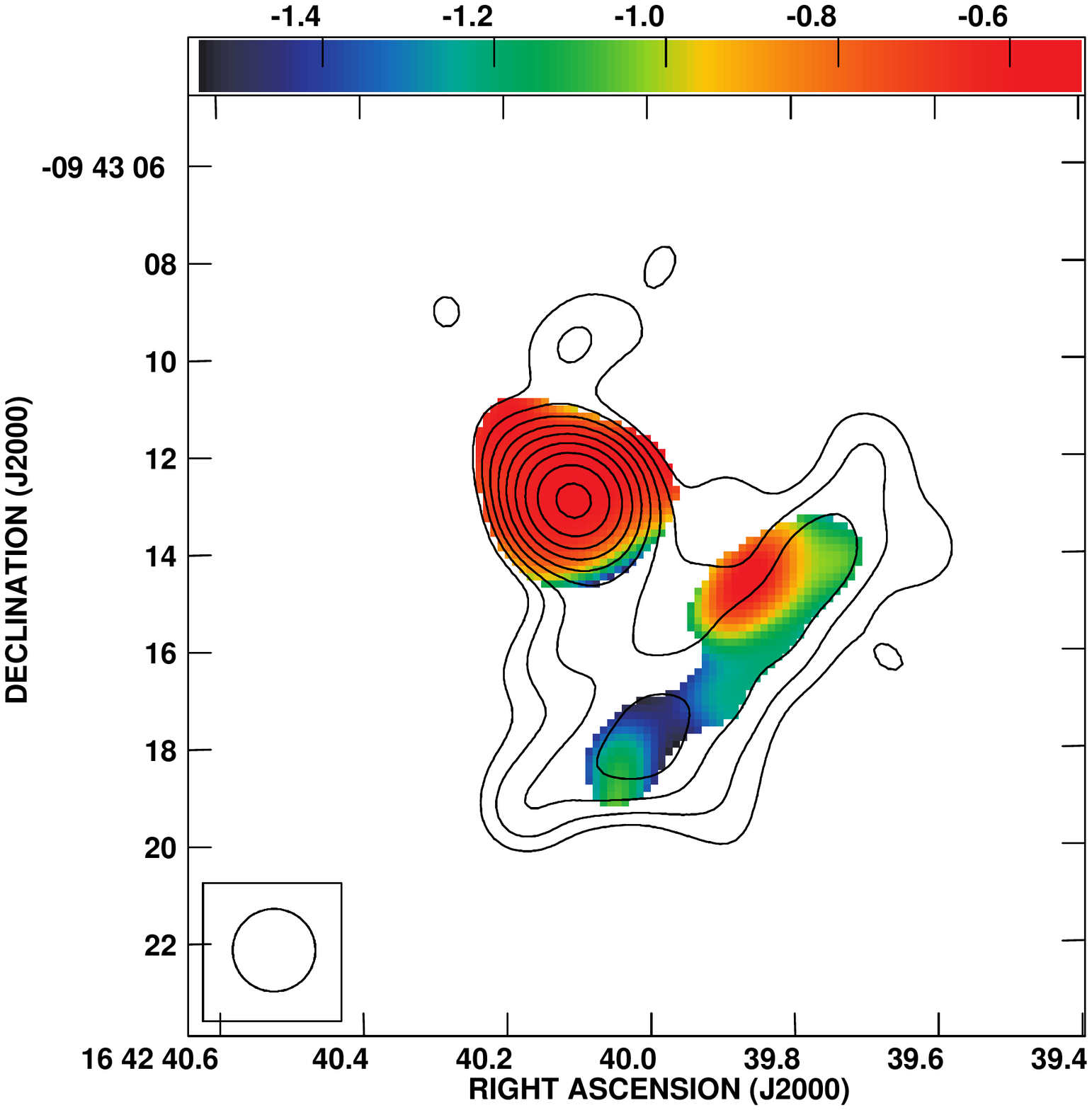}
\caption{Radio spectral index image of IRAS16399$-$0937, derived from 1.49 and 4.9 GHz VLA images (see text).
The 1.49 GHz emission contours are superposed. The contour levels are given in percentage of 
the peak intensity ($I_{peak}$ = 12.9 mJy~beam$^{-1}$) and increase by a factor $\sqrt2$, 
with the lowest contour levels being equal to $\pm$5.6\%.
The convolved beam-size is $1.7\arcsec\times1.7\arcsec$.}
\label{fig:radio_contour}
\end{figure*}

\begin{figure*}
\centering
\includegraphics[scale=0.75]{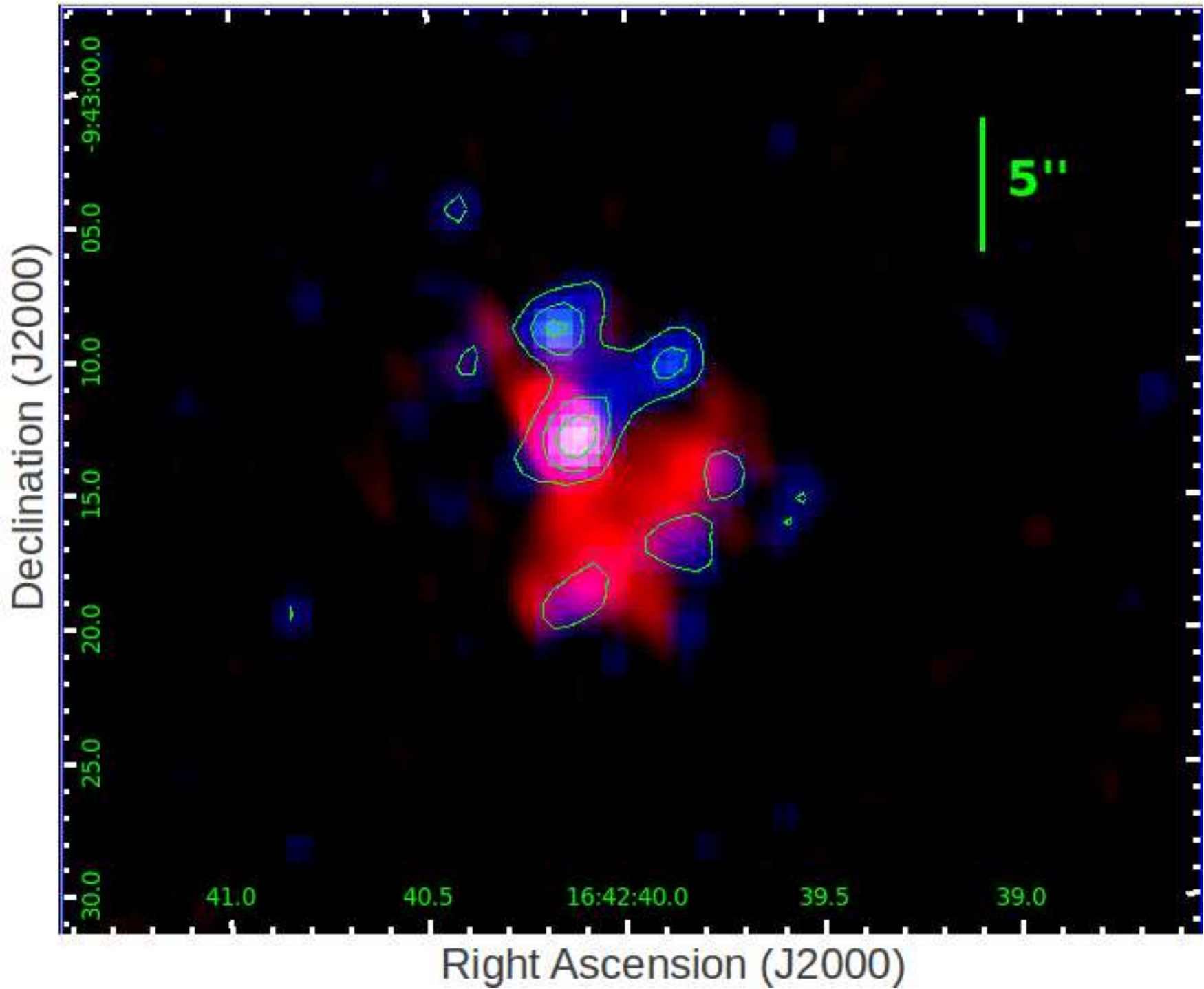}
\caption{Panel shows a composite image of IRAS16399-0937. The red and blue channels represent
the 1.49 GHz VLA and Chandra X-ray, respectively. The contours show the X-ray
emission (contours are linear and stepped in 25\% increments of the peak). The
X-ray image was smoothed with a Gaussian function with $\sigma$ = 3 pixels.}
\label{fig:radio_xray}
\end{figure*}

\begin{figure*}
\centering
\begin{tabular}{c}
\includegraphics[scale=0.32]{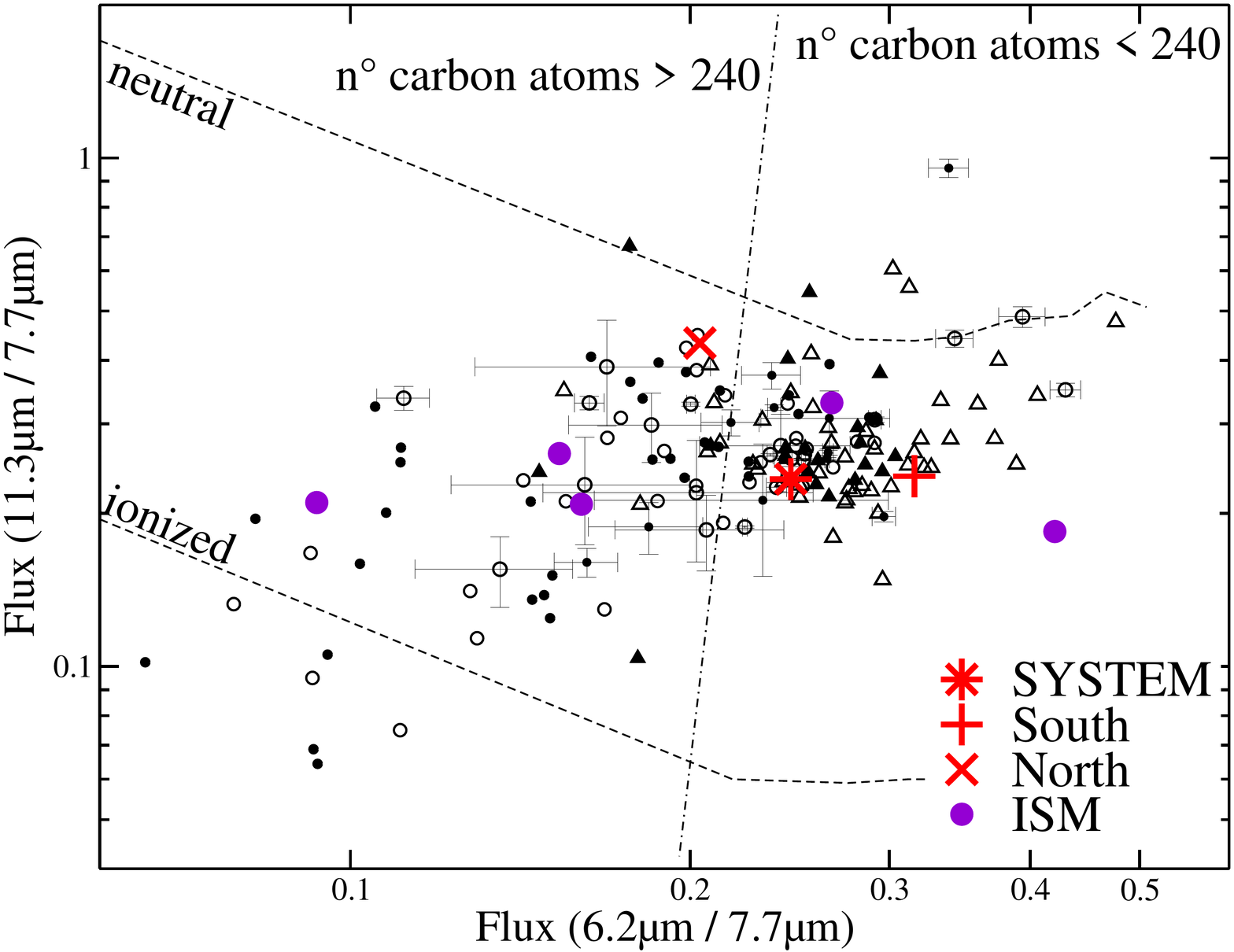}\\
\includegraphics[scale=0.32]{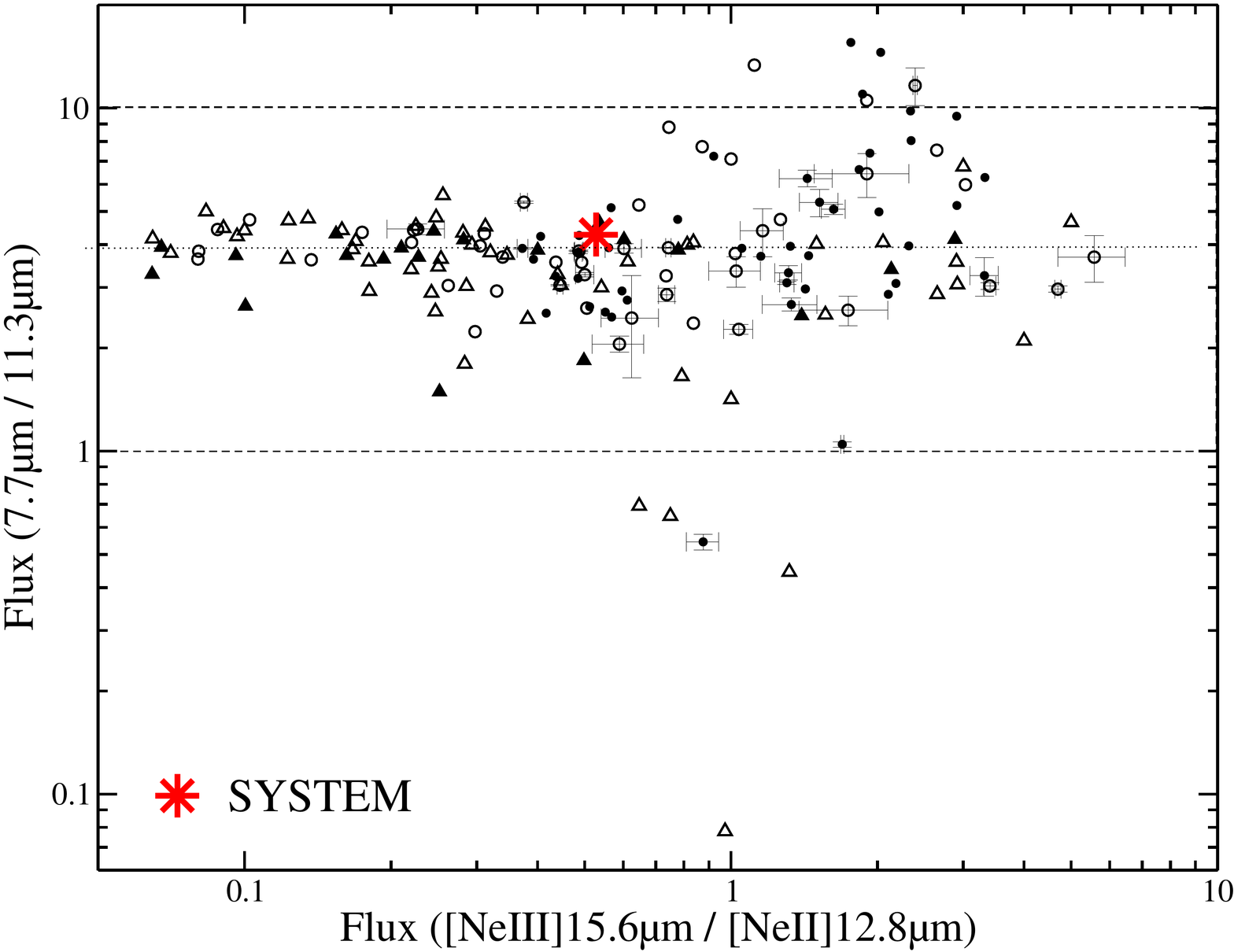}\\
\end{tabular}
\caption{Comparison of IRAS16399-0937 regions with galaxies exhibiting starburst activity or AGN
in mid-IR diagnostic diagrams. Top panel: 6.2$\mu$m/7.7$\mu$m vs. 11.3$\mu$m/7.7$\mu$m PAH flux ratios. 
The dashed lines represent theoretical locii for neutral and ionized PAHs \citep{Draine2001}. 
The dot-dash line shows the locus of molecules formed by 240 carbon atoms. Points representing 
the integrated spectrum of the IRAS16399-0937 system, the IRAS16399N and IRAS16399S nuclei and the ISM sub-regions 
are plotted using the symbols shown in the key. Bottom panel: 7.7$\mu$m/11.3$\mu$m PAH flux 
ratio plotted vs. [Ne{\sc\,iii]\,}15.5/[Ne{\sc\,ii]\,}12.8$\mu$m. This diagram is used as an 
indicator of the hardness of the radiation field \citep[e.g.][]{Smith2007,Sales2010,Baum2010}.
As the [Ne{\sc\,iii]\,}15.5$\mu$m line falls in the IRS LL spectrum, which does not resolve 
the two nuclei, or the ISM sub-regions, only the point representing the integrated spectrum 
of IRAS16399-0937 is plotted in this diagram. In both panels, the galaxy data are taken from 
the sample compiled by \citet{Sales2010} \citep[see also][]{Smith2007,Brandl2006,Gallimore2010}: 
empty triangles represent H{\sc ii} galaxies and LINERs; filled triangles are starburst galaxies; 
filled and empty circles are Seyfert 1 and 2 galaxies, respectively.}
\label{fig:pah}
\end{figure*}

\begin{figure*}
\centering
\includegraphics[scale=0.55]{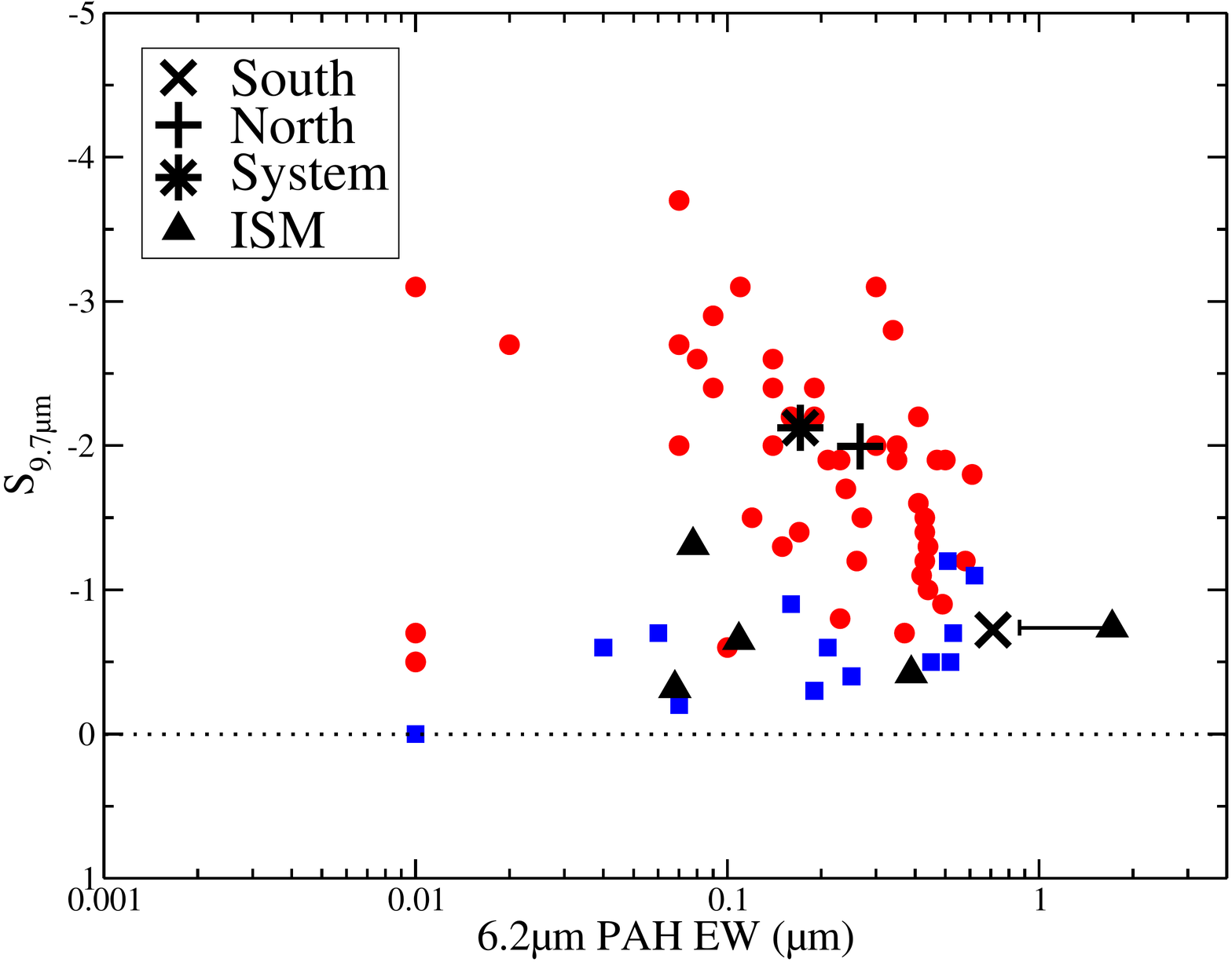}
\caption{Comparison of IRAS16399-0937 regions in ``fork'' diagram of 6.2$\mu$m PAH EW versus 
the silicate optical depth at 9.7$\mu$m. Points representing the integrated spectrum of the 
IRAS16399-0937 system, the IRAS16399N and IRAS16399S nuclei and the ISM sub-regions are 
plotted using the symbols shown in the key. OHMGs from the sample of \citet{Willett2011a} are 
shown in red circles and the non-OHMGs are shown in blue squares. For derived errors see 
Table \ref{tab:mirlines_sns} and \ref{tab:mirlines_ism}.}
\label{fig:fork_diagram}
\end{figure*}

Isophotes of the resulting ISM PAH-dust $8\mu m$ emission are plotted in Figure~\ref{fig:rgb}, 
which also shows the continuum-subtracted H$\alpha$ + [N{\sc ii}] emission and the 1.49 GHz radio image.  

We estimated SFRs from the ISM PAH-dust 8$\mu$m emission, using the relation given by
 \citet{Wu2005},

\begin{equation}
SFR_{(8\mu m, dust)}\,=\,\frac{\nu\,L_{\nu}(8\mu m, dust)}{1.39\,\times\,10^9\,L_{\odot}}~~~~[M_{\odot}\,yr^{-1}].
\end{equation}
The measured values of F$_{(8\mu m,dust)}$ and estimated SFR$_{(8\mu m, dust)}$ for the 
whole IRAS16399-0937 system (using the flux measured within an aperture of radius 7.0\arcsec) 
and both the IRAS16399N and IRAS16399S nuclei (using apertures of radius 2.02\arcsec) are given in Table \ref{tab:fluxes} and \ref{tab:sfr}, respectively.

The continuum emission in the MIPS 24$\mu$m band can be attributed to very small grains
with effective radii in the 15\,-\,40\AA~range. Such grains efficiently convert  
stellar UV ($\sim 6-13.6$\,eV) photons into $\sim24\mu$m continuum emission \citep[see][]{Li2001,Draine2007a} and thus, 
24$\mu$m emission is considered to be a good tracer of star formation \citep{Calzetti2005,Draine2001,Draine2003}.
On the other hand, 24$\mu$m emission can also arise due to thermal emission from large dust grains
heated by AGN \citep{Bendo2006a,Bendo2006b,Smith2007}. 
Although the double nucleus of  IRAS16399-0937 is unresolved in the MIPS 24$\mu$m image,
\citet{Haan2011} inferred that the N component contributes $\approx 90$\% of the 24$\mu$m 
flux, suggesting that the AGN may be dominating
the excitation of small dust grains, in agreement with results obtained by
\citet{Brandl2006}, who claim that if the galaxy hosts a AGN, the dust heated by the AGN 
might dominate the continuum emission at 24$\mu$m \citep{Brandl2006,Diamond-Stanic2010}.

Nevertheless, if we assume that the 24$\mu$m emission is due to star formation rather than AGN heating,
the SFR can be estimated from the correlation proposed by \citet{Rieke2009}, 
\begin{equation} 
SFR_{24\mu m}\,=\,7.8\times10^{-10}\,\nu\,L_{\nu}(24\mu m, L_{\odot})\times[7.76\times10^{-11}\,\nu\,L_{\nu}(24\mu m, L_{\odot})]^{0.048}
~~~~~[M_{\odot}\,yr^{-1}].
\end{equation}
Measuring the 24$\mu$m flux in a circular aperture of radius 7\arcsec\,
(see Table \ref{tab:fluxes}), we find a SFR of 23.18\,M$_{\odot}\,yr^{-1}$ (Table \ref{tab:sfr}).
%to the entire IRAS16399-0937 merger.

While their similar H-band luminosities suggest that the two nuclei also have similar 
stellar masses, the IRAS16399N nucleus is the dominant source of mid-IR dust emission. 
As can be seen in Figure~\ref{fig:rgb}, this is also 
the case for PAH emission: the dominant peak is associated with the IRAS16399N nucleus, with a spur 
approximately coincident with the NE plume. The weaker PAH emission peak associated with 
the IRAS16399S nucleus is offset relative to the H{\sc ii} regions, which border it to the south. PAH 
emission also extends north-west and appears to be associated with the complex of dust lanes 
bounded by the western arc connecting the IRAS16399S nucleus to the NE plume.

\subsection{Radio Source}\label{sec:result_radio}

The radio source at  1.49\,GHz exhibits a morphology very similar to that of the $8\mu$m
PAH-dust emission, with a bright core associated with the IRAS16399N nucleus, which is elongated to
the NE, a secondary peak associated with the IRAS16399S nucleus that coincides with the PAH
emission peak and a knotty extension to the NW overlapping the dust lane complex
(Figures~\ref{fig:multi_wave} and~\ref{fig:rgb}). This striking morphological similarity
strongly suggests that the extended radio emission is due to star formation, rather than
an AGN jet. 

%IRAS16399-0937 exhibits two bright nuclei embedded within a diffuse shell-like structure 
%similar to the multiple structures observed in the Spitzer 8$\mu$m image (see Fig.~\ref{fig:multi_wave}). Since the
%radio and PAH images can be used as an indicator of the star formation, and the extended emission
%in both the images matched the distribution of the H{\sc ii} regions present in the H$\alpha$ 
%image, such regions can be due to photoionization by stars of the host galaxies of the 
%interacting galaxy pair.

Similarly, \citet{Baan2006} classified the nuclear radio source as starburst-dominated
based on consideration of three criteria: spectral index, brightness temperature and the
FIR--radio luminosity ratio ($q$).

The spectral index map (Figure~\ref{fig:radio_contour}) shows that  the bright, compact core
component associated with the IRAS16399N nucleus has a steep spectrum, $\alpha=-0.75\pm0.08$, 
consistent with optically thin synchrotron radiation and characteristic of normal spirals,
extended starbursts and compact starbursts in ULIRGs  \citep{Gioia1982, Condon1983,
Condon1991}. However, steep spectrum radio cores are also commonly observed in Seyfert
galaxies \citep[e.g.,][]{Sadler1995,Roy2000,Kharb2010}, so this does not necessarily
preclude a significant AGN contribution.

We estimated the brightness temperature, ($T_{B}$), of the core component using the relation, 
\begin{equation}
T_{B}=1.8\times10^{9}~(1+z)~(\frac{S_{\nu}}{1~mJy})~(\frac{\nu} {1~GHz})^{-2}~(\frac{\theta_{1}\theta_{2}}{mas^{2}})^{-1}~K, 
\end{equation}
where $z$ is the redshift, $\theta_{1}$ and $\theta_{2}$ are the major and minor axes of the 
source \citep[e.g.,][]{Ulvestad2005}. For a ``deconvolved'' core of size $= 1\farcs1\times0\farcs5$, 
and an integrated flux density = 15.3 mJy as derived from the AIPS task JMFIT, $T_B\sim2.3\times10^4$~K
at 1.49 GHz. For the Southern nucleus, $T_{B}\sim878$~K,
for an  integrated flux density of 6.8 mJy and a ``deconvolved'' core size of $2\farcs9\times2\farcs2$.
At 1.49 GHz, the maximum brightness temperature for supernovae synchrotron emission in a 
starburst is $T_{B} < 10^{5}$\,K \citep[see][equation~8]{Condon1991}, which is consistent % not sure this is useful! T_B for HII regions?
with the value derived for the IRAS16399S nucleus. In contrast,
radio core sources in Seyferts, even those exhibiting steep spectra, typically have brightness temperatures in the range
$10^5\lesssim T_{B }\lesssim 10^7$\,K \citep{Kukula1999,Giroletti2009}. 
% modified per JG comments 7/30/14 - AR
However,
we note that in nearby Seyferts, high brightness temperatures  ($T_{B} \gtrsim 10^{5}$\,K)  are observed only in radio knots and cores;
the extended radio emission typically having a much lower brightness temperature. At the distance of IRAS $16399-0937$, the high and low
brightness temperature regions would be unresolved within the core, leading to a measured brightness temperature below the range quoted above.
Therefore, while the brightness temperature of
the core radio component is therefore consistent with star formation,  we cannot exclude the possibility
of an AGN contribution

The third criterion used by \citeauthor{Baan2006} is based on the well-established
correlation between the FIR and radio luminosites; for a given FIR luminosity,
galaxies hosting AGN radio sources are expected to have a higher radio luminosity than
pure starbursts and hence a lower value of $q$ \citep[e.g.,][]{Condon1991,Yun2001}.
\citeauthor{Baan2006} found that $q\approx 3$ for IRAS16399$-$0937, placing it firmly in
the starburst regime. However, whereas the radio core has a size $\sim 1$\arcsec, the IRAS
FIR fluxes used to determine $q$ are characteristic of the entire $\sim 15$\arcsec\ system
(including both nuclei and the envelope), which is unresolved in IRAS data. It is not clear what
fraction of the FIR luminosity arises from the IRAS16399N nucleus alone, but it is evident from the
H$\alpha$ + [N{\sc ii}] and the ISM PAH-dust $8\mu m$ images that star formation is
widespread throughout the system, with the IRAS16399N nucleus contributing, for example, only
$\approx 20$\% of the ISM PAH-dust $8\mu m$ emission (Table \ref{tab:fluxes}). The values of $q$
estimated by \citeauthor{Baan2006} should therefore be regarded as a high upper limit for
the radio core in the IRAS16399N nucleus.

Unfortunately, as the IRAS16399N nucleus cannot be isolated in MIPS images, it is not possible to
determine a more representative value of $q$ using Spitzer data.

In summary, therefore, the clearest clue to the origin of the compact radio core
associated with the IRAS16399N nucleus is provided by the brightness temperature, which suggests
that it is dominated by a compact starburst rather than a radio-bright AGN.

Assuming that the 1.49 GHz radio emission is entirely due to star formation activity, we estimate SFRs for the IRAS16399N and IRAS16399S nuclei
using  equation~27 of \citet{Condon2002}
\begin{equation}
SFR_{1.49GHz}(M_{\odot}\,yr^{-1}) \approx \frac{L_{1.49GHz}(W\,Hz^{-1})}{4.6\times10^{21}},
\end{equation}
 These turn out to be 
SFR$_{1.49GHz}$ = 6.04\,M$_\sun$~yr$^{-1}$ for the IRAS16399N nucleus and SFR$_{1.49GHz}$ = 2.68\,M$_\sun$~yr$^{-1}$ for the IRAS16399S nucleus, 
respectively (Table \ref{tab:sfr}).
%For comparison, we also derived SFRs using relationship given by \citet{Rieke2009}
%\begin{equation}
%SFR_{1.49GHz}(M_{\odot}\,yr^{-1}) = 1.13\times10^{-4}L(1.49GHz, L_{\odot}).
%\end{equation}

%In this case, we find SFR$_{1.49GHz}$ = 12.12\,M$_\sun$~yr$^{-1}$, and SFR$_{1.49GHz}$ = 5.39\,M$_\sun$~yr$^{-1}$ for the IRAS16399N and IRAS16399S nuclei, 
%respectively (Table \ref{tab:sfr}).
% Averaging the two values for each nucleus, we infer 
%$\langle$SFR$_{1.49GHz}$$\rangle$ = 5.5\,M$_\sun$~yr$^{-1}$
%for the IRAS16399N nucleus and  $\langle$SFR$_{1.49GHz}$$\rangle$ = 1.8\,M$_\sun$~yr$^{-1}$ for the IRAS16399S nucleus.

\subsection{X-ray Source}\label{sec:result_xray}

The peak of the compact (unresolved) X-ray source is clearly associated with
IRAS16399N (Figure \ref{fig:radio_xray}). However, there are also fainter
knots of X-ray emission located approximately 4'' to the N and NE of IRAS16399N, which border the
radio source and appear to be associated with the amorphous region of B-band and H$\alpha$+[N{\sc ii}] 
emission. Yet fainter blobs of X-ray emission are associated with IRAS16399S and extend 
along the western arc. 

The luminosity of the compact source associated with IRAS16399N is 
$\sim 4.69\times 10^{40}$\,erg\,s$^{-1}$, much lower than is typical for Seyfert 1 nuclei 
(which have $42\lesssim \log{L_{2-10\,keV} }\lesssim 44$; e.g., \citealt{Georgantopoulos2010,Jin2012})
but comparable with nearby low luminosity AGN \citep{Ho2009}.

% This repeats part of the first paragraph
%\sout{The IRAS16399S nucleus reveals a faint X-ray emission with extended
%string of H{\sc ii} regions along the ``western arc''. Surprisingly, we can also see two
%faint knots associated with the H$\alpha$ emission in the direction of the northwest of the
%IRAS16399N nucleus that seems to be bordering the radio emission.}

\citet{Ranalli2003} argue that star formation accounts for the most of the hard
X-ray emission of low luminous AGNs. Based on X-ray and IR data of a sample of local 
star-forming galaxies, these authors derived an empirical relation between SFR and X-ray 
luminosity emission,
\begin{equation}
SFR^{0.5-2keV}_{X-ray}\,(M_{\odot}\,yr^{-1}) = 2.2 \times 10^{-40} L^{0.5-2keV}_{X-ray}
%SFR_{2-10keV}(M_{\odot}\,yr^{-1}) = 2.0 \times 10^{-40} L_{2-10keV}.
\end{equation}

Assuming that the X-ray emission from IRAS16399-0937 is produced entirely by star formation, 
we find SFR$^{0.5-2keV}_{X-ray}$ = 10.32\,M$_\sun$~yr$^{-1}$ for the IRAS16399N 
nucleus and SFR$_{X-ray}^{0.5-2keV}$ = 2.46\,M$_\sun$~yr$^{-1}$ for the IRAS16399S nucleus, 
respectively (Table \ref{tab:sfr}).

\section{Spectral Energy Distribution Modeling}\label{sec:modeling}

The flux density measurements summarized in Table~\ref{tab:fluxes} were combined with the
extracted IRS SL and LL spectrum to construct spectral energy distributions for the
IRAS16399-0937 system as well as both nuclei. Flux densities obtained from 7\arcsec\ (or
36\arcsec\ in the case of the Herschel data) apertures were used to form the $0.4-500\mu$m
SED for the system as whole. SED's covering the range $0.4-14\mu$m were also constructed
for each nucleus, using measurements from 2\arcsec\ apertures and the corresponding
extractions from the IRS SL spectrum. The SEDs of both nuclei were simultaneously fit as
described below using a new Markov Chain Monte Carlo code. The nuclei are not resolved in 
the IRS LL spectrum, or in the MIPS or Herschel data, i.e., for wavelengths $> 14\mu$m. These 
data were included in the SED fits but treated as upper limits (see Appendix~\ref{sec:sedfitting} for details).

%We have used the {\sc pahfit} code to fit and subtract from the Spitzer IRAS16399-0937 spectrum
%the contribution of the ionic emission lines, as well as hydrogen molecular lines. Their 
%fluxes and EWs can be seen in Table \ref{tab:mirlines_sns}. This method was adopted in 
%order to fit the multiwavelength IRAS16399-0937 SED describe bellow.

\subsection{SED Model Components}\label{sec:sed-model-components}

We decomposed the infrared spectra and SEDs %(hereafter, simply SEDs)
using a customized version of our code clumpyDREAM (Gallimore, in
preparation). clumpyDREAM makes the usual assumption that the infrared SEDs
of galaxies can be decomposed into a sum of contributions from stars
(photospheric emission), interstellar dust and PAHs (the ISM
component), and reradiation from the dusty torus surrounding an AGN
\citep{Walcher2011,Genzel2000}. The code also allows for contributions
from very hot ($\sim 1500$~K), presumably carbonaceous dust grains and
AGN-heated dust in an NLR. However, the hot dust component is found
only in QSOs and Type 1 AGNs, and the NLR contribution is required
only to fit silicate emission profiles \citep[cf][]{Mor2009}. Since 
IRAS16399$-$0937 is neither Type 1 nor shows silicate in emission, the hot 
dust and NLR components are not included in the SED fits.

We used GRASIL \citep{Silva1998} models of elliptical galaxies to
fit the stellar contribution. These models simulate a short ($\sim
1$~Gyr) burst of star-formation followed by passive evolution, and the
model grids span a range of metallicity and age of the stellar
population. Since stars contribute significantly only to our
near-infrared and optical broadband ($< 5.0 \mu$m) measurements, there
is insufficient information to constrain the metallicity of the
stellar population. 
% JG text added 30/7/14 - AR 
The metallicity of ULIRG nuclei varies, with some evidence for 
sub-solar oxygen abundance and super-solar neon abundance; on average
a typical ULIRG probably has slightly super-solar abundance 
\citep[$Z ~\approx 1.1 Z_{\odot}$;][]{Clemens2008,Veilleux2009,Rupke2008}. Therefore, we fixed the metallicity to solar, which should
provide a reasonable approximation, but allowed the age of the stellar population to vary.

% references to be added
% Clemens, M.S., Vega, O., Bressan, A., Granato, G.L., Silva, L., Panuzzo, P., 2008, A&A, 477, 95
% Veilleux, S. et al. 2009 ApJS, 182, 628
% Rupke, D. S. N., Veilleux, S., & Baker, A. J. 2008, ApJ, 674, 172). 

We used  the models of \citet[hereafter, DL07]{Draine2007a} to fit the ISM % citation should be Draine & Li (2007).
component. The primary strengths of the DL07 models are (1) they have
been successfully employed to model the infrared SED of galaxies with
a range of star-forming properties, including nearby quiescent and
star-forming spiral galaxies \citep[][]{Draine2007b}, 
cold dust galaxies \citep{Willmer2009}, and more distant, luminous 
star-forming galaxies \citep{Magdis2012}; and (2) they permit a range 
of PAH feature strengths by tuning the
PAH mass fraction. The models consist of a population of dust grains
with common composition and size distribution from \citet{Weingartner2001}. 
The grains are heated by a scaled interstellar
radiation field (ISRF), $u_{\nu} = U u_{\nu, {\rm solar}}$, where
$u_{\nu}$ is the energy density of the radiation field, $U$ is a
dimensionless scaling factor, and $u_{\nu, {\rm solar}}$ is the ISRF
for the solar neighborhood \citep{Mathis1983}. Two heating
environments are considered, (1) the diffuse ISM, in which grains are
heated primarily by the diffuse ISRF, and (2) dust associated with
star-forming regions, where the heating is dominated by direct
starlight. The diffuse ISM is heated by a fixed energy density
radiation field, $U = u_{\min}$, where $u_{\min}$ is a free parameter
of the model.  Star-forming regions are modeled by considering a power
law distribution for $U$: $dM/dU \propto U^{-2}$ over the range $U =
u_{\rm min}$ to $U = 10^6$, where $M$ is the dust mass.

In the DL07 models, PAHs are modeled by adopting cross-sections (per C
atom) and effective particle sizes chosen to mimic the emission
observed in the nuclear regions of SINGS galaxies \citep{Smith2007}. 
The strength of the PAH spectral features relative to
continuum is determined in part by the parameter $q_{PAH}$, the
fraction of the total dust mass that is in PAHs. In practice, average
PAH compositions and ionization states will vary from galaxy to galaxy \citep{Smith2007,Galliano2008},
and so we expect {\em a priori} imperfect
fits to the PAH features in detail for any particular galaxy
SED. Nonetheless, examination of residual, misfit PAH features provides
a means to compare individual galaxy PAH features to an
effective SINGS mean PAH spectrum \citep{Smith2007}.

In sum, the spectral shapes of the ISM components are determined by two
free parameters: $u_{\rm min}$ and $q_{\rm PAH}$. For simplicity and
per the recommendations of DL07, we adopt the same values of $u_{\rm
min}$ and $q_{\rm PAH}$ for both the diffuse ISM and star-forming
regions. The diffuse ISM and star-forming regions are otherwise
treated as separate spectral components for the purpose of determining
the best fit. In the present analysis, different values of $u_{\rm
min}$ and $q_{\rm PAH}$ were permitted for each nucleus of
IRAS16399-0937. 
% JG text added 30/7/14 - AR 
The  relative fraction of the SFR component provides a measure of the
dust luminosity that arises from SFRs; we use the prescription of
DL07 (their eq. 29) and calculate $f_{PDR}$, the fraction of the dust
luminosity that comes from photodissociation regions with $U > 100.$

The AGN torus component was fit using the clumpy torus models of
\citet{Nenkova2008a,Nenkova2008b}\footnote{More information is available at https://www.clumpy.org/.}. 
In these models, the AGN, represented
by a broken power law SED, illuminates a pc-scale distribution of
dusty clouds, which then re-radiate at infrared wavelengths. The number of
obscuring clouds along a given sight-line is given by,
\begin{equation}\label{eq:n0}
N = N_0 \exp\left(-(90^{\circ} - i)^2/\sigma^2 \right), 
\end{equation}
where $i$ is the inclination of the observer view relative to the torus axis, $\sigma$ is the angular
scale height, and $N_0$ is the average number of clouds expected along
the equatorial ray of the torus; note that the geometry is essentially a
flared disc with a surface blurred by a Gaussian like distribution of
clouds. For simplicity, each cloud is assumed to have the same
optical depth, parameterized by $\tau_V$.

The radial extent of the torus is set by sublimation at the inner edge
and the parameter $Y = r_{\rm out} / r_{\rm in}$. From \citet{Nenkova2008a}, 
the sublimation radius was set to 0.4~pc $\left(L_{\rm AGN} / 
10^{45}\right)^{1/2}$~ergs~s$^{-1}$, the radius where the adopted grain mixture
and grain size distribution would survive. The radial distribution of
clouds follows a power law, 
\begin{equation}
N_C \propto \left(r / r_{\rm in}\right)^q ,
\end{equation}
where $N_C$ is the number of clouds per unit length, and the
normalization is determined by Eq.~\ref{eq:n0}. 
% replaced with JG text 7/30/14 - AR
\citet{Nenkova2008a,Nenkova2008b} show that the
mass of hydrogen gas for a clumpy torus model is $M_{torus} = m_H N_H
\int N_C dV$, where $N_H$ is the hydrogen column density through a
single cloud. The column density $N_H$ was converted from $\tau_V$ using the
\citet{Bohlin1978} conversion factor for $R_V = 3.1$. Nenkova et
al. give simple formulas for a sharp-edged distribution of clouds and
integer values of $q$. Since we are instead using a gaussian
distribution of clouds and non-integer values of $q$, we instead
calculate the hydrogen mass by numerical integration.

To summarize, the shape of the torus SED is determined by six
parameters: $(\sigma, Y, N_0, q, \tau_V, i)$. The AGN luminosity
determines the normalization.

Finally, clumpyDREAM also allows for partially covering, intrinsic
foreground extinction. For this purpose, 
we used the isothermal turbulent foreground dust screen model of
\citet{Fischera2005}. For simplicity
we used the $R_V = 3.1$ Milky Way extinction model, which provides
reasonable fits to the silicate absorption profile and extinction of the
stellar component. The extinction and covering fraction of the
stellar, ISM, and AGN components are fitted independently, and the AGN
covering fraction due to the foreground extinction is taken to be unity. We present detailed description of 
our applied methodology to fit the IRAS16399-0937 SED in the Appendix \ref{sec:sedfitting}.

\subsection{SED Modeling Results}

%The AGN unified model suggests that the central engine is surrounded by the predicted dusty,
%optically thick, torus geometry. This dusty structure is believed to re-emit the absorbed 
%radiation into the MIR SED and the torus dimensions and geometry, the density distribution 
%and the dust grain properties can be well constrained from the silicate feature pick around 
%9.7$\mu$m and 18$\mu$m \citep[e.g.,][]{Krolik1988,Antonucci1993,Pier1992,Nenkova2002,Fritz2006,
%Ramos-Almeida2009,Sales2011}. For our purposes we present a primary run of the IRAS16399-0937 
%SED including a AGN component through the clump torus models of \citet[][see Section 
%\ref{sec:sed-model-components}]{Nenkova2008a,Nenkova2008b} only for the North nucleus. 

The mid-IR continuum of AGN (with the exception of blazars) is believed to be dominated by
thermal dust emission from the torus, the dust grains being heated by absorption of
UV-optical radiation emitted by the accretion disk. In merging systems like
IRAS16399-0937, the AGN, if present, may have been recently triggered by inflows of
molecular gas, leading to enhanced SMBH accretion rates. However, it is unclear how the
formation of the dusty torus structure is related to the triggering of AGN activity.
Perhaps the molecular gas inflows  form the torus, which then acts as a fuel reservoir for
the accretion disk. However, it has also been suggested that the torus is formed by a
hydromagnetic accretion disk wind \citep{Elitzur2006}, in which case the onset of AGN
activity would precede torus formation. As already noted (Sec.~\ref{sec:intro}) there is
evidence in some objects that the OHMs are distributed in a disk-like structure that may
be associated with the torus. There is no evidence that such a structure has formed in
IRAS16399-0937, whose OHM have not been mapped at high spatial resolution. Nevertheless,
for our purposes, the \citet{Nenkova2008a,Nenkova2008b} clumpy torus model is sufficiently
general that it can represent a range of configurations for the distribution of AGN heated
dust clouds, ranging from a thin disk  to quasi-spherical.

The evidence that the IRAS16399N nucleus of IRAS16399-0937 contains an AGN is ambiguous. The
optical spectrum is characteristic of LINERs, suggesting that it is a low luminosity AGN.
On the other hand, IRAS16399-0937 lacks the high ionization mid-IR emission lines that are
believed to be AGN signatures, while the properties of the compact radio source associated
with the IRAS16399N nucleus are consistent with a stellar origin, rather than an AGN. Nevertheless,
the steep rise in the mid-IR continuum suggests excess warm dust emission that is not
present in the IRAS16399S nucleus.

clumpyDREAM was used to perform simultaneous fits to the SED's of the IRAS16399N and IRAS16399S 
nuclei. Two sets of fits were run utilizing different models for the IRAS16399N nucleus in 
order to test for the presence of an AGN contribution. The two nuclei have to be fit 
simultaneously to accommodate data taken in large apertures that encompass both nuclei. We 
need to ensure during the fit that the sum of the model SEDs of the two nuclei does not exceed 
the upper limits in large apertures.

In the first set, the SED model for the IRAS16399N nucleus includes the clumpy torus
component (that is, the AGN heated dust component) in addition to the stellar and ISM
components. In the second, the torus component is turned off and only the latter two
components contribute to the SED. The SED model for the IRAS16399S nucleus includes only the
stellar and ISM components in both cases.

\begin{figure*}
\centering
\includegraphics[scale=0.45]{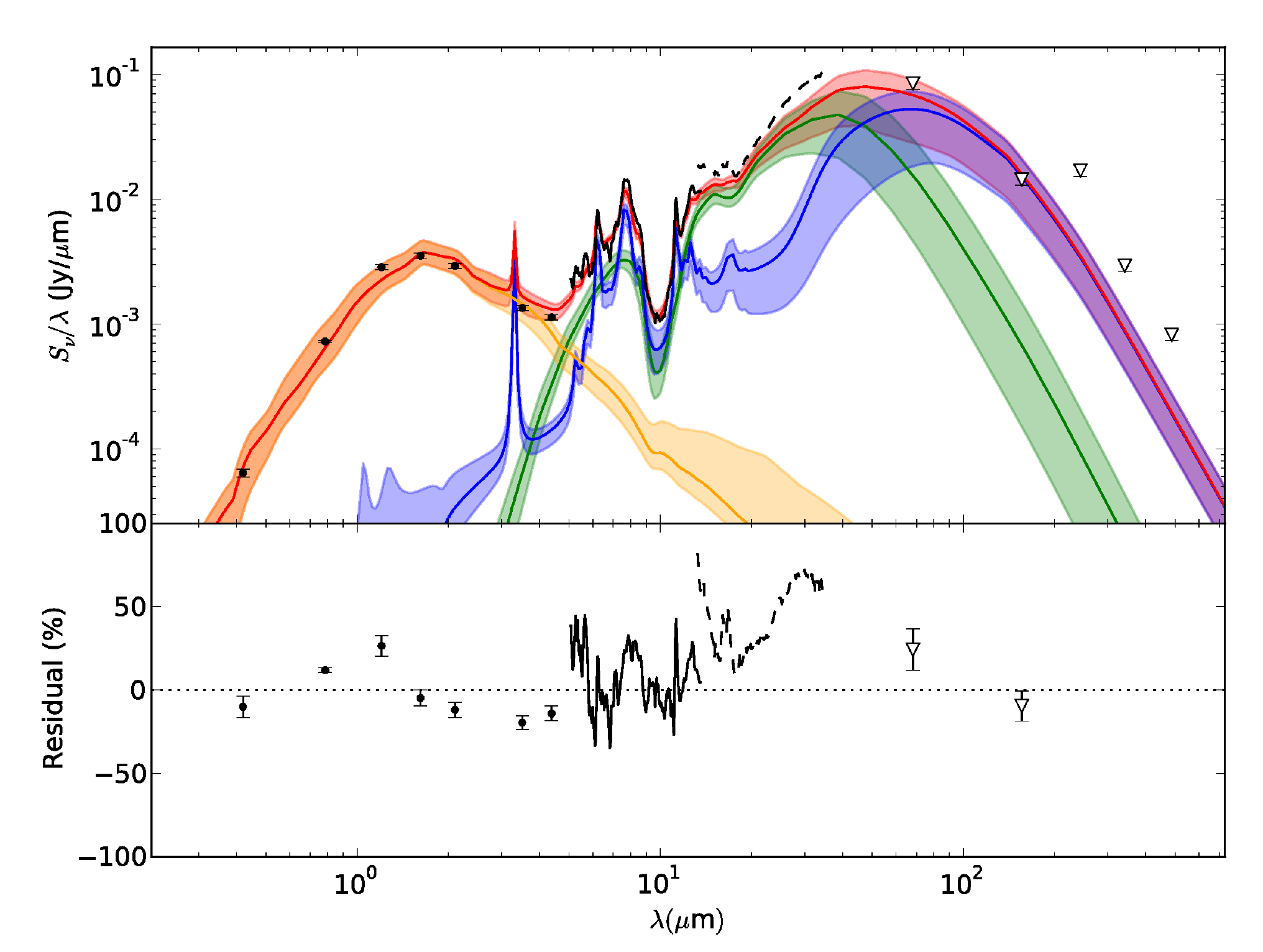}\\
\includegraphics[scale=0.45]{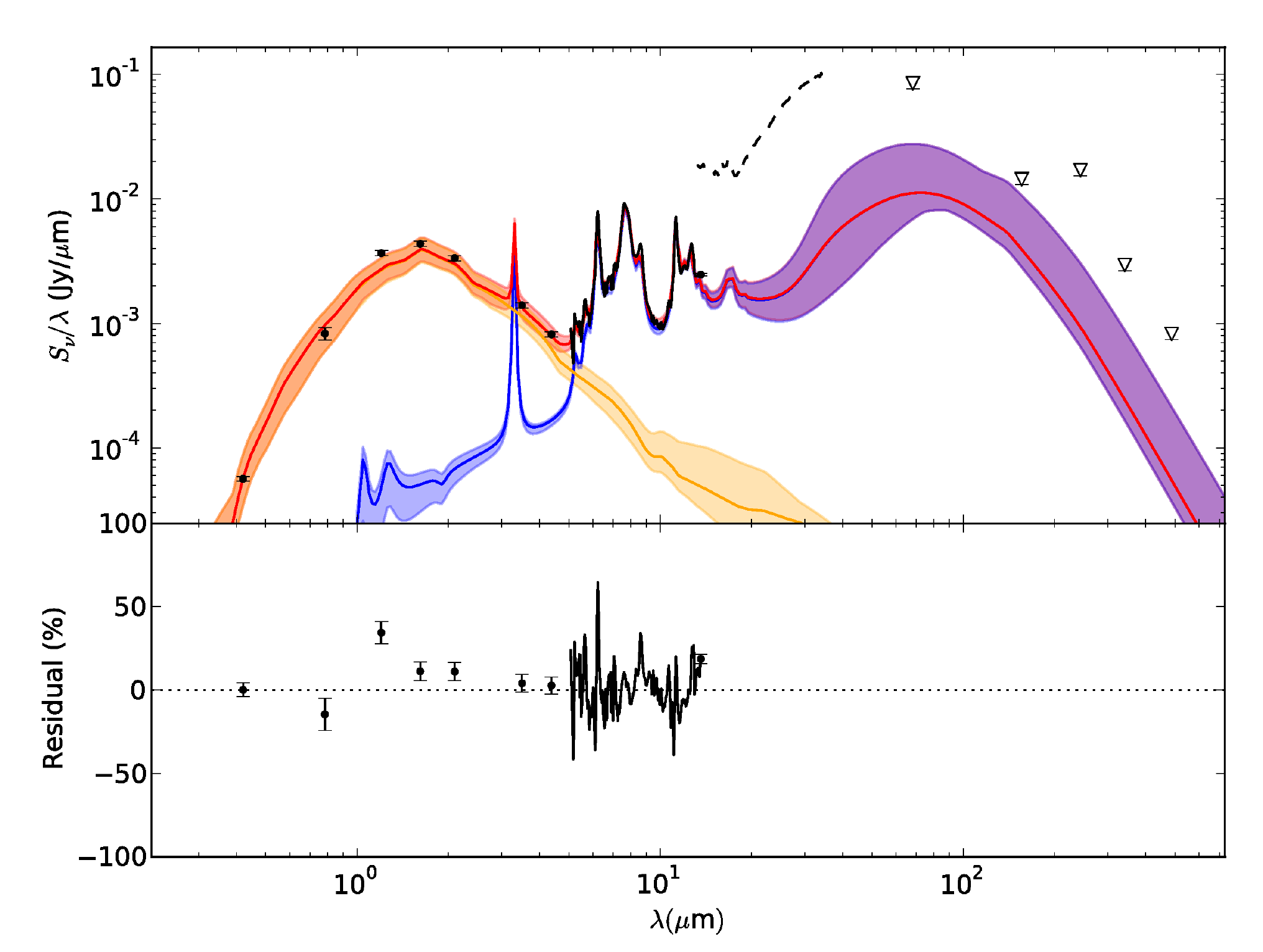}
\caption{Fits to the $0.4 -  500\mu$m spectral energy distributions of the North (top) and South (bottom) nuclei of IRAS16399-0937.
The symbols with error bars represent photometry from HST, 2MASS, Spitzer and Herschel (see Table~\ref{tab:fluxes}); the black lines
represent IRS SL (solid) and IRS LL (dashed) spectra. The overall fit, shown by the red band, is the sum of 
three components: the GRASIL elliptical galaxy models \citep[orange band,][]{Silva1998}, the ISM dust + PAH 
models of DL07 (blue band) and the clumpy AGN torus model of \citet{Nenkova2008a,Nenkova2008b} (green band). 
The fits to both nuclei were carried out simultaneously using the  clumpyDREAM code (see text).  
The width of each band represents, at a given wavelength, the full range of flux density permitted by the converged fit.
In this model, it is assumed that the IRAS16399N nucleus contains an AGN, whereas the IRAS16399S nucleus is a pure starburst. Therefore,
the fit to the IRAS16399N nucleus (top panel) includes all three components, whereas the  fit to
the IRAS16399S nucleus (bottom panel) includes only the stellar and ISM components. 
The two nuclei are not resolved in the LL IRS spectrum, or in the Spitzer MIPS and Herschel SPIRE images. 
Therefore the LL spectrum (dashed line) and photometry data points for $\lambda \ge 30\mu$m (inverted triangles) 
refer to large aperture (7\arcsec) measurements of IRAS16399-0937 system and were treated as upper limits in the fits.}\label{fig:sed_agn}
\end{figure*}

\begin{figure*}
\centering
\includegraphics[scale=0.6]{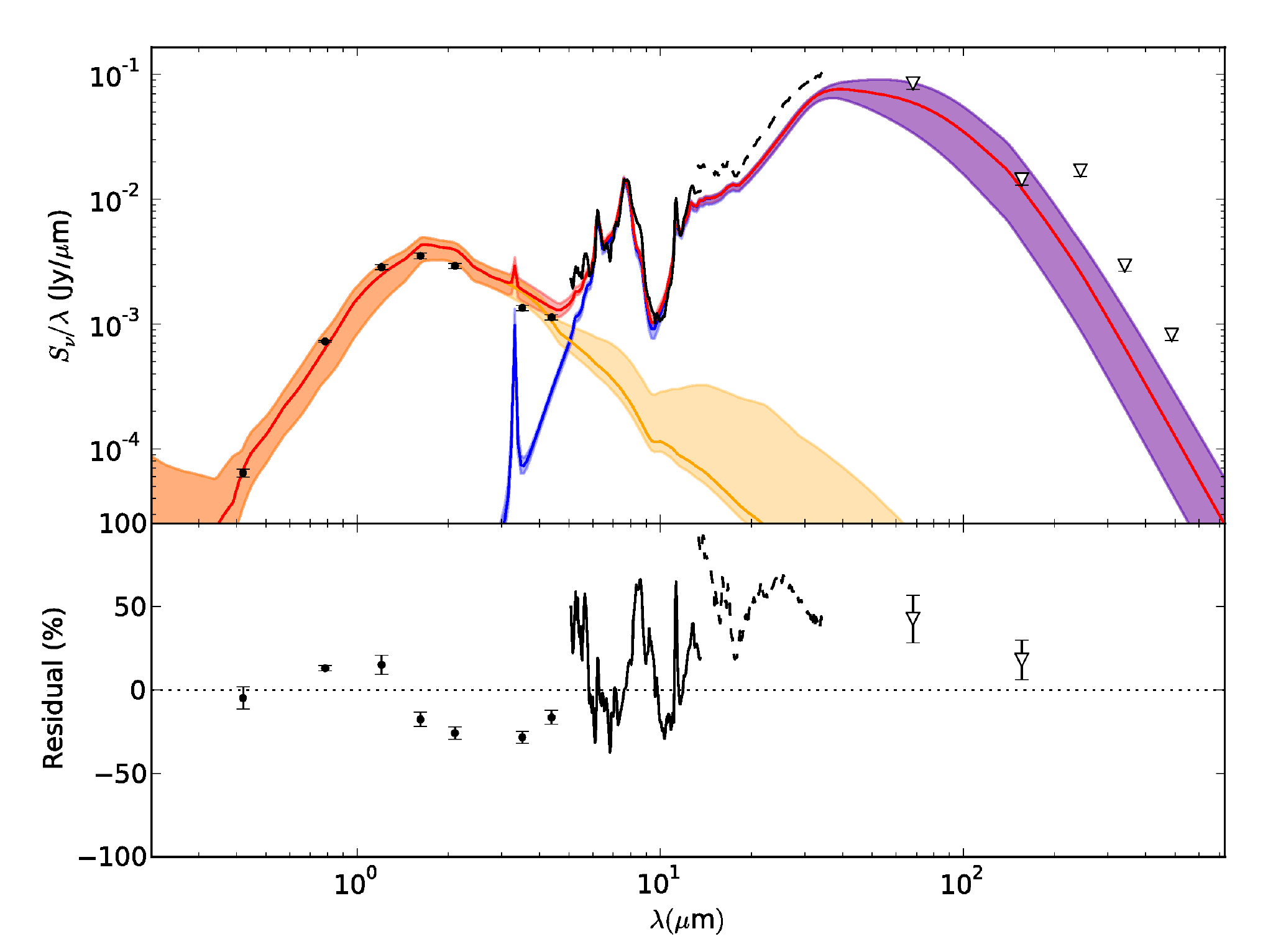}
\includegraphics[scale=0.6]{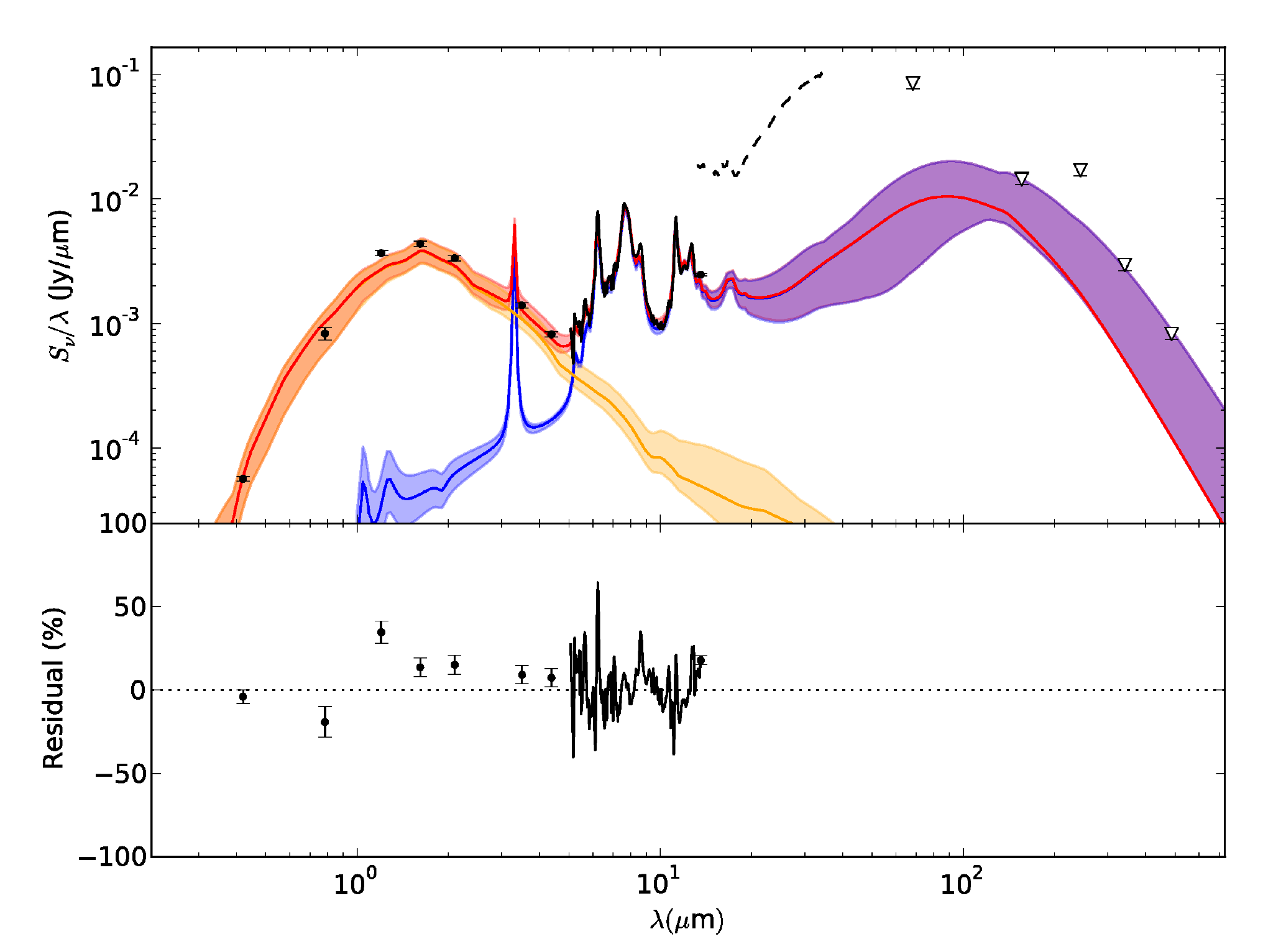}
\caption{As Fig.~\ref{fig:sed_agn} for the case in which the SED fits to both the IRAS16399N (top) 
and IRAS16399S (bottom) nuclei include only stellar (orange band) and ISM (blue band) components.}
\label{fig:sed_sb}
\end{figure*}

\begin{figure*}
\centering
\includegraphics[scale=0.55]{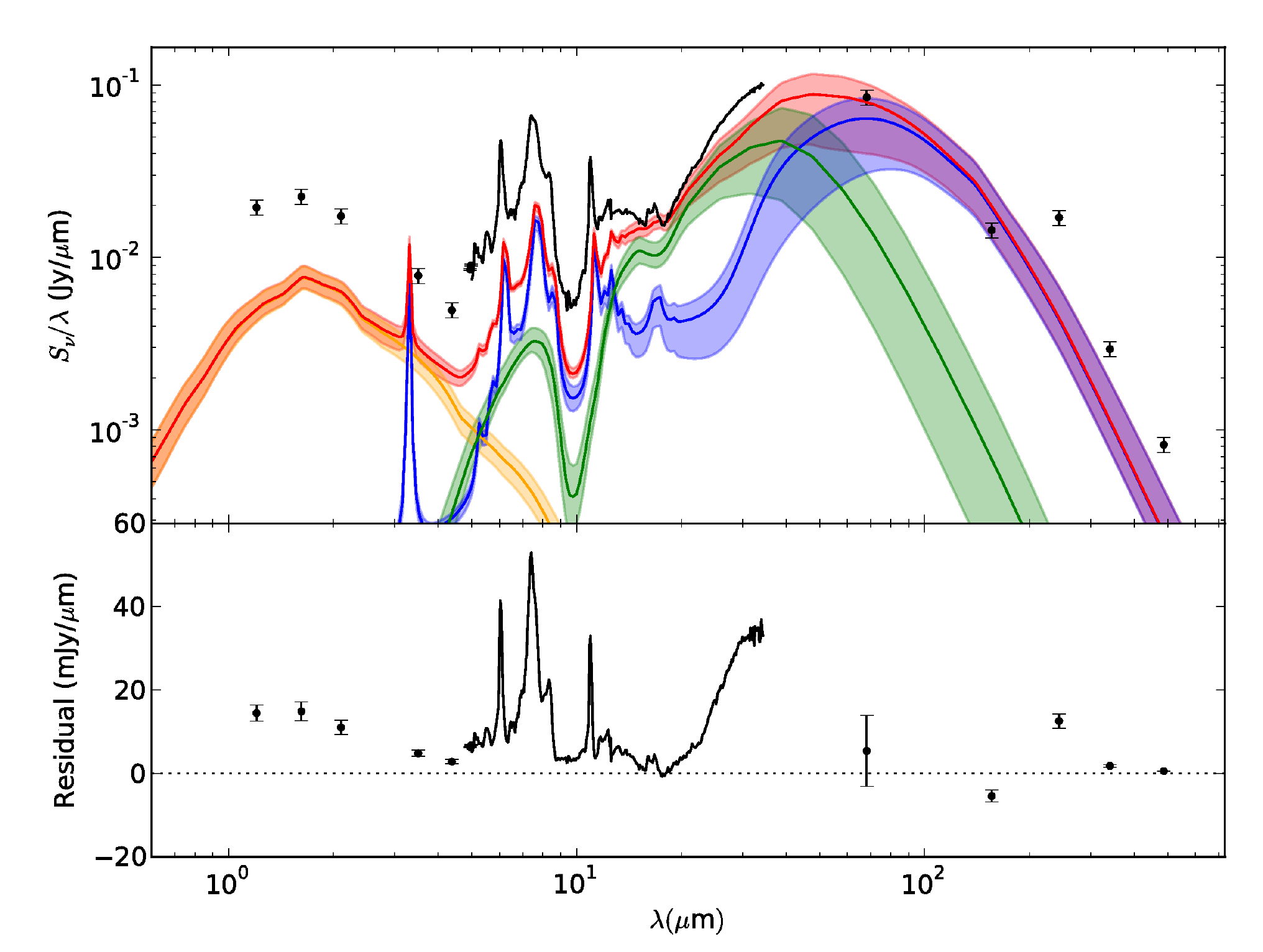}
\includegraphics[scale=0.55]{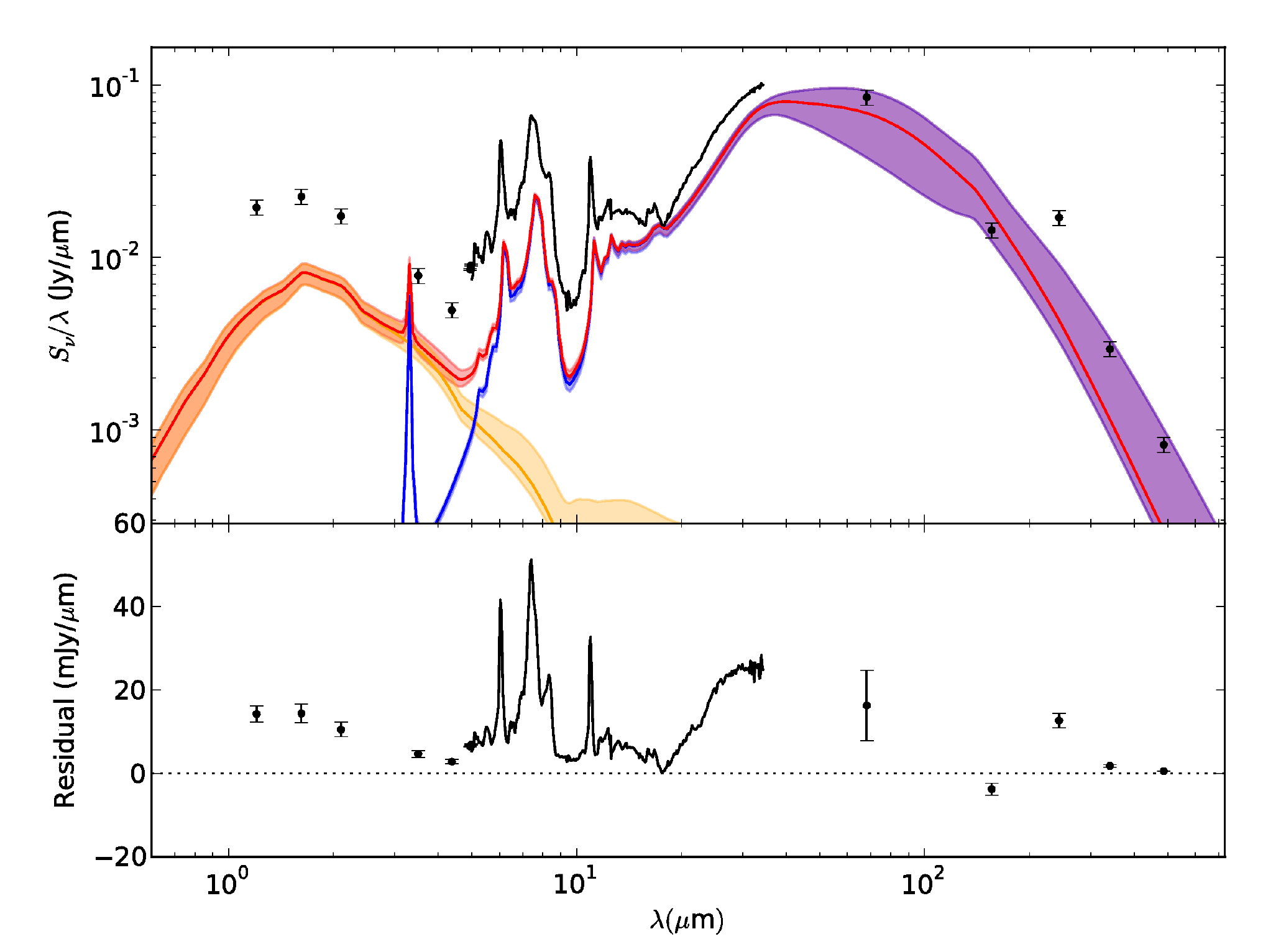}
\caption{Comparison of the summed SED fits of Figs~\ref{fig:sed_agn} (top) and~\ref{fig:sed_sb} 
(bottom) to the SED of the IRAS16399-0937 system obtained from large aperture (7\arcsec) 
photometry (inverted triangles) and IRS SL and LL spectroscopy (dashed line). The summed 
small aperture (2.02\arcsec) fluxes of the IRAS16399N and IRAS16399S nuclei are also plotted 
for wavelengths $<14\mu$m, where the nuclei are resolved (filled points and solid line). 
The top panel shows the sum of the SED models fitted to the IRAS16399N and IRAS16399S nuclei, 
for the case in which an AGN torus is included in the fit to the IRAS16399N nucleus. The 
bottom panel shows the summed SED fit for the case in which both IRAS16399N and IRAS16399S 
nuclei were fitted using only stellar and ISM components. Note that the sum of the SED models 
does not exceed the upper limits imposed by the large aperture measurements.}
\label{fig:sed_system}
\end{figure*}

The results of the fits are illustrated in Figure~\ref{fig:sed_agn} for the model that
includes a torus contribution to the IRAS16399N nucleus and Figure~\ref{fig:sed_sb}, for the model
which excludes the torus component. In both figures, the fits to the IRAS16399N and IRAS16399S nuclei are
shown in the upper and lower panel, respectively. As a consistency check the sum of the
fitted SED models (i.e., the sum of the fits to each nucleus) is compared to the SED of
the IRAS16399-0937 system derived from  large apertures (Table~\ref{tab:fluxes}), for both
sets of fits (Figure \ref{fig:sed_system}). 
% replaced with JG text 7/30/14 - AR
The combined model SED does
not exceed the observed system SED by more than $\sim 10$\%, comparable to the
accuracy of the SED models and the absolute calibration accuracy
for the far-infrared data points.
% JG - (Are these PACS points? I think the
% absolute photometric accuracy is ~ 5%, although color corrections
% worsen the accuracy.)
% The combined model SED does not exceed the
% observed system SED, which represents an upper limit to the flux densities.

Estimators of the probability densities of the fitted parameters for both nuclei in both SED models are
shown in Figures \ref{fig:torus_agn1} to \ref{fig:torus_sb2}.  The median, 5$^{th}$ and
95$^{th}$ percentile values of the parameters are listed in Tables \ref{tab:linerfittorus}
and \ref{tab:linersbfit}, along with the derived star formation rates.

\subsubsection{AGN and starburst nucleus}\label{sec:torus_model}

\begin{figure*}
\centering
\begin{tabular}{c}
\includegraphics[scale=0.7]{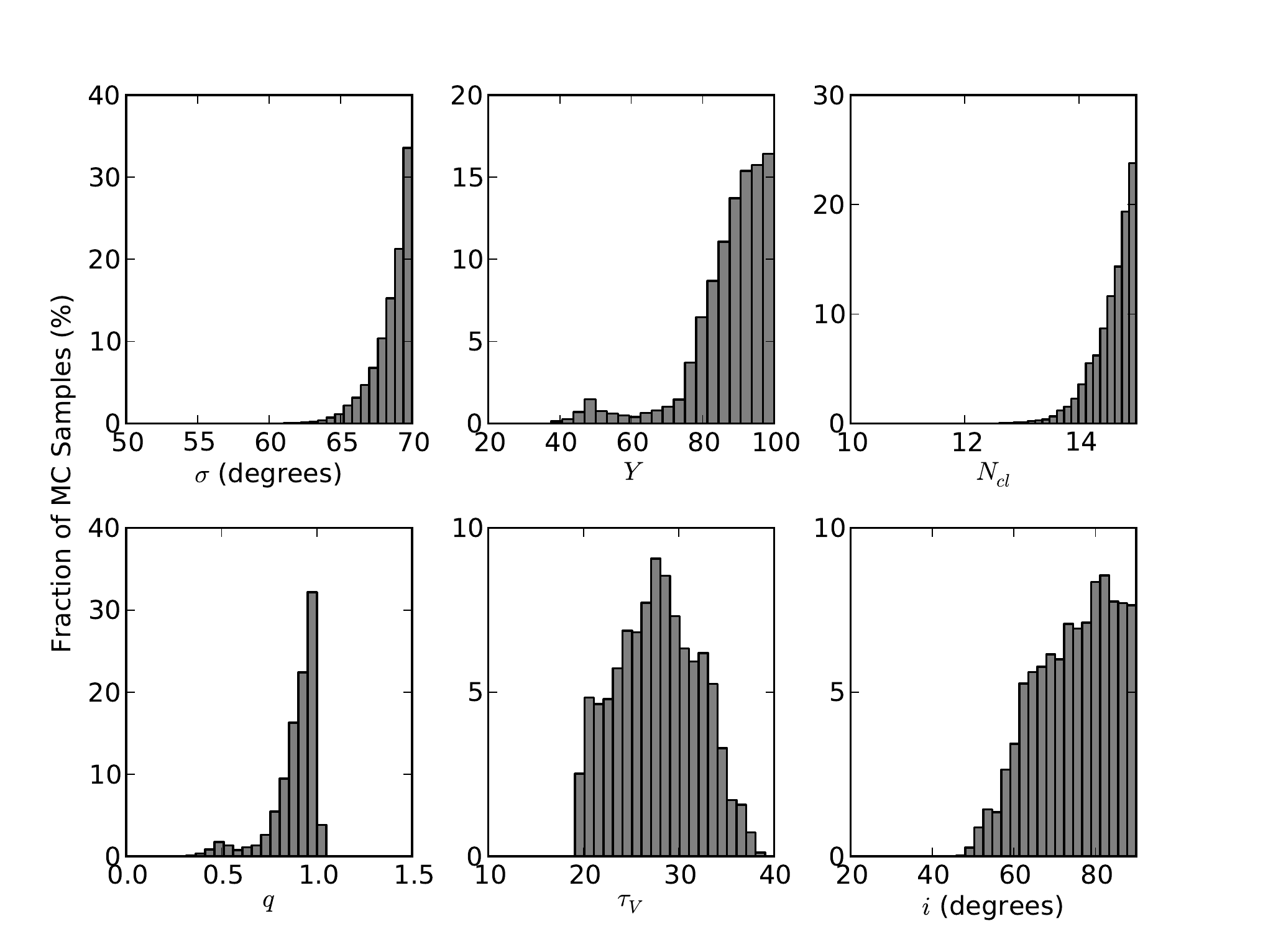}\\
\includegraphics[scale=0.65]{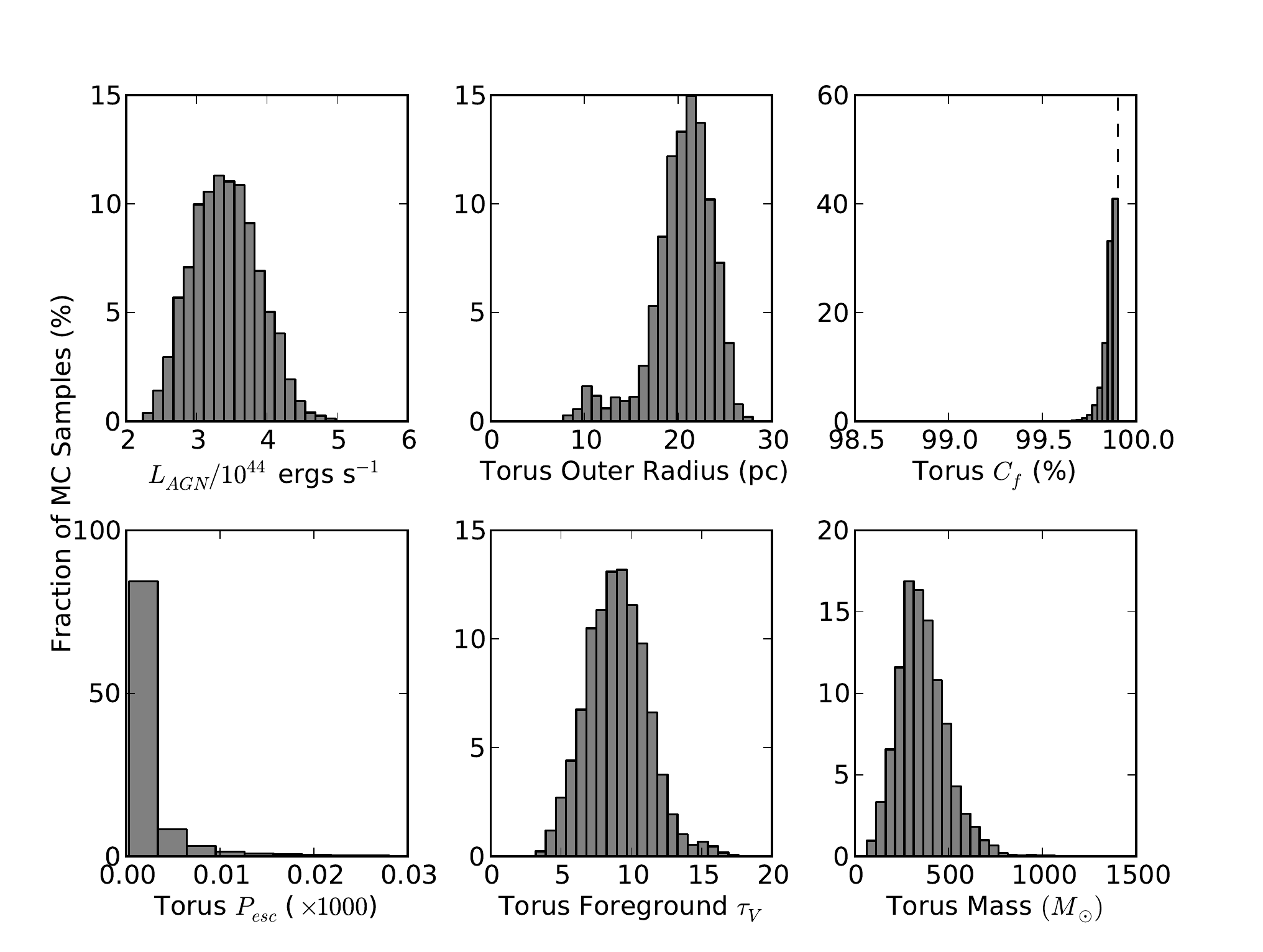}\\
\end{tabular}
\caption{
Marginalized posterior probability densities estimated from the MCMC samples
for the parameters derived from the clumpyDREAM SED fits to IRAS16399-0937.
Torus parameters for the IRAS16399N nucleus in the AGN+starburst model. }
\label{fig:torus_agn1}
\end{figure*}

\begin{figure*}
\centering
\begin{tabular}{c}
\includegraphics[scale=0.8]{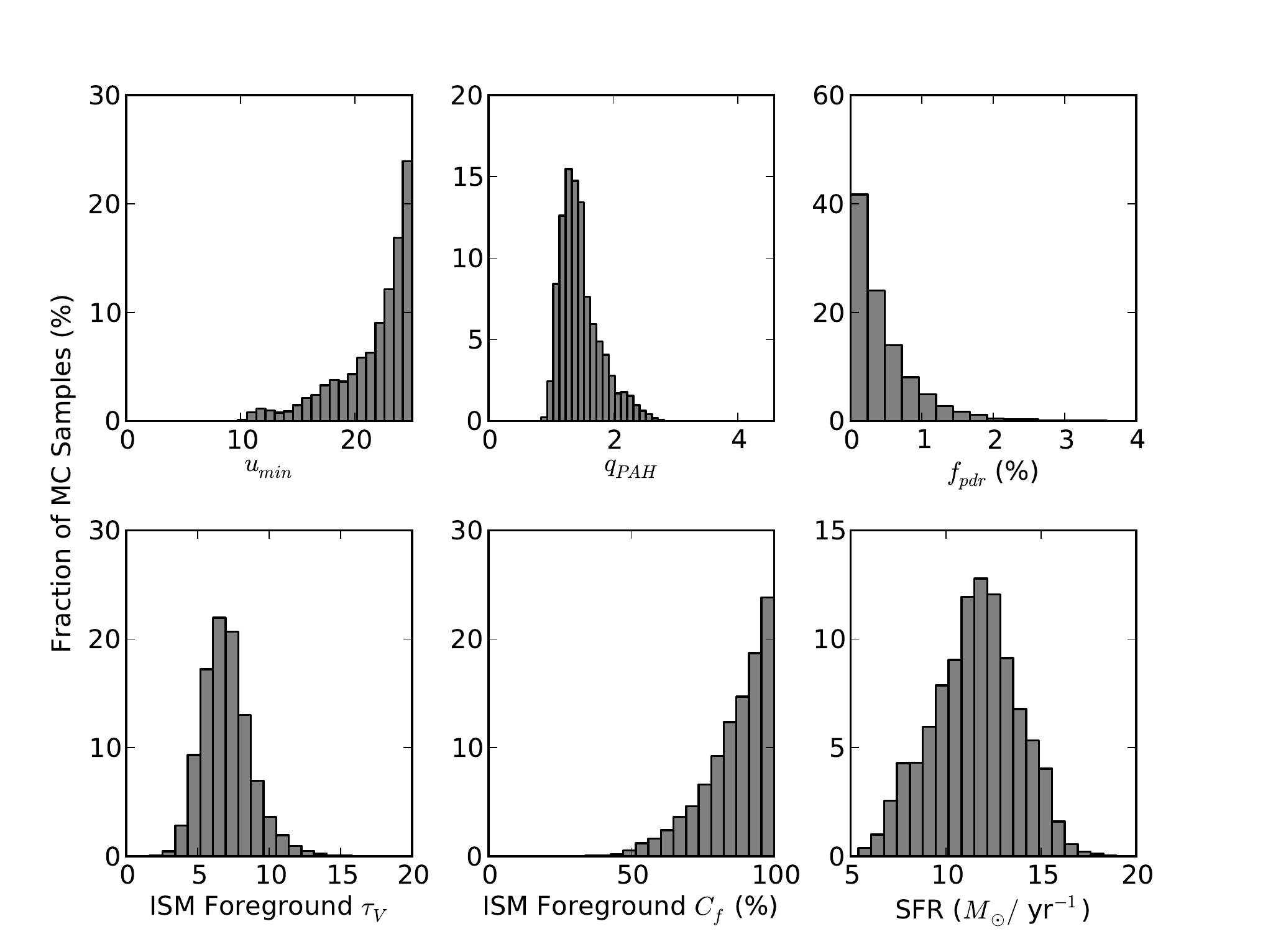}\\
\includegraphics[scale=0.8]{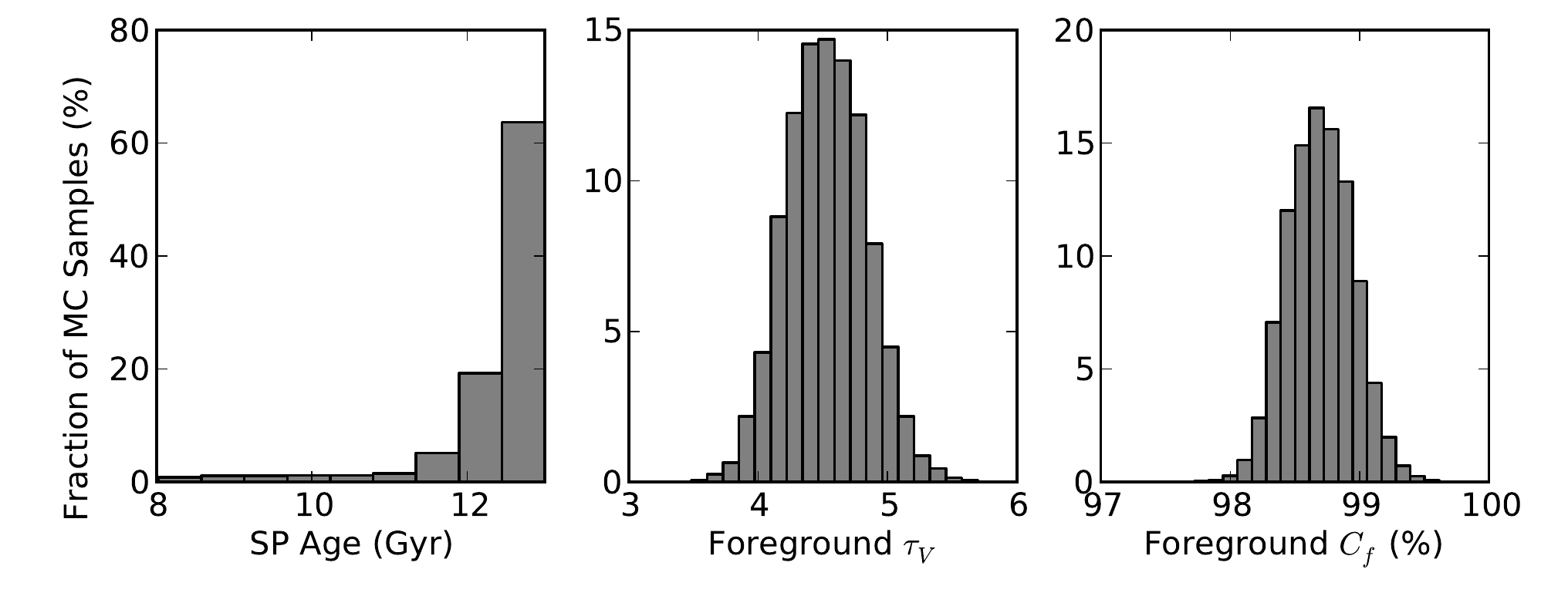}\\
\end{tabular}
\caption{As Fig.~\ref{fig:torus_agn1}, for the parameters of the stellar and ISM components.}
\label{fig:torus_agn2}
\end{figure*}

\begin{figure*}
\centering
\begin{tabular}{c}
\includegraphics[scale=0.8]{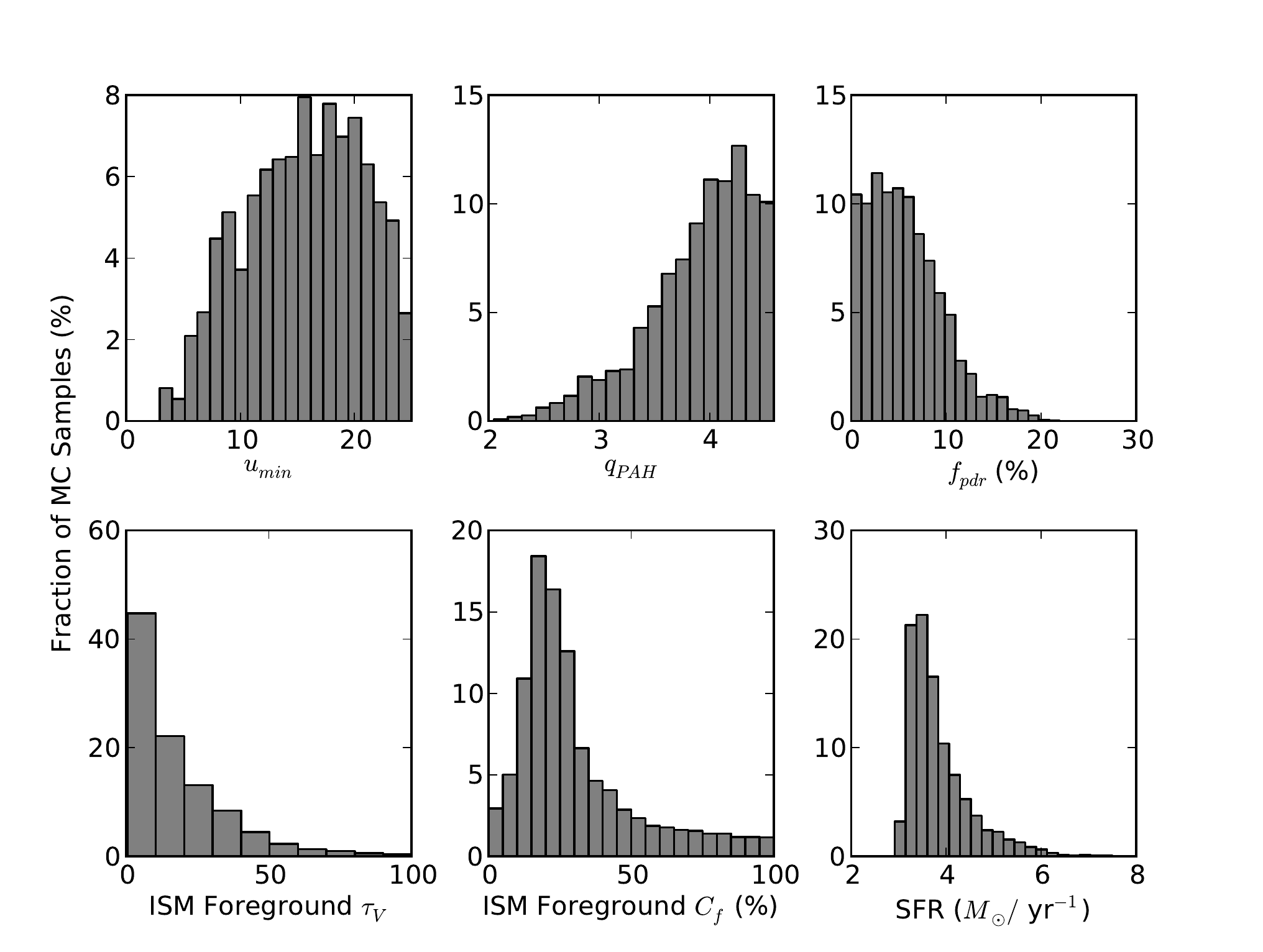}\\
\includegraphics[scale=0.8]{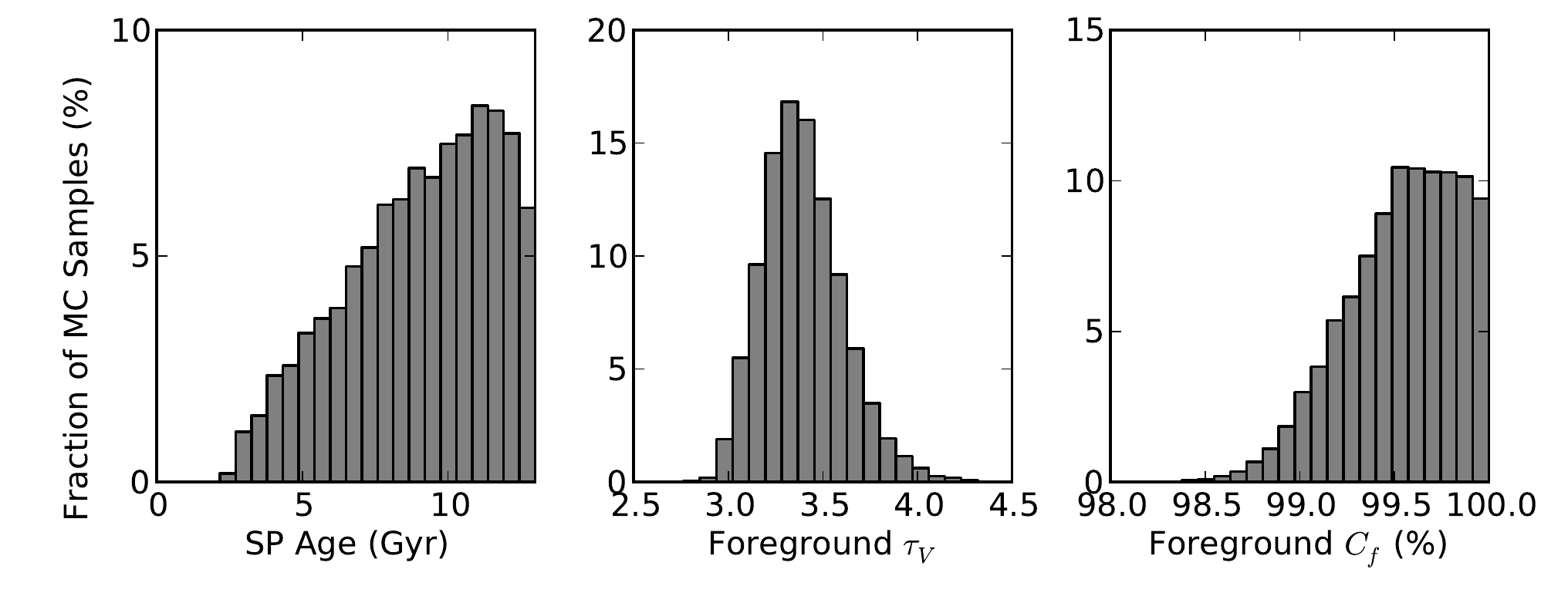}\\
\end{tabular}
\caption{As Fig.~\ref{fig:torus_agn2}, for the IRAS16399S nucleus.}
\label{fig:torus_sb1}
\end{figure*}

\begin{figure*}
\centering
\begin{tabular}{c}
\includegraphics[scale=0.8]{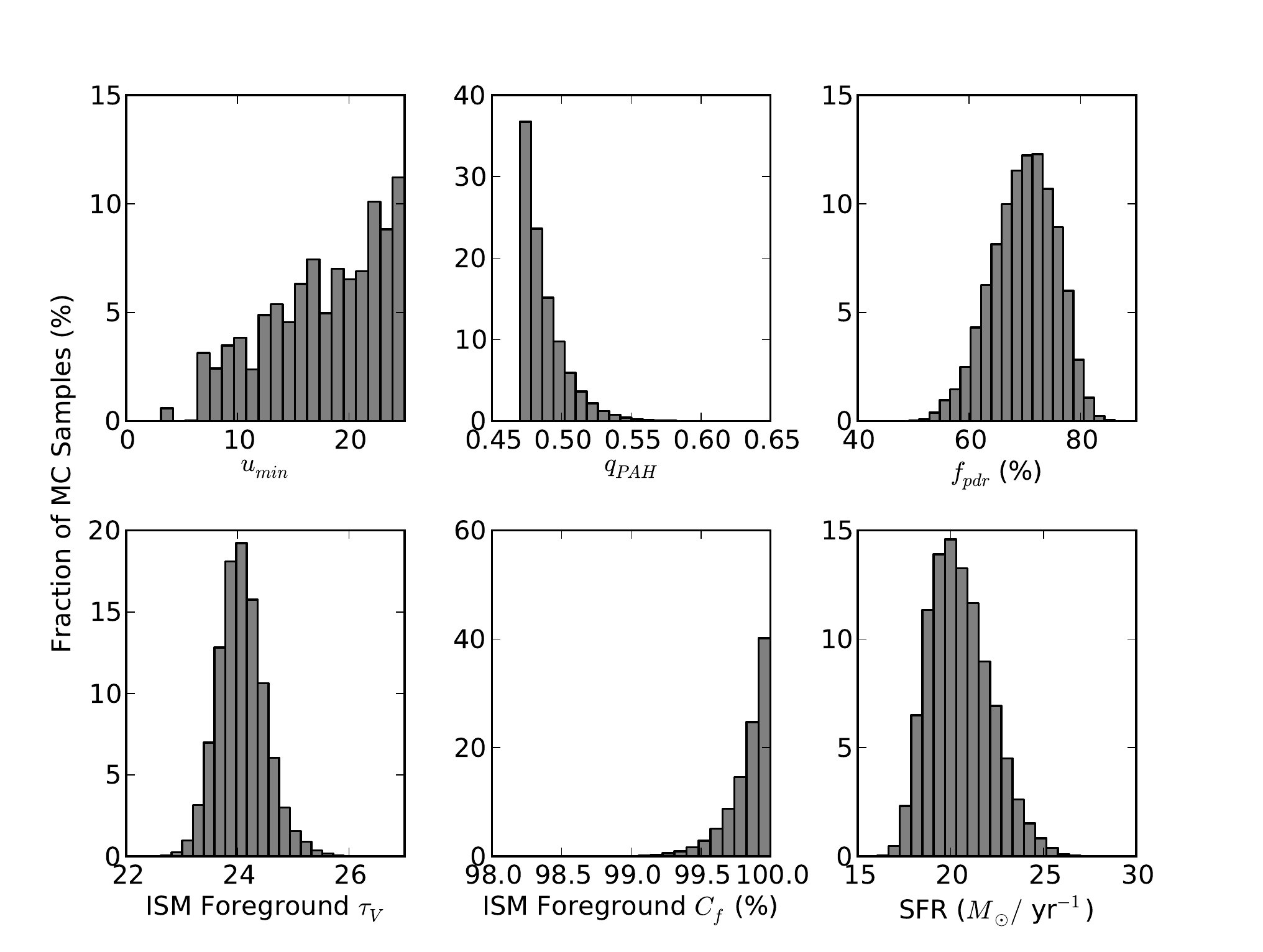}\\
\includegraphics[scale=0.8]{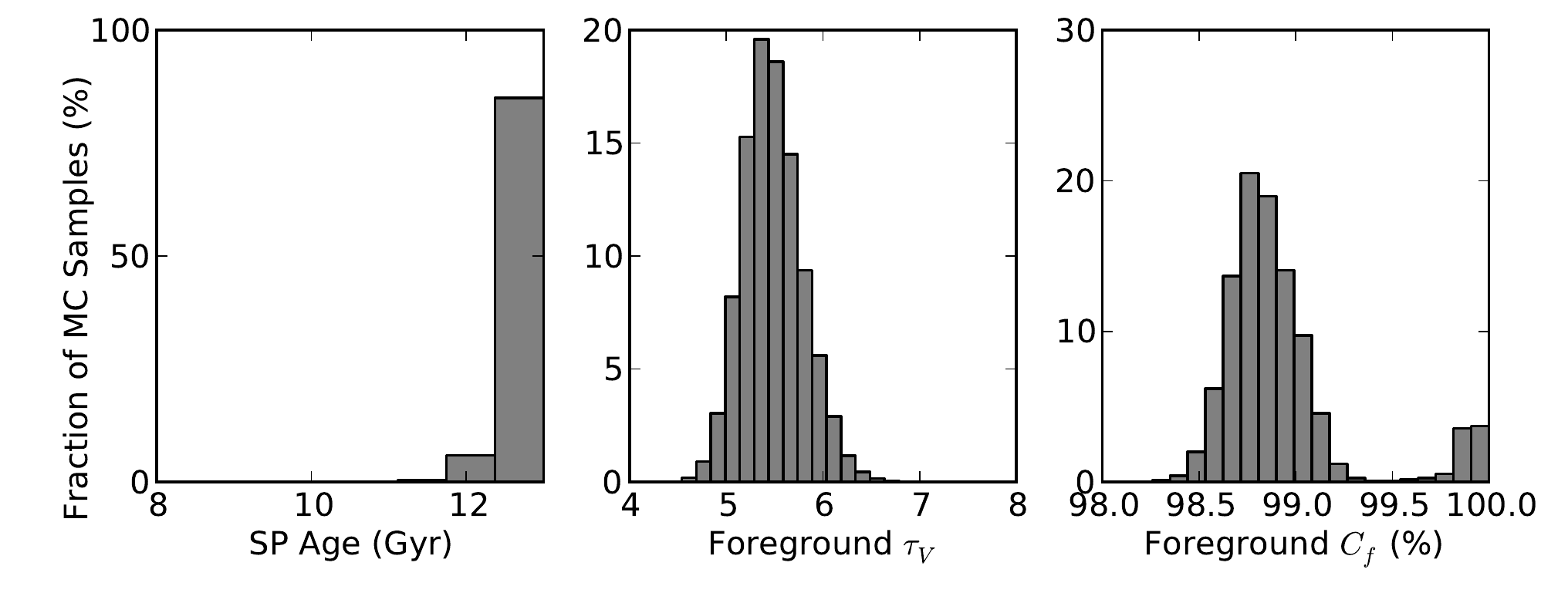}\\
\end{tabular}
\caption{
Probability density distributions of the parameters derived from the clumpyDREAM SED fits to IRAS16399-0937.
Parameters of the stellar and ISM components for the IRAS16399N nucleus in the starburst+starburst model.}
\label{fig:torus_agn3}
\end{figure*}

\begin{figure*}
\centering
\begin{tabular}{c}
\includegraphics[scale=0.8]{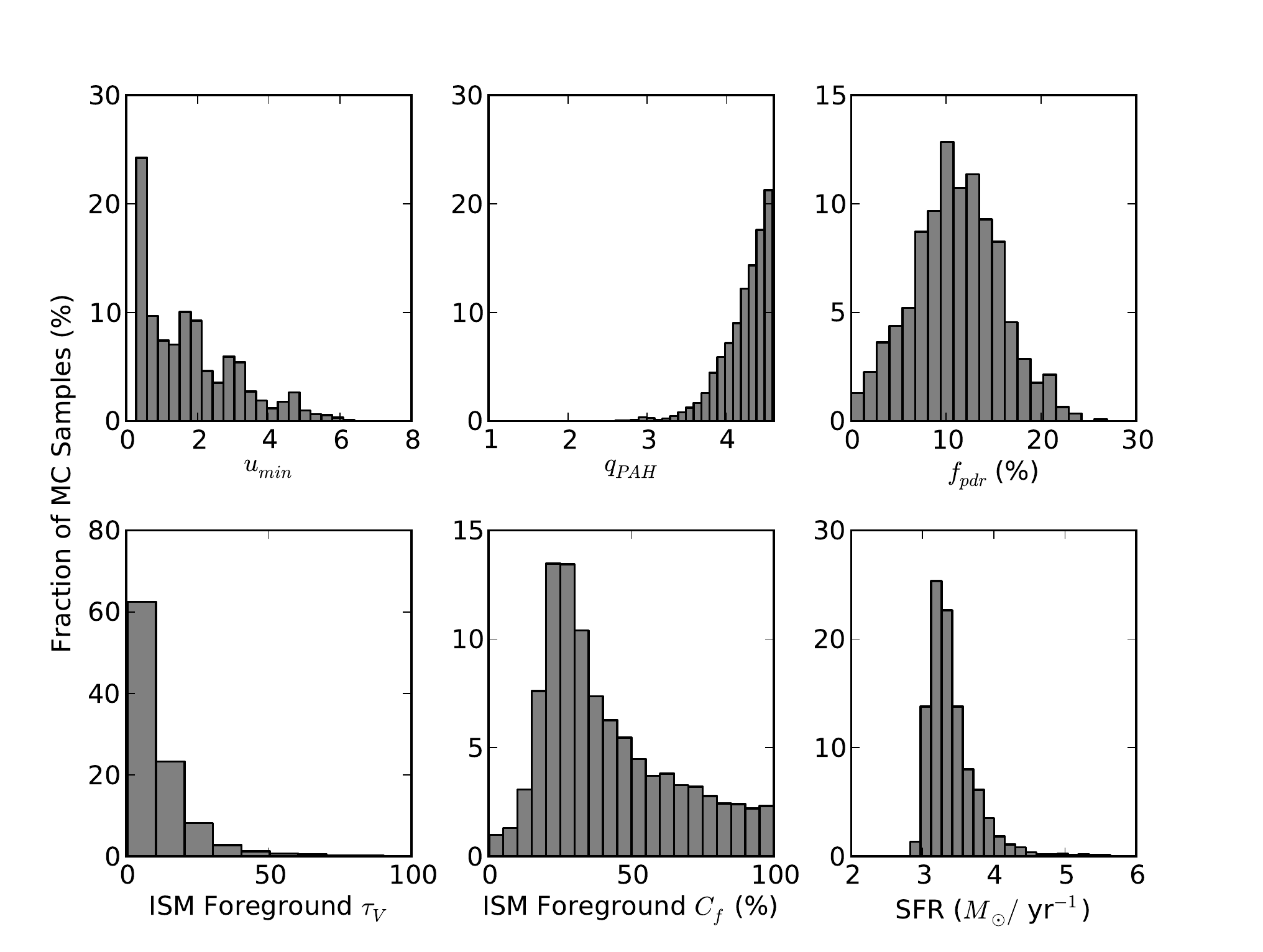}\\
\includegraphics[scale=0.8]{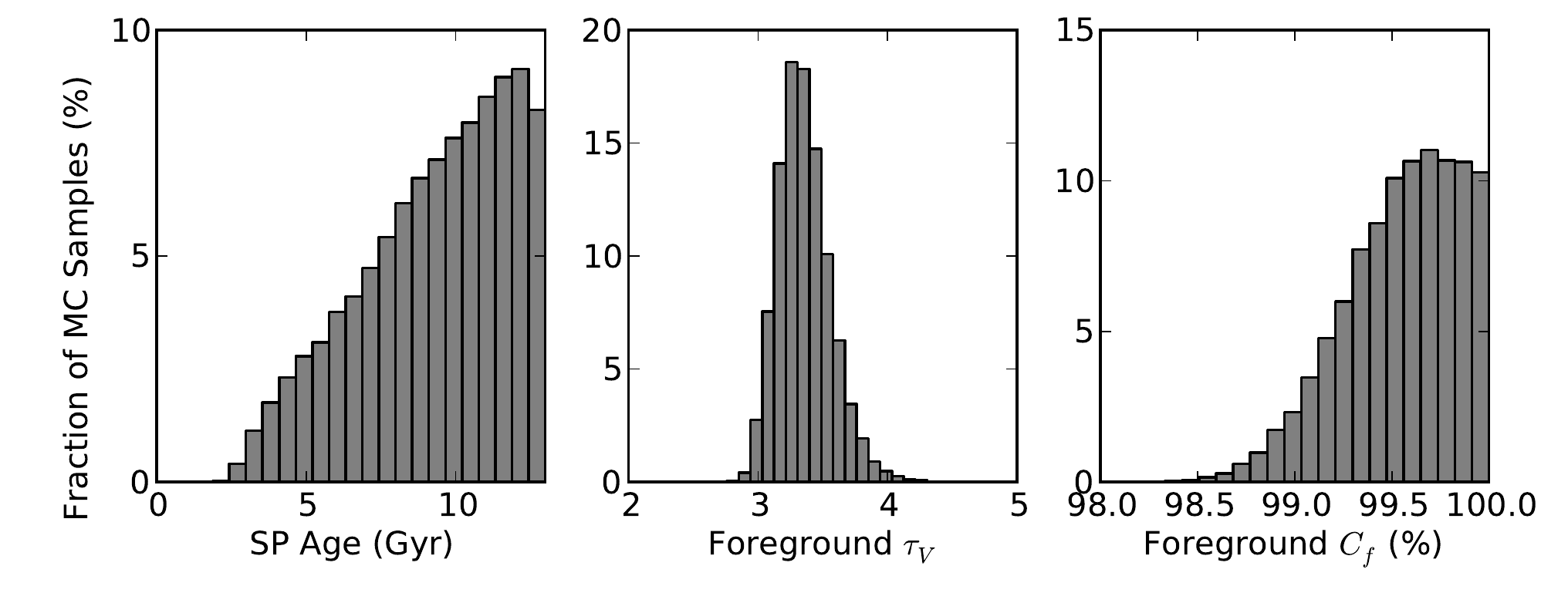}\\
\end{tabular}
\caption{As Fig.~\ref{fig:torus_agn3} for the IRAS16399S nucleus.}
\label{fig:torus_sb2}
\end{figure*}

Here we summarize the results of the SED fit which includes a clumpy torus contribution to
the IRAS16399N nucleus. The fits to the IRAS16399N and IRAS16399S nuclei are shown in Figure~\ref{fig:sed_agn}. The
probability density estimators for the fitted parameters are shown in Figure~\ref{fig:torus_agn1} and
the parameter values given in Table \ref{tab:linerfittorus}. It can be seen that the
observed SED of the IRAS16399N nucleus is well fit by the combination of the three components. The
optical--NIR continuum below $3.5$\micron\ is well fit by an evolved stellar population
with an age of log $\approx 12.6$\,Gyr. The torus and ISM components are comparable in strength
below $\approx 10$\micron\ but the torus emission peaks around 40\micron\ and is the main
contributor to the mid-to-far infrared continuum between 10 and 70\micron. The ISM
components dominates at wavelengths $\gtrsim 70$\micron.

% parameter values given as lower limits as per JG comments 7/30/14 - AR
The strength of the torus component in the SED fit rather tightly constrains the
bolometric luminosity of the AGN, L$_{AGN}\approx 3.4\,\times\,10^{44}$\,ergs/s. However,
although this implies that the IRAS16399N nucleus hosts a moderately luminous AGN (comparable with
Type 1 Seyfert galaxies), the torus parameters derived from the fit indicate that it is
deeply embedded within a quasi-spherical distribution of optically thick clouds. The torus
has a large angular scale height, $\sigma\gtrsim 66^{\circ}$, and is also radially extended,
$Y\gtrsim \,60$. The average number of clouds along an equatorial ray is $N_0\gtrsim 14$. Note that the
probability distributions for these parameters are truncated at the maximum values sampled by the model
grid (Figure~\ref{fig:torus_agn1}), and the recovered median values given in Table \ref{tab:linerfittorus} may therefore be underestimates. Individual clouds
have an optical depth of $20\lesssim \tau_V\lesssim 35$mag. The torus inclination to the line of sight is not well constrained,
but is probably $i\gtrsim 60^{\circ}$. Taking the median values of $i$, $\sigma$ and $N_0$, we estimate 
that there are $\sim 14$ clouds along the observer's line of sight (eq.~\ref{eq:n0}).
The line-of-sight escape probability for AGN photons is $\sim 10^{-6}$. Clearly, in
this model, the AGN is very heavily obscured along the line of sight. However, the angular
height of the torus is such that, even for a line of sight along the axis ($i=0$), we
predict $\approx 3 $ intervening clouds (corresponding to an escape probability $\sim 0.1$).
It should be noted that increasing $N_{0}$ 
and $\sigma$ would likely not improve the fit, since the covering fraction is already unity. 
We could conclude that we have lower limits for $N_{0}$ and $\sigma$.

% above values correct for new *median* values of fit parameters in Table 7 - AR 14/7/14; updated 7/28/14
%\sout{It should be noted that the SED fit likely produced artificially tight constraints on the torus inclination, $i$,
%and angular scale height $\sigma$. IRASF16399-0937 shows strong silicate absorption, and we use 
%the prior constraint that foreground (non-torus) dust extinction must be $A_V < 7$ (see 
%Section \ref{sec:result_Ha_phot}). In our model, the remaining silicate absorption depth must be 
%made up by the torus, pushing the fit to high torus covering fraction. As a result, $N_{0}$ 
%and $\sigma$ ran to the limits of the parameter grid, $N_{0} \leq 15$ and $\sigma \leq
%70^{\circ}$. The parameters $\sigma$, $N_{0}$, and $i$ are highly covariant for a required 
%silicate absorption depth, and so the grid limits on $N_{0}$ and $\sigma$ resulted in artificially 
%narrow constraints on $i$.}

Considering the global energetics, the covering fraction due to the clumpy torus is
$C_{tor}\gtrsim 0.998$, implying that $\lesssim 0.2$\% of the AGN photons escape from the torus without
being absorbed. This is consistent with the absence of signs of powerful AGN activity in
the optical and Spitzer mid-IR spectrum (see Section \ref{sec:mir_spectra}). As already
noted, the IRAS16399N nucleus has an optical spectrum characteristic of LINERs, which are generally
regarded as low luminosity AGN. The emission lines in LINERs are often attributed to
photoionization by a weak AGN radiation field, although alternative models invoking shock
ionization have also been advocated. Photon counting arguments show that the
embedded AGN in IRAS16399-0937 cannot account for the observed line emission from IRAS16399N. 
Given the bolometric luminosity determined by the fit, integrating over the AGN
spectrum adopted by \citet[][as defined by eq.~13]{Nenkova2008a} for the clumpy torus
model yields an ionizing photon luminosity $Q_{AGN}\approx 3.6\times
10^{54}$\,photons\,s$^{-1}$. % corrected 10/3/14 - DS
Taking the median value $C_{tor} = 0.999$, the number escaping the torus is
$Q_{esc}=(1-C_{tor})Q_{AGN}\approx 3.6 10^{53}$\,photons\,s$^{-1}$. % corrected 10/03/14 - DS
Assuming that the escaping
photons are absorbed by photoionizing hydrogen in the surrounding interstellar medium, the
H$\alpha$ luminosity resulting from recombination will be $L_{H\alpha, pred}\approx
C_{ISM}p_{H\alpha}h\nu_{H\alpha}Q_{esc}$, where $C_{ISM}$ is the covering fraction of the
surrounding gas and $p_{H\alpha}\approx 0.45$, is the fraction of H$\alpha$ photons
emitted per H recombination (assuming an electron density $n_e = 10^2$\,cm$^{-3}$ and
temperature $T_e=10^4$\,K). If $C_{ISM}=1$, $L_{H\alpha, pred}\approx
4.9\times10^{39}$\,erg\,s$^{-1}$, % corrected 10/03/14 - DS
which is $\lesssim 2$\% of the measured luminosity for
IRAS16399N ($L_{H\alpha}\approx 3.27\times 10^{41}$\,erg\,s$^{-1}$;
Sec.~\ref{sec:result_opt_nir}).  Thus, even if all the escaping AGN photons are absorbed
by the gas ($C_{ISM}=1$), the attenuated AGN radiation field cannot account for the
observed H$\alpha$ luminosity. This implies that the LINER spectrum cannot result from
photoionization of the surrounding ISM  by the embedded AGN.

The properties of the ISM component of the IRAS16399N nucleus are also constrained by the 
SED fit. In particular, the inferred ISM luminosity is L$_{ISM}\approx 3\times10^{44}$\,erg\,s$^{-1}$, 
implying an SFR of $\sim12$\,M$_{\odot}$/yr. The torus (by definition), stellar and ISM components 
are all essentially completely covered by foreground extinction of varying optical depth 
($\tau_V \approx9, 4.5$ and $7$\,mag, respectively), suggesting that the IRAS16399N nucleus
 is submerged in a gas and dust rich environment.

%which turn up an old stellar population with age of $10^{12}$\,yr, an SFR of $\sim3$\,M$_{\odot}$/yr.
%The energy density of the ISM radiation field is $u_{\min}\sim1.4$ and the fraction of the
%PAHs relative to total dust mass is $q_{PAH}\sim4.5$\%. 

As IRAS16399S was spectroscopically classified as a starburst \citep{Baan1998},  only
the stellar and ISM components were included in the SED fit. The SED is well reproduced
with this combination (Figure \ref{fig:sed_sb}). 
%\sout{The parameter values derived from the fit
%suggest that the ISM component has {\bf different} properties in the two nuclei.
%The ISM radiation field parameter, $u_{\min}\approx1.4$, and the PAH mass fraction
%$q_{PAH}\approx 4.33$\%, respectively.} % Not sure why the values of u_min are so different for N & S....
The luminosity of the ISM component in IRAS16399S is a factor $\approx 3$ smaller
than that of IRAS16399N, L$_{ISM}\approx 9\times10^{43}$\,ergs/s, % corrected 7/14/14 - AR
corresponding to an SFR $\sim3.6$\,M$_{\odot}$/yr. 
%\sout{However, the IRAS16399S nucleus ISM component
%has a {\bf little larger} foreground extinction ($\tau_V\sim8.8$\,mag) and therefore appears to
%be {\bf more} embedded than its IRAS16399N nucleus counterpart. } % if you look at the 5 & 95% values, there's no significant difference 
The fit also suggests that
the IRAS16399S nucleus has a somewhat younger stellar population, with an age of log 10\,Gyr, 
and the fraction of the dust luminosity coming from photodissociation regions
($U>$ 100), f$_{\rm PDR}$, is a factor $\sim 17$ larger than in the IRAS16399N nucleus, perhaps also indicating 
a younger stellar population than that of
IRAS16399N.

%\sout{This scenario corroborates to the Starburst classification from optical 
%wavelength range \citep{Baan1998}.\,\,\,  } % don't think we need to spell this out...

In summary, the results of the SED-fitting suggest that IRAS16399N nucleus contains an
embedded AGN but also hosts a star forming region which has luminosity and SFR roughly 3x 
that of the IRAS16399S ``starburst'' nucleus.

%\sout{The ISM component in the IRAS16399N nucleus is also
%affected by much greater foreground extinction, which is consistent with the presence of
%deeper ice and HAC diffuse bands (Section \ref{sec:mir_spectra}) in the mid-IR spectrum
%and the presence of the dust lane that bisects the IRAS16399N nucleus (Figure~\ref{fig:rgb}).}

%This scenario could suggest that the OH megamasers emission of IRAS16399-0937 maybe a 
%signpost of the already established circum-nuclear gas disk surrounding an enshrouded AGN 
%in the process of being cleared away along the rotation axis of the maser disk. 

\subsubsection{Dual starburst nuclei}

In order to test the alternative scenario in which the IRAS16399N nucleus does not contain an AGN, we also fitted the SEDs 
using only stellar and ISM components for both nuclei. Not surprisingly, the fit to the IRAS16399S nucleus is almost indistinguishable
from that obtained in the previous case and very similar parameter values are recovered for the ISM component
(Fig.~\ref{fig:sed_sb}, Table~\ref{tab:linersbfit}). In particular, there is no significant 
difference in the luminosity of the ISM component and hence the SFR.
%\sout{as well as in the  age of the stellar component. % don't think we need to mention this
%The age of the stellar component increases to
%11\,Gyr, but this parameter is relatively weakly constrained.}
 
The overall fit to the IRAS16399N nucleus SED using only the stellar and ISM components is
qualitatively similar to that obtained with the torus component (Figure
\ref{fig:sed_agn}), although, formally, this solution is not as good according to
the Bayes information criterion (BIC) statistic\footnote{The BIC is $-2 \cdot ln \hat{L} + k \cdot ln(n)$, 
which is estimated by $\chi^2 + k \cdot ln(n)$, where $\hat{L}$ is the maximized value of 
the likelihood function of the model, $k$ is the number of free parameters and n is the 
number of data points. The smaller BIC value indicates the preferred model \citep{Schwarz1978}.}, which has 
values of 987 and 1225 for the models with and without the torus, respectively. % put this in footnote

In order to compensate for the absence of the torus contribution to the mid-IR, 
the fraction of the ISM luminosity coming from photodissociation region dust increases radically, from f$_{\rm PDR} < 1.4\%$
to  $\approx70\%$. Removing the torus effectively creates a gap in the mid-infrared, 
and the no-torus model compensates by requiring a greater contribution from the hot dust associated with PDRs. % revised 14/7/14 - AR

%\sout{component becomes radically stronger, both its
%luminosity and the strength of the ISM radiation field increasing by nearly an order of
%magnitude ($L_{ISM}$ increases from $6.5\times 10^{43}$ to $4.4\times
%10^{44}$\,erg\,s$^{-1}$; $u_{min}$ from 1.4 to 12.9).}

The PAH mass fraction in the IRAS16399N nucleus ($q_{PAH}\approx 0.5$) is a factor $\sim 3$ smaller in the absence
of the torus component and is therefore also an order magnitude smaller than that in the IRAS16399S
nucleus (which is similar in the two models). This is a consequence of the fact that when the
torus is included, the relatively low observed equivalent width of the PAH  features
results from dilution by the torus hot dust continuum.
%\sout{ {\bf from the photodissociation regions}}. % Don't think this can be correct; PDR's are present in both models, it is the torus contribution that is missing 
%\sout{and a PAH mass fraction very similar to that of the IRAS16399S nucleus is recovered.} 
When the torus is omitted, the PAH equivalent widths are matched mainly by decreasing $q_{PAH}$.  
It is also notable that the foreground extinction to the ISM component is greatly
increased in the fit without the torus component, $\tau_{ISM,V}\approx 24$, compared to
$\approx 6.9$ with the torus. This is mainly driven by the strength of the 9.7\micron\
silicate absorption feature. When the torus is present, this feature is produced by
combination of  torus and ISM extinction. Without the torus, the silicate absorption is
produced exclusively by foreground extinction due to the ISM component, resulting in an
increased $\tau_{ISM,V}$.

As the ``no torus'' model requires a much stronger ISM component, it also implies a
higher SFR for the IRAS16399N nucleus, since star forming regions are the main heating source
for the ISM. Thus, we obtain an SFR$\sim20$\,M$_{\odot}$/yr, nearly a factor 2 greater than
derived for the model including the AGN torus (see Tables~\ref{tab:linerfittorus} and
\ref{tab:linersbfit}).

% The second fit also retrieves the picture
%that the North nucleus is embedded in rich dust components from ISM and stars with their
%covering fraction almost 100\%.

%However, 
%the mass fraction of the PAH component (q$_ {PAH}$) of dust grains is $\sim$\,3 times less than in the
%previous fitting. Our finding is suggesting that when we take out the torus models, we effectively 
%take away the MIR contribution of hot grains, and its was replaced in the ISM model by raising 
%the ISRF ($u_{\min}$), which caused a dilution of the PAH material (q$_ {PAH}$).
% Note low q_pah hard to reconcile with fluxes.

\section{Discussion}\label{sec:discussion}

Our multiwavelength study of IRAS16399-0937 combines new HST ACS I-band and emission line 
images, with archival Spitzer IRAC, IRS and MIPS data, Herschel photometry and VLA images at 1.49 and 4.9\,GHz. 
These data provide a comprehensive picture of the overall morphology of this 
system and the spatial distribution of the ionized gas, dust and radio emission. We have 
also compiled a broad-band spectral energy distribution for the system, spanning $0.45 - 500\micron$, 
which, importantly, resolves the two nuclei at wavelengths below $14\micron$.

\subsection{IRAS16399-0937 as an Individual Object}

According to \citet{Haan2011}'s study of the nuclear structures in nearby LIRGs, 
merging systems can be classified into six different merger
stages: (1) separate galaxies, but disks symmetric (intact) and no tidal tails, (2) progenitor 
galaxies distinguishable with disks asymmetric or amorphous and/or tidal tails, (3) two nuclei 
in common envelope, (4) double nuclei plus tidal tail, (5) single or obscured nucleus with 
disturbed central morphology and short faint tails. Using their scheme IRAS16399-0937 is 
classified as type 3 (nuclei in common envelope) and this group constitutes 21.6\% of a
sample of 60 systems. However, the majority of their sample is classified as
type 2 (25\%), while 5 (8.3\%) systems are classified as type 1, 8 (13.3\%) are type 4, 12 
(20\%) are type 5, 1 (1.6\%) is type 6 and 6 (10\%) are classified as type 0.

\citet{Rigopoulou1999} classified the ULIRG mergers as follows: (1) fully 
relaxed systems, in which we see only a single pointlike nucleus with relatively little or 
no tail; (2) systems in which the merger is completed; in such systems we see a single nucleus, 
but with significant residual structure or tail formation; and (3) interacting pairs in which 
the interacting nuclei are clearly visible and can be found in a variety of separations.
Of 27 targets they studied 68\% appear to have double nuclei (similar to IRAS16399-0937), 
with a wide range of projected separations between the nuclear components.

In terms of separation, IRAS16399-0937 is at the lower end of the distribution for objects 
that have resolved double nuclei and 
also has a relatively low infrared luminosity \citep[L$_{IR}$ = 11.63 L$_{\odot}$, see table 1 of][]{Haan2011}
compared with other (U)LIRGs \citep[see Fig. 11 of][]{Rigopoulou1999}. In \citet{Haan2011}'s 
sample, IRAS16399-0937 is slightly below the median projected separation \citep[7.2\,kpc, 
see section 3.1 and Fig. 4 of][]{Haan2011} and the bulge luminosity derived from H-band image
is significantly lower than that of the majority of the (U)LIRGs studied in this sample. 
These studies lead us conclude 
that IRAS16399-0937 is not unusual in terms of its merger stage among (U)LIRGs and it also 
fits in with \citet{Haan2011}'s finding that non-merging LIRGs have larger bulge 
masses than merging LIRGs \citep[see Fig. 8 of][]{Haan2011}.

According to \citet{Yuan2010}, IRAS16399-0937 is classified as a ``close binary'' 
(i.e., early stage merger) and systems turn out to be about 12\% of the Southern Warm Infrared 
Galaxy sample \citep[hereafter SWIG,][]{Kewley2001}, but about 30\% of the merging/pair systems. 
The infrared luminosity of IRAS16399-0937 is very close to the mean for the combined LIRG sample selected from
SWIG and the IRAS Bright Galaxy Survey \citep[hereafter BGS][]{Veilleux1999}. However, it is 
somewhat unusual in terms of its spectroscopic properties 
since the majority of the ``close binary'' LIRG systems exhibit 
 H{\sc ii} or composite spectra while IRAS16399-0937 has a
LINER+starburst double nucleus \citep[see Fig. 12 of][]{Yuan2010}.

As noted in Section~\ref{sec:liner}, \citet{Yuan2010} also argue that there are few ``genuine'' LINERs among (U)LIRGs. They 
reclassify most previously identified LINERs as composite nuclei and note 
that the few (9) remaining objects falling in the 
LINER region of the diagnostic diagrams are close to the Seyfert/LINER borderline (this includes 
IRAS16399-0937), or in the few remaining cases (particularly, the well-studied NGC6240 and Arp220) are 
either starburst superwinds or shocks driven by galaxy collisions, rather than being low accretion 
rate AGN \citep[see Fig. 15 and Section 4.1 of][]{Yuan2010}.

Our picture for IRAS16399-0937 is very similar to what is proposed for the double nucleus NGC6240 (LIRG)
and Arp220 (ULIRG) galaxies, except that they both have higher infrared luminosities and smaller 
separations ($\sim$0.7 kpc for NGC6240 and $\sim$0.3 kpc for Arp220). In general, IRAS16399-0937 
is a fairly typical example of a LIRG merger, but one in which the AGN is deeply embedded, 
with the LINER spectrum arising through shocks rather than AGN photoionization.

\subsection{Star formation}\label{sec:sfr}

% revised AR 7/27/14
The ISM-dust 8.0$\mu$m PAH, mid-IR ($24\micron$) and radio luminosities indicate a star formation
rate $\sim 20$\,M$_{\odot}/yr$ for the entire system. The SFR derived from the X-ray luminosity is higher, nearly $50$\,M$_{\odot}/yr$, but this
is subject to large uncertainties. Nevertheless, in all wavebands  in which the nuclei are resolved, 
the system SFR is notably larger than the sum of the rates
derived for the two nuclei, which account for $\sim 30-60$\% of the total
SFR, depending on the waveband. Evidently, star formation activity is not especially localized to the nuclei, 
The X-ray, PAH and radio luminosities all imply a higher SFR for
IRAS16399N than for IRAS16399S. For IRAS16399S, the SFRs derived from those 3 bands are all consistent with a rate of $\approx 3$\,M$_{\odot}/yr$.
There is a much larger spread for IRAS16399N, with the X-ray and radio luminosities indicating higher rates ($\approx 6-10$\,M$_{\odot}/yr$) than 
the PAH luminosity ($\approx 4$\,M$_{\odot}/yr$, but it is possible that the AGN contributes to the X-ray and radio emission.

The star formation rates derived from the H$\alpha$ emission are approximately a factor 3 smaller
than those derived from the radio and mid-IR emission for both nuclei. The H$\alpha$ SFR
for the system is uncertain by at least a factor 3, since the extinction is poorly
constrained. The upper limit on SFR$_{H\alpha}$ (determined using the reddening of the IRAS16399S
nucleus) approaches the values obtained for longer wavelengths. Nevertheless,
morphological evidence suggests that even outside the nuclei, most of the current star
formation occurring in IRAS16399-0937 is heavily obscured by dust -- for example, the
radio and PAH images indicate that star formation is occurring in the dust lane complex
between and to the east of the nuclei, a region which is devoid of H$\alpha$+[N{\sc ii}]
emission.

\subsection{Does IRAS16399-0937 host an AGN?}\label{sec:agn}
% edited by AR 7/28/14
Classifications of nuclear activity in OHMG based 
on optical spectroscopy \citep{Darling2006,Baan1998} imply a much higher incidence of AGN 
than results from radio, mid-IR and X-ray data \citep{Baan2006,Vignali2005}.  In the case of 
IRAS16399-0937, \citet{Baan1998} used emission line diagnostic diagrams pioneered by \citet{Baldwin1981}  
and extended by \citet{Veilleux1987} (hereafter referred to as VO diagrams) to classify the 
IRAS16399N nucleus as a LINER and the IRAS16399S nucleus as a starburst, respectively. \citet{Yuan2010} arrived 
at essentially the same conclusion, using measurements of the emission line ratios from 
\citet{Kewley2001} and the refined classification scheme developed by \citet{Kewley2006}\footnote{The 
adopted classification for the IRAS16399S nucleus listed in Table 3 of \citet{Yuan2010} is ``cp, composite'', 
that is, a mixture of starburst and AGN spectra. However, IRAS16399S falls well within 
the starburst regions of two of the three diagnostic diagrams ([O{\sc iii}]/H$\beta$ vs 
[S{\sc ii}]/H$\alpha$   and [O{\sc iii}]/H$\beta$ vs [O{\sc i}]/H$\alpha$) and lies near the 
H{\sc ii} region -- composite boundary in the remaining diagram ([O{\sc iii}]/H$\beta$ vs 
[N{\sc ii}]/H$\alpha$).}. Insofar as LINERs can be considered low luminosity and/or low accretion 
rate AGN \citep[for a review, see][]{Ho2008}, the optical spectra appear to clearly indicate 
the presence of a weak AGN in the IRAS16399N nucleus. In other wavebands, however, the evidence is 
ambiguous. The IRAS16399N and IRAS16399S nuclei fall within the regions occupied by Seyferts and starbursts, 
respectively, in the PAH 6.2$\mu$m/7.7$\mu$m $\times$ 11.3$\mu$m/7.7$\mu$m diagram (Figure~\ref{fig:pah}), 
although LINERs are mixed with starburst and H\,{\sc ii} galaxies in this diagram \citep{Sales2010}. 
On the other hand, the mid-IR spectrum lacks the high ionization fine structure lines 
[Ne\,{\sc v}]\,14.3$\mu$m (ionization potential 97.1\,eV) and [O\,{\sc iv}]\,25.8$\mu$m 
(ionization potential 55\,eV), that are generally considered to be unambiguous tracers of 
AGN activity \citep[e.g.,][]{Willett2011a}. However, these lines fall into the wavelength 
band covered by Long-Low module of IRS, which does not resolve the two nuclei, so the AGN 
line emission may be swamped by dust emission from the nuclei and envelope. It is also worth 
noting that such lines are not always detected in the nuclei of
Seyfert~2 galaxies \citep{Sales2010,Gallimore2010}. 
There is both a compact radio source and a compact X-ray source associated with IRAS16399N. However, the radio source 
has a low brightness temperature which seems more consistent with star formation than with an AGN radio jet, 
while the {\em observed} X-ray luminosity is 
is a factor $\sim 10^{2-3}$ lower than expected for a typical Seyfert galaxy and could also be due to star formation. Nevertheless,
we cannot rule out an AGN contribution in either band.
 % (although it is also comparable with that of nearby low luminosity AGNs \citep{Ho2009}).

%In the case of IRAS16399-0937, we were able to resolve the ambiguity as to the nature of the nuclear source
%by carefully decomposing the multiwavelength spectral energy distribution. We have 
%successfully modeled the optical -- FIR SED's of both nuclei, fitting both simultaneously 
%with a combination of elliptical galaxy, ISM and AGN clumpy torus components, with the latter 
%used only for the N nucleus. 
The results of the SED decomposition indicate that a moderate luminosity AGN 
($L_{bol}\sim3.4\times 10^{44}$\,erg\,s$^{-1}$) is indeed present in the IRAS16399N nucleus. An AGN-heated clumpy torus is 
required, in particular, to account for excess mid-IR emission (as compared to the IRAS16399S nucleus) 
that is apparent as a steep rise in the SED for $\lambda > 10\mu$m. 
% text revised as per JG comments 7/30/14 - AR
The SED fit obtained for IRAS16399N when the AGN torus component is significantly better than that 
obtained when the torus is omitted. However, there are also other reasons for 
favoring the torus model. If the torus is omitted, the ISM component must increase in strength 
to compensate, 
%\sout{the ISM component must radically increase in strength 
%to compensate, resulting in a factor $\sim 175$ increases in the ISM dust luminosity from
%photodissociation regions with U$>$100.}
implying an SFR $\sim 20$M$_{\odot}$/yr, which is significantly higher (by factors $\approx 2-5$)  
than the SFRs inferred from the observations (Table~\ref{tab:sfr}).
%\sout{which is a factor $\approx3$ higher 
%than that inferred from the nuclear radio luminosity (Table~\ref{tab:sfr})}. 
On the other hand, for the model including the torus, the derived SFR agrees with 
that derived from the X-ray luminosity and is more consistent with the values inferred from PAH-dust 
8$\mu$m and the radio emission. 
%With the exception of the rates derived from $L_{H\alpha}$ (which is 
%affected by reddening) and the 8.0$\mu$m PAH luminosity, % SFR_PAH = 4.2+/- 0.6 is not statistically consistent.
%these are consistent with the 5--95 percentile range recovered from the SED fit (Tables~\ref{tab:sfr} and~\ref{tab:linerfittorus}).
%\sout{with the value determined from the radio luminosity.} 

Furthermore, without the hot dust 
contribution from the torus, the PAH mass fraction is reduced by a factor $\sim 3$ in order 
to match the observed PAH equivalent widths. In the non-torus models, therefore, $q_{PAH}$ 
is a factor $\sim 9$ smaller for the IRAS16399N nucleus than the IRAS16399S nucleus. This is difficult to 
reconcile with the observed PAH fluxes, which are larger for the IRAS16399N nucleus, as indicated by 
%\sout{ISM PAH-dust $8\mu$m image and} 
the fluxes derived from the IRS spectra (Tables~\ref{tab:fluxes} and~\ref{tab:mirlines_sns}).

% values below updated - AR 7/28/14
The parameters recovered for the torus component suggest that the AGN is embedded in a 
quasi-spherical (angular scale height $\gtrsim 66^{\circ}$), radially extended (outer radius 
$\gtrsim 60\times$ the dust sublimation radius) distribution of highly optically thick clumps 
($\tau_V \approx A_V\sim 27$), with the average number of clumps along a given ray varying 
from $\gtrsim14$ at the equator to $\sim 3$ along the poles ($\sim 14$ clouds 
would intercept the line of sight at $i\sim 60^{\circ}$.%; although it should be noted that this parameter is poorly constrained). 
The torus itself is screened by another $\approx 9$ 
magnitudes of visual extinction. Therefore, the AGN is deeply embedded in an effectively 
complete cocoon of dusty clumps, each of which is individually highly opaque. This is consistent 
with the detection of water ice and HAC absorption bands in the IRAS16399N nucleus; it seems likely 
that these features originate in the torus clumps. 
% and deeper silicate absorption

This quasi-spherical ``cocoon'' of dusty gas would also be highly Compton-thick and 
this seems consistent with the weakness of the X-ray source associated with IRAS16399N,
relative to the inferred bolometric luminosity of the AGN. 
Given a bolometric correction factor $\sim 20$ \citep{Elvis2004,Lusso2012},
we predict an intrinsic 0.5--2\,keV luminosity of $L_{0.5-2\,keV} \approx 1.7\times 10^{43}$\,erg\,s$^{-1}$ for the embedded AGN, 
compared with an observed luminosity of  $\approx 5\times 10^{40}$\,erg\,s$^{-1}$.

% Lusso2012: http://adsabs.harvard.edu/abs/2012MNRAS.425..623L

\subsection{Is the OH Maser emission associated with the AGN?}\label{sec:ohm}

The location of the OH megamaser source(s) in IRAS\,16399-0937 is, unfortunately, unknown, 
since there are no extant high resolution interferometry data at the OH line frequencies. 
However, \citet{Willett2011a,Willett2011b} 
find that, in comparison with a control sample of non-OHMG ULIRGs, the OHMGs preferentially
exhibit H$_{2}$O and HAC bands and (as observed in the IRAS16399N nucleus) steeper mid-IR spectra.
Therefore, it seems reasonable to conclude that the OH masers also reside in the IRAS16399N nucleus, 
presumably in the torus clumps. Indeed, the currently favored model for OH megamaser emission
invokes a clumpy medium, with individual clumps generating low-gain unsaturated emission
while strong compact sources occur when the line of sight intersects multiple clouds \citep{Parra2005}.
The pumping calculations of \citet{Lockett2008} indicate that the masers are strongly radiatively pumped
at 53$\mu$m, with line overlap effects caused by internal cloud turbulence  
$\sim 20$\,km\,s$^{-1}$ also playing an important role. In their model 
maximum maser inversion occurs for dust optical depths $10\lesssim \tau{_V}\lesssim 50$ and
dust temperatures $80\lesssim T_d\lesssim 140$\,K. The structural and physical properties of the torus
component of the SED model fitted to the IRAS16399N nucleus are a  good match to these conditions: the
torus emission peaks in the 40--50$\mu$m range, a typical sight-line intercepts multiple clouds ($\sim 14$), 
individual clouds have optical depths $\tau_V\approx 27$ and for distances
$\gtrsim 10r_{in}$ (where $r_{in}$ is taken to be the dust sublimation radius, Section~\ref{sec:sed-model-components}),
the dust temperature in the cloud interior is $\lesssim 140$\,K \citep{Nenkova2008a}. 
Thus, the IRAS16399N nucleus, and specifically the AGN-heated torus, appears to have all the necessary
ingredients for strong, radiatively pumped OH masers. 
% JG text added 7/30/14
Nonetheless, high resolution radio interferometry
measurements are needed to locate unambiguously determine the location of the OH
masers relative to the double nucleus.

\subsection{The origin of the LINER spectrum}\label{sec:liner}

The overall covering factor of the torus is close to unity, implying that only a very small 
fraction ($\leq 0.2$\%) of the ionizing photons produced by the AGN escape to photoionize 
the surrounding gas. If this is correct, radiation from the AGN itself fails by a factor 
$\gtrsim 66$ to account for the H$\alpha$ flux measured from the IRAS16399N nucleus, even if all the % factor corrected - AR 7/28/14
ionizing photons that escape are absorbed in photoionizing the surrounding gas. Alternatively, 
we would require at least 5\% of the AGN ionizing photons to escape, implying a covering 
factor $C_{tor}\approx 0.95$, which is ruled out by the posterior probability distribution 
for this parameter (Figure~\ref{fig:torus_agn1}). 

It follows that the LINER spectrum of the IRAS16399N nucleus cannot be due to photoionization by the 
AGN. However, optical emission line ratios characteristic of LINERs can also be produced 
by several other mechanisms, including shocks \citep{Dopita1995,Allen2008}, hot evolved 
(post-AGB) stars \citep[e.g.,][]{Binette1994} and in dense environments, young massive stars 
\cite[e.g.,][]{Barth2000}. Any of these alternative ionization mechanism could plausibly 
be operating at some level in the IRAS16399N nucleus of IRAS16399-0937, but given that gravitational torques are 
expected to drive gas flows in merging systems, shock ionization seems the most likely 
candidate. Indeed, inspection of the shock model grids presented by \citet{Allen2008} 
reveals that models including a photoionized precursor match the [O{\sc iii}]/H$\beta$, 
[O{\sc i}]/H$\alpha$, [N{\sc ii}]/H$\alpha$  and [S{\sc ii}]/H$\alpha$ line ratios reported 
by \citet{Baan1998} and \citet{Kewley2001} remarkably well for shock velocities 
$\approx 100 - 200$\,km\,s$^{-1}$. Such shock velocities are comparable with radial inflow 
velocities found in gas dynamical simulations \citep[e.g.,][]{Iono2004}.
Similarly, \citet{Monreal2006} found that extended emission line regions in several ULIRGs have LINER-like ionization
conditions consistent with shock ionization, which they attribute to shocks associated with merger-driven gas flows.
Curiously, therefore, 
the results of the SED fitting support the presence of an AGN in the IRAS16399N nucleus, but also suggest 
that the LINER spectrum, upon which the original classification was based, is not due to AGN 
photoionization. The case of IRAS16399N exemplifies the point that a LINER spectrum 
does not necessarily indicate the presence of an AGN. As 22.5\% and 42\% of the OHMGs studied 
by, respectively, \citet{Baan1998} and \citet{Darling2006} were spectroscopically classified 
as LINERs, misinterpretation of these objects as AGN may partly explain the discrepancy 
between the high  AGN fraction inferred in OHMG from optical spectroscopy ($\sim$\,70\%) 
and the much lower estimates obtained from radio, mid-IR and X-ray studies 
\citep[10 -- 30\%,][]{Willett2011a,Willett2011b,Vignali2005}.

It is worth noting, however, that \citet{Yuan2010} have recently argued that most previous 
LINER identifications in (U)LIRGs should be reclassified as composite starburst/AGN systems, 
based on the refined optical classification scheme introduced by \citet{Kewley2006}. They 
conclude that ``true'' LINERs (in the sense of low accretion rate AGN) rarely occur in IR 
luminous mergers. The IRAS16399N nucleus of IRAS16399-0937 is one of only 9 objects that are classified 
as LINERs in their sample of $\sim 500$ (U)LIRGs. Nevertheless, as it falls near the 
Seyfert-LINER boundary in the VO diagrams, \citet{Yuan2010} suggest that it may be a Seyfert, 
or an intermediate object\footnote{The distribution of the Sloan Digital Sky Survey data in 
these diagrams shows that Seyferts and LINERs occupy distinct branches}. Conversely, our 
results suggest that the LINER spectrum originates from shock ionization and therefore is 
not directly related to the deeply embedded AGN.

\subsection{Nuclear gas flows}\label{sec:gasflows}

Taking the black hole masses ($M_{BH}\approx  10^8$\,M$_{\odot}$) derived from the
\citet{Marconi2003} relation at face value, the inferred AGN luminosity 
($L_{bol}\approx 3.4\times10^{44}$\,erg\,s$^{-1}$)  implies a small Eddington ratio 
$L_{bol}/L_{Edd}\sim 10^{-2}$; where $L_{Edd}\sim 10^{46}$\,erg\,s$^{-1}$ is the Eddington 
luminosity. The black hole in the IRAS16399N nucleus is therefore accreting at well below the 
Eddington rate, while, as there is no evidence for an AGN in the IRAS16399S nucleus, its black hole 
is presumably accreting at an even lower rate. Similarly, the star formation rates in both 
nuclei are relatively modest ($\sim$ 3 - 12 \,M$_{\odot}$\,yr$^{-1}$) and the bulk of the 
current star formation ($\gtrsim 60$\%) is  occurring in extended regions throughout the IRAS16399-0937 
system. Evidently the merger, while triggering widespread star formation has not (yet) 
resulted in the massive gas inflows that would lead to vigorous nuclear starbursts and 
enhanced BH accretion. 

In the case of IRAS16399-0937, therefore, the OHM emission does not appear to be associated
with a transient enhancement in tidally driven gas inflows, as proposed by \citet{Darling2007}.

\section{Summary and Conclusions}\label{sec:summary}

We have presented a multi-wavelength study of the luminous infrared galaxy IRAS16399-0937.  
This system is a gas rich, mid-stage merger containing an OH megamaser and consisting of 
two nuclei separated by $\sim 3.4$\,kpc, situated in a common envelope. In previous works, 
the nuclei have been classified on the basis of optical emission line ratios as a LINER 
(IRAS16399N) and a starburst (IRAS16399S), respectively. The data set analyzed here combines new emission 
line (H$\alpha$+[N{\sc ii}]) and broad-band (FR914M, F814W) imaging obtained with the HST 
ACS Wide Field Camera with archival imaging and spectroscopy in the optical (HST ACS F435W), 
near infrared (HST NICMOS 1.6$\mu$m, 2MASS J and K), mid/far-infrared (Spitzer IRAC, IRS 
and MIPS), sub-mm (Herschel SPIRE) and radio (1.49 and 4.9 GHz VLA). The main results from this 
analysis are summarized below:

\begin{itemize} 

\item The HST images reveal a complex system of dust lanes and emission line knots and 
filaments. Aside from the two nuclei, the main morphological features are a large ($\sim 3$\,kpc), 
diffuse region of star formation extending North and East of the IRAS16399N nucleus and an arc-like 
structure, also including star formation regions (as traced by H$\alpha$+[N{\sc ii}] line 
emission and B-band continuum) curving around the Western side of the system, which connects 
the IRAS16399S nucleus to the NE diffuse region.  An extensive dust lane complex running E-W separates 
the nuclei while the IRAS16399N nucleus itself is crossed by a dust lane oriented SE-NE. 

\item PAH emission at 8$\mu$m occupies the interior of envelope, with bright peaks 
corresponding to the two nuclei and spurs associated with the NE diffuse region and the 
dust lane complex E of the nuclei. The extended radio source morphology is very similar to 
that of the PAH emission, suggesting that it is due to star formation. Similarly, we find 
that the radio core associated with the IRAS16399N nucleus has a spectral index ($\alpha\,=\,-0.75\pm0.08$) 
and brightness temperature ($\sim\,2.3\,\times\,10^4\,$K) consistent with star formation 
rather than an AGN jet, in agreement with \citet{Baan2006}'s earlier classification. 

\item Star formation activity is widespread throughout the system but mostly heavily obscured. 
The star formation rate for the system as a whole as derived from radio and IR luminosities 
is $\sim 20 $M$_\odot$/yr, a factor $\sim 4$ greater than the rate obtained from the H$\alpha$ 
emission. The two nuclei together account for only $\approx 30-60$\% of the total. Although 
the two nuclei have similar stellar masses (as indicated by their H-band luminosities), the 
IRAS16399N nucleus is brighter in both PAH and radio, indicating a higher star formation rate. 

\item The Spitzer IRS spectrum (5.2\,-\,37$\mu$m) of the whole system is characterized by 
deep silicate absorption, PAH emission features, vibrational lines of H$_2$ and ionic lines 
from species of moderate ionization potential (IP$\,<\,$50\,eV) such as [Ar\,{\sc ii}], [Ne\,{\sc iii}] 
and [S\,{\sc iii}]. Absorption bands of water ice (6.0$\mu$m) and hydrogenated amorphous carbon 
(6.85 and 7.25$\mu$m) are also present. The higher spatial resolution in SL mode also allows 
separate extractions in the wavelength range 5.2 -- 14$\mu$m for the two nuclei and for 
several locations along the slit sampling the ISM of the extended envelope. The two nuclei 
show similar spectral features but the IRAS16399N nucleus exhibits the deepest ice and HAC absorption 
bands and is distinguished by a steeply rising continuum at wavelengths $>12\mu$m. The H$_2$O 
and HAC bands indicate the presence of compact and cold material, while the steep continuum is 
likely due to warm dust from an AGN-heated torus. However, we do not detect the high 
ionization fine structure lines (e.g., [Ne\,{\sc v}]\,14.3$\mu$m), that are widely regarded 
as tracers of AGN activity, in either nucleus. In general, the ISM mid-IR spectra in the 
5.2\,-\,14$\mu$m band show prominent PAH features, but the H$_2$O and HAC absorption bands 
and ionic emission lines are weaker or absent. 

\item Decomposition of the optical -- FIR SED of the two nuclei of IRAS16399-0937 into stellar, 
dusty ISM and AGN torus components confirms the presence of a moderately luminous AGN 
($L_{bol}\approx 3.4\times10^{44}$\,erg\,s$^{-1}$) in the IRAS16399N nucleus. However, the torus 
parameters recovered from the fit indicate that the AGN is completely embedded in a 
quasi-spherical distribution of optically thick clumps. This structure has a covering factor 
close to unity, implying that only a very small fraction ($\leq 0.2$\%) of the photons 
produced by the AGN escape. The SED fits are consistent with star formation rates of  
$\sim 12$ and 3\,M$_{\odot}$\,yr$^{-1}$ for the IRAS16399N and IRAS16399S nucleus, respectively. 
These values are consistent with that inferred from the X-ray emission and are
(within a factor $\sim 2-3$) of the SFR's derived from the ISM PAH-dust 8$\mu$m emission 
and the radio emission. However, if the AGN torus is omitted from the SED model, the 
SFR inferred for IRAS16399N increases to $\sim$20\,M$_{\odot}$\,yr$^{-1}$, which is 
comparable with the value determined for 
the entire system. The properties of the clumpy ``torus'' are consistent with the conditions
required for strong, radiatively pumped OH maser emission. % values corrected - DS 10/06/2014

\item The small fraction of ionizing photons that escapes the torus is insufficient to 
produce the observed H$\alpha$ luminosity from the IRAS16399N nucleus by a factor $\sim 66$. It 
follows that the LINER emission line spectrum cannot be attributed to AGN photoionization. 
We propose instead that the line emission from the IRAS16399N nucleus is due to shocks with velocities 
$\approx 100 - 200$\,km\,s$^{-1}$, perhaps associated with gas flows induced by gravitational 
torques generated by the merger.

\item If the IRAS16399N nucleus follows the \citeauthor{Marconi2003} scaling relationship between 
H-band luminosity and black hole mass, the luminosity derived for the embedded AGN implies 
that the $\sim  10^8$\,M$_{\odot}$ black hole is accreting at a small fraction ($\sim 1$\%) 
of the Eddington rate. As the star formation rates in both nuclei are also relatively modest, 
it seems clear that the massive gas inflows predicted by merger simulations have yet to 
fully develop in this system.
\end{itemize}

The picture that emerges of IRAS16399-0937 is that of a gas-rich merger of mass ratio 
$\approx 1:1$. The merger has triggered widespread star formation but massive gas flows 
into the still distinct nuclei have not yet fully developed. The IRAS16399N nucleus harbors an embedded 
AGN of relatively modest luminosity and is also the likely source of the OH megamaser emission,
while the IRAS16399S nucleus is starburst dominated. The AGN is apparent almost entirely 
through the contribution of AGN-heated dust to the IR SED. The LINER optical spectrum upon 
which the original classification was based is probably due to shocks, rather than AGN photoionization. 
This work also highlights the value of multiwavelength data and spatial resolution in 
establishing the nature of the buried power sources in (U)LIRGs. 

%The estimate of the SFR 
%in the IRAS16399N nucleus obtained from VLA data provided the key to confirming the presence of the 
%AGN, by placing an independent constraint on the SED fit and thus ruling out a pure starburst 
%model. 

\acknowledgments{
We thank the anonymous referee for useful comments that improved the clarity of the paper.
Support for program \#11604 was provided by NASA through a grant from the Space Telescope 
Science Institute, which is operated by the Association of Universities for Research in 
Astronomy, Inc., under NASA contract NAS 5-26555. This material is based upon work partly 
supported by the National Aeronautics and Space Administration under Grant No. NNX11AI03G 
issued through the Science Mission Directorate. D. A. Sales gratefully acknowledge for partial
financial support received from FAPERGS/CAPES n.05/2013.} %{\em Facility  acknowledgements...}

%\begin{figure*}
%\centering
%\includegraphics[angle=-90,scale=1.2]{iras16399_opt_pah_radio_black.pdf}
%\caption{.}
%\label{profile}
%\end{figure*}

%\begin{figure*}
%\centering
%\begin{tabular}{cc}
%\includegraphics[scale=3.1]{iras16399_goals.pdf}&
%\includegraphics[scale=3.1]{iras16399_goals1.pdf}\\
%\includegraphics[scale=3]{iras16399_goals2.pdf}&
%\includegraphics[scale=3]{iras16399_goals3.pdf}\\
%\includegraphics[scale=3]{iras16399_goals4.pdf}&
%\end{tabular}
%\caption{Haan et al. 2011. HST NICMOS imaging of the GOAL sample. I am showing such images, 
%but they will not include in the paper.}
%\label{fig:haan}
%\end{figure*}

%\begin{figure}
%\centering
%\includegraphics[width=6cm]{Halpha_EWPAHcontorno.pdf}
%\caption{I overploted the PAH emission contours to HST H$\alpha$ emission.}
%\label{profile}
%\end{figure}

%\begin{figure*}
%\centering
%\includegraphics[scale=0.8]{IRAS16399_AGN_SB_ISM.pdf}
%\includegraphics[scale=0.8]{IRAS16399_AGN_SB_ISM_const.pdf}
%\caption{.}
%\label{fig:spec_agn_sb_ism}
%\end{figure*}

%\begin{figure*}
%\centering
%\begin{tabular}{c}
%\includegraphics[scale=0.35,angle=-90]{Willett_fork_diagram.pdf}\\
%\includegraphics[scale=0.35,angle=-90]{Willett_PAH_slope_diagram.pdf}\\
%\end{tabular}
%\caption{.}
%\label{fig:willett}
%\end{figure*}

%\begin{figure*}
%\centering
%\begin{tabular}{cc}
%\includegraphics[scale=0.4]{pahpahpahpahcerro.pdf}&
%\includegraphics[scale=0.6]{pahnecerro_syst.pdf}\\
%\end{tabular}
%\caption{}
%\label{profile}
%\end{figure*}

%\begin{figure*}
%\centering
%\includegraphics[scale=0.8]{EW7-11xne_syst.pdf}
%\includegraphics[scale=0.8]{position_spec_extraction.pdf}
%\caption{.}
%\label{fig:pahxne}
%\end{figure*}

% SED-fitting (red lines) using an AGN (top panel) and Starburst Nuclei
%(bottom panel).

%\begin{figure*}
%\centering
%\includegraphics[scale=0.7]{hist_sfr_darling2000.pdf}
%\caption{{\bf This histogram is just to show the SFR value find in OHMG. It will not be in our paper!!}}
%\end{figure*}

% Revised tables inserted 7/10/13

\clearpage

\begin{deluxetable}{lccc}
\footnotesize
\tablecaption{Photometric properties of IRAS16399-0937}
\tablehead{\colhead{Filter} & \colhead{System} & \colhead{IRAS16399N} & \colhead{IRAS16399S} \\
\colhead{} & \colhead{aperture r = 7\arcsec\ } & \colhead{aperture r = 2\arcsec} & \colhead{aperture r = 2\arcsec}}
\startdata
F$_{X-ray}^{0.5-2keV}$ (erg/s/cm$^2$)	 & 1.18$\pm$0.23$\times10^{-13}$  & 2.58$\pm$0.93$\times10^{-14}$ & 6.15$\pm$4.02$\times10^{-15}$	\\
H$\alpha$ (erg/s/cm$^2$) &    $5.79-17.4\times10^{-13}$ &  1.80$\pm$0.26$\times10^{-13}$ &  1.00$\pm$0.14$\times10^{-13}$  \\
F435W (erg/s/cm$^2$/\AA)	& 6.57$\pm$0.98$\times10^{-16}$   &  8.54$\pm$1.27$\times10^{-17}$  &  1.26$\pm$0.18$\times10^{-16}$  \\
F814W (erg/s/cm$^2$/\AA)	& 2.33$\pm$0.34$\times10^{-15}$   &  3.84$\pm$0.57$\times10^{-16}$  &  4.50$\pm$0.67$\times10^{-16}$  \\
F914M (erg/s/cm$^2$/\AA)	& 2.33$\pm$0.35$\times10^{-15}$   &  3.78$\pm$0.56$\times10^{-16}$  &  4.82$\pm$0.72$\times10^{-16}$  \\
1.2$\mu$m (mJy) 2MASS & 23.4$\pm$3.5   &  2.53$\pm$0.37  &  5.23$\pm$0.78  \\
1.6$\mu$m (mJy) NICMOS& 36.4$\pm$5.44   &  4.23$\pm$0.63  &  5.17$\pm$0.77  \\
2.1$\mu$m (mJy) 2MASS & 36.5$\pm$5.46   &  4.60$\pm$0.68  &  5.15$\pm$0.77  \\
3.6$\mu$m (mJy) IRAC  &  18.42$\pm$2.75  &  3.85$\pm$0.57  &  3.88$\pm$0.58  \\
4.5$\mu$m (mJy) IRAC  &  15.27$\pm$2.28  &  4.14$\pm$0.61  &  2.82$\pm$0.42  \\
5.8$\mu$m (mJy) IRAC  &  49.78$\pm$7.44  &  12.55$\pm$1.87  &  7.56$\pm$1.13  \\
8.0$\mu$m (mJy) IRAC  &  156.9$\pm$23.47  &  34.03$\pm$5.09  & 24.14$\pm$3.61   \\
ISM-dust 8.0$\mu$m (mJy) &  152.6$\pm$22.82  &  33.14$\pm$4.95  & 23.23$\pm$3.47   \\
24$\mu$m (mJy) MIPS   &  486.33$\pm$72.75  &  -  &  -  \\
70$\mu$m (Jy) MIPS    &  5.78$\pm$0.86   &  -  &  -  \\
160$\mu$m (Jy) MIPS   &  2.241$\pm$0.33  &  -  &  -  \\
250$\mu$m (Jy) SPIRE\tablenotemark{a} &  4.14$\pm$0.3  &  -  &  -  \\
350$\mu$m (Jy) SPIRE\tablenotemark{a} &  1.00$\pm$0.18  &  -  &  -  \\
500$\mu$m (Jy) SPIRE\tablenotemark{a} &  0.39$\pm$0.13  &  -  &  -  \\
1.49GHz (Jy) \tablenotemark{b, c} & 3.47$\pm$0.02$\times10^{-2}$ & 1.53$\pm$0.01$\times10^{-2}$   &  6.8$\pm$0.32$\times10^{-3}$  \\
\enddata
\tablecomments{IRAS16399N and System apertures were centered at RA:16h42m40.18s and DEC:-09d43m13.26s. IRAS16399S 
was taken as RA:16h42m40.15s and DEC:-09d43m19.02s. $^a$ The photometry aperture used to measure the fluxes 
taken from Herschel/SPIRE images were r = 36\arcsec. $^b$ The AIPS Gaussian-fitting task JMFIT was used to obtain these values.
The beam-deconvolved sizes of the IRAS16399N and IRAS16399S components were 1\farcs1 x 0\farcs5 and 2\farcs07 x 1\farcs3, respectively.
$^c$ The AIPS verbs TVWIN+IMSTAT were used to obtain the total flux density value for a box region of size 14\arcsec x 14\arcsec.}
%\tablenotetext{a}{The photometry aperture used to measure the fluxes were r = 36\arcsec.}
\label{tab:fluxes}
\end{deluxetable}

\clearpage

\begin{deluxetable}{lccc}
\footnotesize
\tablecaption{Luminosity in IRAS16399-0937}
\tablehead{\colhead{Filter} & \colhead{System} & \colhead{IRAS16399N} & \colhead{IRAS16399S} \\
\colhead{} & \colhead{aperture r = 7\arcsec\ } & \colhead{aperture r = 2\arcsec} & \colhead{aperture r = 2\arcsec}}
\startdata
L$_{X-ray}^{0.5-2keV}$ (erg/s) & 2.14$\pm$0.41$\times10^{41}$ & 4.69$\pm$1.69$\times10^{40}$ & 1.12$\pm$0.73$\times10^{40}$	\\
L$_{H\alpha}$ (erg/s) & $1.05 - 3.16\times10^{42}$ & 3.27$\pm$0.47$\times10^{41}$ & 1.81$\pm$0.25$\times10^{41}$ \\
%SFR$_{H\alpha}$ (M$_{\odot}/yr$) &  4.78$-$14.4  &  1.48  &  0.83  \\
$\nu L_{8.0\mu\,m,dust}$ (erg/s) & 1.04$\pm$0.15$\times10^{44}$ & 2.25$\pm$0.37$\times10^{43}$ & 1.58$\pm$0.23$\times10^{43}$\\
%SFR$_{8.0\mu\,m,dust}$ (M$_{\odot}/yr$)&  16.03  & 3.48  &  2.44  \\
$\nu L_{24\mu\,m,dust}$ (erg/s) & 1.10$\pm$0.16$\times10^{44}$& -& -\\
%SFR$_{24\mu\,m,dust}$ (M$_{\odot}/yr$)&  19.00	  &  -  &  -  \\
$\nu L_{1.49GHz}$ (erg/s) & 9.40$\pm$0.54$\times10^{38}$ & 4.14$\pm$0.38$\times10^{38}$ & 1.84$\pm$0	.28$\times10^{38}$ \\
%SFR$_{1.49GHz}$ (M$_{\odot}/yr$)\tablenotemark{a}&  16.54  &  7.34  &  2.41  \\
%SFR$_{1.49GHz}$ (M$_{\odot}/yr$)\tablenotemark{b}&  8.24  &  3.66  &  1.20  \\
%$\langle$SFR$_{1.49GHz}$$\rangle$ (M$_{\odot}/yr$)\tablenotemark{c}&  12.39  &  5.5  &  1.8  \\
\enddata
%\tablenotetext{a}{Estimated using equation 17 of \citet{Rieke2009}.}
%\tablenotetext{b}{Estimated using equation 27 of \citet{Condon2002}.}
%\tablenotetext{c}{Average of values estimated using \citet{Condon2002}'s and \citet{Rieke2009}'s indicators.}
\label{tab:luminosity}
%\tablecomments{.}
\end{deluxetable}

\begin{deluxetable}{lccc}
\footnotesize
\tablecaption{Star formation rates in IRAS16399-0937}
\tablehead{\colhead{Filter} & \colhead{System} & \colhead{IRAS16399N} & \colhead{IRAS16399S} \\
\colhead{} & \colhead{aperture r = 7\arcsec\ } & \colhead{aperture r = 2\arcsec} & \colhead{aperture r = 2\arcsec}}
\startdata
SFR$_{X-ray}^{0.5-2keV}$ (M$_{\odot}/yr$) &  47.08$\pm$9.02 & 10.32$\pm$3.72 &  2.46$\pm$1.60  \\
%L$_{H\alpha}$ (erg/s) & $9.0 - 27.1\times10^{41}$ & 2.8$\times10^{41}$ & 1.56$\times10^{41}$ \\
SFR$_{H\alpha}$ (M$_{\odot}/yr$) &  5.58$-$16.76  &  1.73$\pm$0.25  &  0.96$\pm$0.14  \\
%$\nu L_{8.0\mu\,m,dust}$ (erg/s) & 8.60$\times10^{43}$ & 1.86$\times10^{43}$ & 1.31$\times10^{43}$\\
SFR$_{8.0\mu\,m,dust}$ (M$_{\odot}/yr$)&  19.37$\pm$2.89  & 4.21$\pm$0.63  &  2.95$\pm$0.44  \\
%$\nu L_{24\mu\,m,dust}$ (erg/s) & 9.13$\times10^{43}$& -& -\\
SFR$_{24\mu\,m,dust}$ (M$_{\odot}/yr$)&  23.18$\pm$3.16	  &  -  &  -  \\
%$\nu L_{1.49GHz}$ (erg/s) & 5.65$\times10^{38}$ & 2.51$\times10^{38}$ & 8.27$\times10^{37}$ \\
%SFR$_{1.49GHz}$ (M$_{\odot}/yr$)\tablenotemark{a}&  27.51$\pm$2.94  &  12.12$\pm$1.30  &  5.39$\pm$0.42  \\
SFR$_{1.49GHz}$ (M$_{\odot}/yr$)&  13.72$\pm$1.46  &  6.04$\pm$0.65  &  2.68$\pm$0.21  \\
%$\langle$SFR$_{1.49GHz}$$\rangle$ (M$_{\odot}/yr$)\tablenotemark{c}&  12.39  &  5.5  &  1.8  \\
\enddata
%\tablenotetext{a}{Estimated using equation 17 of \citet{Rieke2009}.}
%\tablenotetext{b}{Estimated using equation 27 of \citet{Condon2002}.}
%\tablenotetext{c}{Average of values estimated using \citet{Condon2002}'s and \citet{Rieke2009}'s indicators.}
\label{tab:sfr}
%\tablecomments{.}
\end{deluxetable}

%\clearpage

%\begin{landscape}
%\begin{table}
%\renewcommand{\tabcolsep}{0.7mm}
%\small
%\caption{IRAS16399-0937 MIR Emission Line Strengths (10$^{-16}$ W\,m$^2$) and Equivalent Widths ($\mu$m)}
%\begin{tabular}{lcccccc}
%\noalign{\smallskip}
%\hline\hline
%Line & \multicolumn{2}{c}{System} & \multicolumn{2}{c}{North} & \multicolumn{2}{c}{South} \\
%& Flux & EW & Flux & EW & Flux & EW \\
%\noalign{\smallskip}
%\hline\hline
%\hline
%\end{tabular}
%\\ \tablenotemark{a}{$7.7$$\mu$m Complex.}
%\\ \tablenotemark{b}{11.3$\mu$m Complex.}
%\\ \tablenotemark{c}{12.7$\mu$m Complex.}
%\\ \tablenotemark{d}{17$\mu$m Complex.}
%\label{tab:mirlines_sns}
%\end{table}
%\end{landscape}

\clearpage

\begin{table}
\renewcommand{\tabcolsep}{0.3cm}
\footnotesize
\caption{Mid infrared emission line strengths (10$^{-17}$ W\,m$^2$) and equivalent widths ($\mu$m)}
\begin{tabular}{lcccccc}
\noalign{\smallskip}
\hline\hline
Line & \multicolumn{2}{c}{System}&\multicolumn{2}{c}{IRAS16399N} & \multicolumn{2}{c}{IRAS16399S} \\
& Flux & EW & Flux & EW & Flux & EW \\
\noalign{\smallskip}
\hline\hline 			    												    
{}H$_2$ S(7) 5.5$\mu$m & 10.7$\,\pm\,$5.13 &	 0.073 &  12.7$\,\pm\,$0.733 &    0.165 & 1.57$\,\pm\,$0.275 &    0.106  \\
{}H$_2$ S(5) 6.9$\mu$m & $<2.65$ & 0.016 &  4.06$\,\pm\,$0.406 &   0.035 & 3.41$\,\pm\,$0.220 &    0.169  \\
{}[Ar{\sc\,ii}] 6.99$\mu$m & $<5.37$ &	0.032 &  8.50$\,\pm\,$0.471 &   0.073 & 5.47$\,\pm\,$ 0.216 &    0.268  \\
{}H$_2$ S(4) 8.0$\mu$m & $21.7\,\pm\,18.9$ & 0.140 &  15.0$\,\pm\,$0.5.97 &    0.132 &	- &     -  \\
{}[Ar{\sc\,iii}] 8.9$\mu$m & $<8.73$ &	  0.129 &  3.51$\,\pm\,$0.994 &   0.084 & 1.10$\,\pm\,$0.151 &   0.052  \\
{}H$_2$ S(3) 9.6$\mu$m & 42.4$\,\pm\,$10.4 &	  0.917 &  38.0$\,\pm\,$9.56 &     1.41 & 0.799$\,\pm\,$0.09 &   0.038  \\
{}[S{\sc\,iv}] 10.5$\mu$m & $<3.43$ & 0.047 &  4.59$\,\pm\,$0.744 &    0.116 & 0.751$\,\pm\,$0.107 &   0.035  \\
{}H$_2$ S(2) 12.2$\mu$m & $<4.77$ & 0.017 &  3.98$\,\pm\,$0.217 &   0.024 & 1.03$\,\pm\,$0.115 &   0.033  \\
{}[Ne{\sc\,ii}] 12.8$\mu$m & 41.6$\,\pm\,$16.1 &	  0.132 &  24.9$\,\pm\,$3.15 &    0.122 & 9.62$\,\pm\,$1.67 &    0.261  \\
{}[Ne{\sc\,v}] 14.2$\mu$m & $<4.92$ & 0.013 &  - &  -  & - &   -  \\
{}[Ne{\sc\,iii}] 15.5$\mu$m & 9.29$\,\pm\,$1.35 &	 0.029 &  - &   - & - &   -  \\
{}H$_2$ S(1) 17.0$\mu$m & 55.0$\,\pm\,$28.6 &	  - &  - &   - & - &  -  \\
{}[S{\sc\,iii}] 18.7$\mu$m & 29.4$\,\pm\,$19.6 &	  0.134 &  - &  - & - &  -  \\
{}[O{\sc\,iv}] 25.8$\mu$m & $<9.77$ &	 0.015 &  - &  - & - &  -  \\
{}[Fe{\sc\,ii}] 26.0$\mu$m & $<9.05$ &	- & - & - & - & -  \\
{}H$_2$ S(0) 28.1$\mu$m & $<16.4$ & 0.023 & - & - & - & - \\
{}[S{\sc\,iii}] 33.4$\mu$m & $<16.2$ & 0.019 & - & - & - & - \\
{}[Si{\sc\,ii}] 34.7$\mu$m & $<4.51$ &	0.052 & - & - & - & - \\
\hline
\multicolumn{7}{c}{PAH Features}\\
\hline																	 
%5.3$\mu$m &  66.7$\,\pm\,$24.4  &   0.316 &  73.9$\,\pm\,$3.75 &  0.707 &  11.3$\,\pm\,$1.828 &  0.575 \\
%5.7$\mu$m &  6.28e-10 $\,\pm\,$  2.01e-10  &   0.288 &  7.57e-10$\,\pm\,$  2.70e-11 &  0.625 &  1.18e-10 $\,\pm\,$ 8.87e-12 &  0.507 \\
6.2$\mu$m &  217$\,\pm\,$23.2  &   0.953 &  183$\,\pm\,$23.2 &   1.28 &  92.5$\,\pm\,$9.23 &   3.45 \\
%6.7$\mu$m &  57.3$\,\pm\,$30.8  &   0.242 &  8.51e-11$\,\pm\,$  4.12e-11 & 0.0525 &  2.59e-10 $\,\pm\,$ 1.33e-11 &  0.890 \\
7.4$\mu$m\tablenotemark{a} & 365$\,\pm\,$107  &    1.52 &  188$\,\pm\,$10.9 &   1.07 &  61.9$\,\pm\,$3.34 &   2.00 \\
7.6$\mu$m\tablenotemark{a} & 361$\,\pm\,$95.6  &    1.49 &  281$\,\pm\,$3.63 &   1.57 &  121$\,\pm\,$1.55 &   3.90 \\
7.8$\mu$m\tablenotemark{a} & 575$\,\pm\,$9.58  &    2.43 &  428$\,\pm\,$31.4 &   2.45 &  110$\,\pm\,$11.0 &   3.50 \\
8.3$\mu$m &  117$\,\pm\,$63.2  &   0.603 &  112$\,\pm\,$26.8 &  0.827 &  19.5$\,\pm\,$1.748 &  0.624 \\
8.6$\mu$m &  322$\,\pm\,$53.2  &    2.07 &  209$\,\pm\,$26.0 &   2.12 &  54.3$\,\pm\,$6.18 &   1.75 \\
11.2$\mu$m\tablenotemark{b} & 242$\,\pm\,$51.4  &    1.43 &  180$\,\pm\,$16.0 &   1.97 &  21.7$\,\pm\,$3.35 &  0.636 \\
11.3$\mu$m\tablenotemark{b} & 205$\,\pm\,$64.5  &    1.12 &  208$\,\pm\,$25.4 &   2.11 &  47.6$\,\pm\,$6.71 &   1.37 \\
12.0$\mu$m & $<17.2$ & 0.053 &  19.4$\,\pm\,$1.58 &  0.105 &  24.7$\,\pm\,$7.79 &  0.599 \\
12.6$\mu$m\tablenotemark{c} & 230$\,\pm\,$78.5  &   0.520 &  124$\,\pm\,$12.6 &  0.440 &  47.7$\,\pm\,$7.82 &  0.932 \\
12.7$\mu$m\tablenotemark{c} & $<0.93$ & 0.002 & 	- &   - & 	- &   - \\
16.4$\mu$m\tablenotemark{d} & 12.5$\,\pm\,$3.72  &  0.032 & 	- &   - & 	- &   - \\
17.0$\mu$m\tablenotemark{d} & 377$\,\pm\,$149  &    1.12 & 	- &   - & 	- &   - \\
17.4$\mu$m\tablenotemark{d} & 78.9$\,\pm\,$44.1  &   0.253 & 	- &   - & 	- &   - \\
17.9$\mu$m\tablenotemark{d} & 84.2$\,\pm\,$38.1  &   0.282 & 	- &   - & 	- &   - \\
\hline
$S_{sil}$\tablenotemark{e} & \multicolumn{2}{c}{-2.12$\,\pm\,$0.376}&\multicolumn{2}{c}{-2.06$\,\pm\,$0.366} & \multicolumn{2}{c}{-0.75$\,\pm\,$0.133} \\
EW$_{6.2\mu\,m}$\tablenotemark{f} & \multicolumn{2}{c}{0.171\,$\pm$\,0.01}&\multicolumn{2}{c}{0.266\,$\pm$\,0.01}&\multicolumn{2}{c}{0.711\,$\pm$\,0.012} \\
$\tau_{9.7\mu\,m}$\tablenotemark{g} & \multicolumn{2}{c}{5.59$\,\pm\,$0.017}&\multicolumn{2}{c}{5.1$\,\pm\,$0.018} & \multicolumn{2}{c}{0.35$\,\pm\,$0.002} \\
$\tau_{ice}\,\,6.0\mu\,$m & \multicolumn{2}{c}{0.729$\,\pm\,$0.108}&\multicolumn{2}{c}{0.548$\,\pm\,$0.081} & \multicolumn{2}{c}{0.279$\,\pm\,$0.041} \\
$\tau_{HAC}\,\,6.85\mu\,$m & \multicolumn{2}{c}{0.909$\,\pm\,$0.135}&\multicolumn{2}{c}{1.072$\,\pm\,$0.160} & \multicolumn{2}{c}{0.723$\,\pm\,$0.107} \\
$\tau_{HAC}\,\,7.25\mu\,$m & \multicolumn{2}{c}{0.519$\,\pm\,$0.077}&\multicolumn{2}{c}{0.670$\,\pm\,$0.100} & \multicolumn{2}{c}{0.685$\,\pm\,$0.102} \\
\hline\hline
\end{tabular}
\\ \tablenotemark{a}{7.7$\mu$m Complex.}
 \tablenotemark{b}{11.3$\mu$m Complex.}
 \tablenotemark{c}{12.7$\mu$m Complex.}
 \tablenotemark{d}{17$\mu$m Complex.}
\\ \tablenotemark{e}{Silicate apparent strength, $S_{sil}\,=\,ln\,(f_{obs}/f_{cont}$), was derived using the method of \citet{Spoon2007,Willett2011a}.}
\\ \tablenotemark{f}{Equivalent Width of 6.2$\mu$m PAH derived using the same method of \citet{Spoon2007,Willett2011a}.}
\\ \tablenotemark{g}{Silicate optical depth derived using {\sc pahfit} \citep{Smith2007}.}
\label{tab:mirlines_sns}
\end{table}

\clearpage

\begin{landscape}
\begin{table}
\renewcommand{\tabcolsep}{0.2cm}
\footnotesize
\caption{MIR Emission Line Strengths (10$^{-17}$ W\,m$^2$) and Equivalent Widths ($\mu$m)}
\begin{tabular}{lcccccccccc}
\noalign{\smallskip}
\hline\hline
Line & \multicolumn{2}{c}{ISM1}&\multicolumn{2}{c}{ISM2} & \multicolumn{2}{c}{ISM3} & \multicolumn{2}{c}{ISM4} & \multicolumn{2}{c}{ISM5}\\
& Flux & EW & Flux & EW & Flux & EW & Flux & EW & Flux & EW \\
\noalign{\smallskip}
\hline\hline
{}H$_2$ S(5) 5.5$\mu$m	  &     -	&	-	&	-	&	-	&	-	&	-	&	-	&	-	&	0.70	$\,\pm\,$	0.37	&	1.65	\\
{}[Ar{\sc\, ii}] 6.9$\mu$m  &  -	&	-	&	-	&	-	&	-	&	-	&	$<$0.04	&	0.008	&	0.33	$\,\pm\,$	0.28	&	0.218	\\
{}H$_2$ S(4) 8.0$\mu$m & 0.62	$\,\pm\,$	0.23	&	0.040	&	0.62	$\,\pm\,$	0.23	&	0.040	&	$<$0.007	&	0.001	&	$<$0.12		&	0.023	&	0.53	$\,\pm\,$	0.41	&	0.256	\\
{}[Ar{\sc\, iii}] 8.9$\mu$m & 1.34	$\,\pm\,$	0.19	&	0.114	&	1.34	$\,\pm\,$	0.19	&	0.114	&	0.37	$\,\pm\,$	0.17	&	0.074	&	1.05	$\,\pm\,$	0.13	&	0.215	&	-	&	-	\\
{}H$_2$ S(3) 9.6$\mu$m 	  & 1.14	$\,\pm\,$	0.14	&	0.121	&	1.14	$\,\pm\,$	0.14	&	0.121	&	-	&	-	&	0.39	$\,\pm\,$	0.08	&	0.084	&	0.92	$\,\pm\,$	0.55	&	0.977	\\
{}[S{\sc\, iv}] 10.5$\mu$m  & 0.26	$\,\pm\,$	0.15	&	0.024	&	0.26	$\,\pm\,$	0.15	&	0.024	&	0.39	$\,\pm\,$	0.14	&	0.075	&	0.50	$\,\pm\,$	0.11	&	0.122	&	0.42	$\,\pm\,$	0.36	&	0.343	\\
{}H$_2$ S(2) 12.27$\mu$m  & 0.75	$\,\pm\,$	0.15	&	0.039	&	0.75	$\,\pm\,$	0.15	&	0.039	&	0.23	$\,\pm\,$	0.14	&	0.040	&	0.38	$\,\pm\,$	0.13	&	0.121	&	$<$0.06	&	0.025	\\
{}[Ne{\sc\, ii}] 12.8$\mu$m & 4.21	$\,\pm\,$	0.21	&	0.179	&	4.21	$\,\pm\,$	0.21	&	0.179	&	0.36	$\,\pm\,$	0.17	&	0.065	&	3.15	$\,\pm\,$	0.17	&	1.08	&	0.21	$\,\pm\,$	0.20	&	0.086	\\
%{}[Ne{\sc iii}] 15.6$\mu$m& 8.10e-12 $\pm$    0.00 & 0.0131 & 4.18e-12 $\pm$    0.00 & 0.117 &3.76e-12 $\pm$    0.00 &0.0868&3.00e-12$\pm$     0.00 &   0.119 & 6.84e-13 $\pm$	0.00&0.0424\\
%{}H$_2$ S(1) 17.03$\mu$m  & 6.76e-12 $\pm$    0.00 &0.00748 & 3.49e-12 $\pm$    0.00 & 0.114 &3.13e-12 $\pm$    0.00 &0.0881&2.50e-12$\pm$     0.00 &  0.0485 & 5.70e-13 $\pm$	0.00&0.0563\\
%{}[S{\sc iii}] 18.7$\mu$m & 5.60e-12 $\pm$    0.00 &0.00423 & 2.89e-12 $\pm$    0.00 & 0.114 &2.60e-12 $\pm$    0.00 &0.0894&2.07e-12$\pm$     0.00 &  0.0114 & 4.73e-13 $\pm$	0.00&0.0633\\
%{}[O{\sc iv}] 25.9$\mu$m  & 7.09e-12 $\pm$    0.00 &0.00209 & 3.66e-12 $\pm$    0.00 & 0.351 &3.29e-12 $\pm$    0.00 & 0.262&2.62e-12$\pm$     0.00 &0.000335 & 5.99e-13 $\pm$	0.00& 0.153\\
%{}[S{\sc iii}] 33$\mu$m   & 4.25e-12 $\pm$    0.00 &0.00114 & 2.19e-12 $\pm$    0.00 & 0.527 &1.97e-12 $\pm$    0.00 & 0.398&1.57e-12$\pm$     0.00 &3.34e-05 & 3.59e-13 $\pm$	0.00& 0.262\\
%{}[Si{\sc ii}] 34.8$\mu$m & 3.93e-12 $\pm$    0.00 &0.00108 & 2.03e-12 $\pm$    0.00 & 0.569 &1.82e-12 $\pm$    0.00 & 0.430&1.45e-12$\pm$     0.00 &2.53e-05 & 3.32e-13 $\pm$	0.00& 0.288\\
%\hline
%\multicolumn{3}{c}{Hydrogen Molecular Lines}\\
%\hline
\hline
\multicolumn{11}{c}{PAH Features}\\
\hline
6.2$\mu$m 		   &15.3	$\,\pm\,$	0.85	&	0.645	&	15.3	$\,\pm\,$	0.85	&	0.645	&	8.88	$\,\pm\,$	0.88	&	1.98	&	11.0	$\,\pm\,$	0.74	&	1.46	&	3.31	$\,\pm\,$	1.11	&	2.49	\\
%6.7$\mu$m 		   &	  - & -& 57.3$\pm$30.8 & 0.242 & 5.22 $\pm$ 1.40 & 0.213 & 2.71$\pm$1.44 & 0.292 &4.50 $\pm$1.63 & 0.834\\ 
7.4$\mu$m\tablenotemark{a} &	 68.4	$\,\pm\,$	3.53	&	2.86	&	68.4	$\,\pm\,$	3.53	&	2.86	&	29.6	$\,\pm\,$	4.77	&	4.46	&	20.7	$\,\pm\,$	3.02	&	2.66	&	-	&	-	\\
7.6$\mu$m\tablenotemark{a} & 65.9	$\,\pm\,$	1.79	&	2.77	&	65.9	$\,\pm\,$	1.79	&	2.77	&	18.7	$\,\pm\,$	1.78	&	2.74	&	43.0	$\,\pm\,$	1.53	&	5.55	&	-	&	-	\\
7.8$\mu$m\tablenotemark{a} &29.5	$\,\pm\,$	1.29	&	1.27	&	29.5	$\,\pm\,$	1.29	&	1.27	&	9.67	$\,\pm\,$	1.36	&	1.35	&	4.96	$\,\pm\,$	1.13	&	0.643	&	7.87	$\,\pm\,$	1.42	&	2.6	\\
8.3$\mu$m 		   &-	&	-	&	-	&	-	&	$<$0.26	&	0.035	&	-	&	-	&	-	&	-	\\
8.6$\mu$m 		   & 33.4	$\,\pm\,$	0.72	&	1.69	&	33.4	$\,\pm\,$	0.72	&	1.69	&	13.7	$\,\pm\,$	0.794	&	1.81	&	16.0	$\,\pm\,$	0.53	&	2.15	&	$<$0.91	&	0.344	\\
11.2$\mu$m\tablenotemark{b}&11.5	$\,\pm\,$	0.45	&	0.612	&	11.5	$\,\pm\,$	0.45	&	0.612	&	5.78	$\,\pm\,$	0.39	&	0.716	&	4.90	$\,\pm\,$	0.33	&	0.889	&	-	&	-	\\
11.3$\mu$m\tablenotemark{b}&22.9	$\,\pm\,$	0.88	&	1.18	&	22.9	$\,\pm\,$	0.88	&	1.18	&	9.42	$\,\pm\,$	0.692	&	1.16	&	9.41	$\,\pm\,$	0.54	&	1.73	&	1.45	$\,\pm\,$	0.96	&	0.555	\\
12.0$\mu$m		   &9.33	$\,\pm\,$	0.90	&	0.374	&	9.33	$\,\pm\,$	0.90	&	0.374	&	3.54	$\,\pm\,$	0.57	&	0.425	&	6.32	$\,\pm\,$	0.45	&	1.29	&	-	&	-	\\
12.6$\mu$m\tablenotemark{c}&21.7	$\,\pm\,$	1.43	&	0.679	&	21.7	$\,\pm\,$	1.43	&	0.679	&	8.25	$\,\pm\,$	1.01	&	1	&	10.0	$\,\pm\,$	0.93	&	2.27	&	$<$0.51	&	0.137	\\
12.7$\mu$m\tablenotemark{c}&1.46	$\,\pm\,$	0.51	&	0.044	&	1.46	$\,\pm\,$	0.51	&	0.044	&	$<$0.17	&	0.0215	&	0.97	$\,\pm\,$	0.43	&	0.222	&	$<$0.55	&	0.146	\\
%16.4$\mu$m\tablenotemark{d}&	0.00$\pm$    - & 0.00& 1.25e-10$\pm$3.72e-10 & 0.032 &     0.00 $\pm$     - &  0.00 &	0.00$\pm$    - &  0.00 &    0.00 $\pm$    - &  0.00\\ 
%17.0$\mu$m\tablenotemark{d}&	0.00$\pm$    - & -& 3.77e-09$\pm$1.49e-09 & 1.12  &     0.00 $\pm$     - &  - &	0.00$\pm$    - &  - &    0.00 $\pm$    - &  -\\ 
%17.4$\mu$m\tablenotemark{d}&	0.00$\pm$    - & 0.00& 7.89e-10$\pm$4.41e-10 & 0.253 &     0.00 $\pm$     - &  0.00 &	0.00$\pm$    - &  0.00 &    0.00 $\pm$    - &  0.00\\ 
%17.9$\mu$m\tablenotemark{d}&	0.00$\pm$    - & 0.00& 8.42e-10$\pm$3.81e-10 & 0.282 &     0.00 $\pm$     - &  0.00 &	0.00$\pm$    - &  0.00 &    0.00 $\pm$    - &  0.00\\ 
\hline
$S_{sil}$\tablenotemark{e}&\multicolumn{2}{c}{-1.31$\pm$0.150}&\multicolumn{2}{c}{-0.41$\pm$0.048}&\multicolumn{2}{c}{-0.31$\pm$0.036}& \multicolumn{2}{c}{-0.65$\pm$0.074} & \multicolumn{2}{c}{-0.73$\pm$0.084} \\
EW$_{6.2\mu\,m}$\tablenotemark{f} & \multicolumn{2}{c}{0.077$\pm$0.008}&\multicolumn{2}{c}{0.388$\pm$0.044}&\multicolumn{2}{c}{0.067$\pm$0.007}&\multicolumn{2}{c}{0.108$\pm$0.012}&\multicolumn{2}{c}{1.715$\pm$0.850}\\
$\tau_{9.7\mu\,m}$\tablenotemark{g} & \multicolumn{2}{c}{0.87$\pm$0.099}&\multicolumn{2}{c}{$<0.01$} & \multicolumn{2}{c}{0.52$\pm$0.059} & \multicolumn{2}{c}{$<0.01$} & \multicolumn{2}{c}{3.94$\pm$1.27} \\
$\tau_{ice}\,\,6.0\mu\,$m & \multicolumn{2}{c}{0.273$\pm$0.031}&\multicolumn{2}{c}{0.282$\pm$0.032} & \multicolumn{2}{c}{0.805$\pm$0.153} & \multicolumn{2}{c}{0.112$\pm$0.02} & \multicolumn{2}{c}{$<0.01$}\\
$\tau_{HAC}\,\,6.85\mu\,$m & \multicolumn{2}{c}{0.137$\pm$0.015}&\multicolumn{2}{c}{$<0.05$} & \multicolumn{2}{c}{1.664$\pm$0.316} & \multicolumn{2}{c}{$<0.1$} & \multicolumn{2}{c}{$<0.01$}\\
$\tau_{HAC}\,\,7.25\mu\,$m & \multicolumn{2}{c}{0.131$\pm$0.015}&\multicolumn{2}{c}{0.125$\pm$0.014} & \multicolumn{2}{c}{0.248$\pm$0.047} & \multicolumn{2}{c}{0.319$\pm$0.040} & \multicolumn{2}{c}{$<0.01$}\\
\hline\hline
\end{tabular}
\\ \tablenotemark{a}{$7.7$$\mu$m Complex.}
 \tablenotemark{b}{11.3$\mu$m Complex.}
 \tablenotemark{c}{12.7$\mu$m Complex.}
 \tablenotemark{d}{17$\mu$m Complex.}
\\ \tablenotemark{e}{Silicate apparent strength, $S_{sil}\,=\,ln\,(f_{obs}/f_{cont}$), was derived using the method of \citet{Spoon2007,Willett2011a}.}
\\ \tablenotemark{f}{Equivalent Width of 6.2$\mu$m PAH derived using the same method of \citet{Spoon2007,Willett2011a}.}
\\ \tablenotemark{g}{Silicate optical depth derived using {\sc pahfit} \citep{Smith2007}.}
\label{tab:mirlines_ism}
\end{table}
\end{landscape}

\clearpage

\begin{deluxetable}{lll}
\tabletypesize{\footnotesize}
\tablecaption{Summary of model parameters used for SED decomposition.}
\tablehead{\colhead{Parameter}&\colhead{Description}&\colhead{Values}}
\startdata
$\sigma$ & Torus angular scale height (degrees) & 15, 20, 25, 30, 35, 40, 45, 50, 55, 60, 65, 70 \\ 
$Y$ & Torus radial extent & 5, 10, 20, 30, 40, 50, 60, 70, 80, 90, 100 \\ 
$N_0$ & Torus number of clouds & 1, 2, 3, 4, 5, 6, 7, 8, 9, 10, 11, 12, 13, 14, 15 \\ 
$q$ & Torus radial power law index & 0.0, 0.5, 1.0, 1.5, 2.0, 2.5, 3.0 \\ 
$\tau_V$ & Torus cloud optical depth & 5, 10, 20, 30, 40, 60, 80, 100, 150 \\ 
$i$ & Torus inclination (degrees) & 0, 10, 20, 30, 40, 50, 60, 70, 80, 90 \\ 
$q_{\rm PAH}$ & ISM PAH mass fraction & 0.47, 1.12, 1.77, 2.5, 3.19, 3.9, 4.58 \\ 
$u_{\rm min}$ & ISM ISRF scaling & 0.1, 0.15, 0.2, 0.3, 0.4, 0.5, 0.7,
0.8, \\
& & 1.0, 1.2, 1.5, 2.0, 2.5, 3.0, 4.0, 5.0, \\
& & 7.0, 8.0, 10.0, 12.0, 15.0, 20.0, 25.0 \\ 
age & Age of stellar population (Gyr) & 0.1, 0.2, 0.4, 0.8, 1.5, 2.0, 3.0, 4.0, 5.0, 8.0, 11.0, 13.0 \\
\enddata
\label{tab:modelParms}
%\tablecomments{}
\end{deluxetable}

\clearpage

\begin{deluxetable}{lcccccc}
\tabletypesize{\footnotesize}
\tablecaption{Parameters derived from SED fits assuming AGN and starburst nuclei}
\tablewidth{0pt}
\tablehead{\colhead{Parameters} & \colhead{} & \colhead{IRAS16399N} &  \colhead{} &  \colhead{} & \colhead{IRAS16399S} & \colhead{}\\
\colhead{}&\colhead{median} & \colhead{5\%} & \colhead{95\%} & \colhead{median} & \colhead{5\%} & \colhead{95\%}}
\startdata
\multicolumn{7}{c}{Clumpy Torus Model} \\
\hline
$\sigma$& 69.0$^{\circ}$ & 65.8$^{\circ}$ & 69.9$^{\circ}$ & - & - & - \\
Y       & 90.1           & 63.4           & 99.2         & - & - & - \\    
N$_0$   & 14.7           & 13.9           & 15.0           & - & - & - \\    
q       & 0.92           & 0.59           & 1.00           & - & - & - \\    
$\tau_V$& 27.7           & 20.6           & 34.8	   & - & - & - \\
$i$        & 75.8$^{\circ}$  & 57.9$^{\circ}$ & 88.5$^{\circ}$  & - & - & - \\
C$_{frac}$ & 99.9\%          & 99.8\%         & 99.9\%		& - & - & -  \\ 
%hoa        & 6.2$^{\circ}$   & 6.1$^{\circ}$  & 6.3$^{\circ}$   & - & - & -  \\
escProb & \multicolumn{3}{c}{$\leq1\,\times\,10^{-6}$}		& - & - & -  \\
Mass (M$_{\odot}$)   & 347	      & 174		& 589  & - & - & -  \\
Size (pc)	  & 20.9	      & 13.5		& 24.5 & - & - & -  \\
\hline
\multicolumn{7}{c}{Interstellar Medium} \\
\hline
u$_{min}$  & 22.8            & 14.7        & 24.9		& 15.9  & 6.9  & 23.3	\\    
q$_{PAH}$  & 1.41 	     & 1.07	   & 2.16		& 4.01 & 2.93 &	4.51	\\    
f$_{PDR}$  & 0.3\%	     & 0.0\%	   & 1.4\%		& 5.1\% & 0.5\%  & 13.0\% 	\\
\hline
\multicolumn{7}{c}{Interstellar Medium Foreground} \\
\hline
$\tau_{ISM,V}$   (mag) & 6.9 	    & 4.5  	  & 10.1		& 14.1  & 1.9  & 48.73	\\    
C$_{frac,ISM}$         & 89.1\%     & 62.8\%      & 99.2\%	& 23.8\% & 7.6\% &	79.7\%	\\    
$\tau_{torus,V}$ (mag) & 9.0 	    & 5.5	  & 12.0	& -    & -    &	-	\\    
$\tau_{stars,V}$ (mag) & 4.5 	    & 4.0 	  & 5.0		& 3.4  & 3.1  &	3.7	\\   
C$_{frac,stars}$ & 98.7\% 	    & 98.3\%	  & 99.1\%	& 99.6\% & 99.0\% &	100\%	\\
\hline
\multicolumn{7}{c}{Simple Stellar Population} \\
\hline
log Age (Gyr)  	       & 12.6       & 7.9	  & 13.0	& 9.4  & 4.3  &	12.6\\
\hline
\multicolumn{7}{c}{Amplitude} \\
\hline
L$_{AGN}$ (ergs/s) & 3.4$\,\times\,10^{44}$ & 2.7$\,\times\,10^{44}$ & 4.2$\,\times\,10^{44}$ & - & - & - \\
L$_{ISM}$ (ergs/s) & 2.9$\,\times\,10^{44}$ & 1.9$\,\times\,10^{44}$ & 3.7$\,\times\,10^{44}$ & 8.9$\,\times\,10^{43}$ & 7.8$\,\times\,10^{43}$ & 1.3$\,\times\,10^{44}$ \\
SFR (M$_{\odot}/yr$) & 11.6 & 7.6 & 15.1 & 3.6 & 3.2 & 5.2 \\
\enddata
\label{tab:linerfittorus}
\tablecomments{It should be noted the $N_{0}$, $Y$ and $\sigma$ have ran to 
the limits of the parameter grid of the clumpy model, consequently, there is no automatic 
detection of upper limits. See marginalized posterior probability densities in Figure \ref{fig:torus_agn1}.}
\end{deluxetable}

\clearpage

\begin{deluxetable}{lcccccc}
\tabletypesize{\footnotesize}
\tablecaption{Parameters derived from SED fits assuming dual starburst nuclei}
\tablewidth{0pt}
\tablehead{\colhead{Parameters} & \colhead{} & \colhead{IRAS16399N} &  \colhead{} &  \colhead{} & \colhead{IRAS16399S} & \colhead{}\\
\colhead{}&\colhead{median} & \colhead{5\%} & \colhead{95\%} & \colhead{median} & \colhead{5\%} & \colhead{95\%}}
\startdata
\multicolumn{7}{c}{Interstellar Medium} \\
\hline
u$_{min}$  & 18.5           & 8.2         & 24.6		& 1.6  & 0.4  & 4.6	\\    
q$_{PAH}$  & 0.48           & 0.47	  & 0.52		& 4.31 & 3.66 &	4.56	\\    
f$_{PDR}$  & 70.2\%	    & 60.0\%	  & 78.4\%		& 10.9\%  & 3.5\%  & 18.9\%	\\
\hline
\multicolumn{7}{c}{Interstellar Medium Foreground} \\
\hline
$\tau_{ISM,V}$ (mag) & 24.0  	    & 23.4  	  & 24.8	& 7.7  & 2.2  & 38.0	\\    
C$_{frac,ISM}$      & 99.9\%        & 99.5\%      & 100\%	& 34.9\% & 14.6\% & 89.1\%	\\    
$\tau_{stars,V}$ (mag)   & 5.5 	    & 5.0 	  & 6.0		& 3.3  & 3.1  &	3.7	\\   
C$_{frac,stars}$    & 98.8\% 	    & 98.6\%	  & 99.9\%	& 99.6\% & 99\% & 100\%	\\
\hline
\multicolumn{7}{c}{Simple Stellar Population} \\
\hline
log Age (Gyr)  & 12.8      & 0.9   & 13.0	& 9.7  & 4.5  &	12.7\\
\hline
\multicolumn{7}{c}{Amplitude} \\
\hline
%L$_{AGN}$ & 1.24\,e44 ergs/s & 1.22\,e44 ergs/s & 1.25\,e44 ergs/s & - & - & - \\
L$_{ISM}$ (erg/s) & 5.06$\,\times\,10^{44}$ & 4.50$\,\times\,10^{44}$ & 5.83$\,\times\,10^{44}$ & 8.18$\,\times\,10^{43}$ & 7.46$\,\times\,10^{43}$ & 9.93$\,\times\,10^{43}$ \\
SFR (M$_{\odot}/yr$)   & 20.32 & 18.11 & 23.44 & 3.32 & 3.04 & 4.02 \\
\enddata
\label{tab:linersbfit}
%\tablecomments{}
\end{deluxetable}

\clearpage

\appendix
\section{Fitting the SED with clumpyDREAM}\label{sec:sedfitting}

The stellar, ISM, and AGN torus components of the SED were computed on
parameter grids; the parameters and their grid values are exhaustively
listed in Table~\ref{tab:modelParms}. The number of possible model
combinations is $2.4 \times 10^9$. One search strategy would be to
sample all of the grid points exhaustively. At each grid point, the
corresponding stellar, ISM, and AGN torus components are generated,
and their normalizations are determined by a fit to the observed
SED. The best fits, determined by minimum $\chi^2$ or
maximum-likelihood, out of all of the grid points would then represent
the best decomposition. Although the model is a linear combination of
SED components, the requirement of non-negative coefficients, the
use of censored data (upper limits from large apertures), and
the inclusion of foreground extinction require non-linear fitting
methods at any set of grid points. At this time, performing
non-linear fits over $10^9$ parameter combinations is computationally
prohibitive.

clumpyDREAM instead performs a Markov Chain Monte Carlo (MCMC) search
over the parameter grid. This approach has become commonplace for SED
fitting \citep[e.g., recently,][]{Pirzkal2012,Serra2011,Asensio-Ramos2009,Ptak2007,Sajina2006} 
because of its economy and efficiency over an exhaustive search
over large parameter spaces. The commonest approach is to use the
random walk Metropolis (RWM) algorithm \citep{Metropolis1953,Hastings1970}. 
However, RWM inefficiently samples complex or multimodal
probability distributions, especially when modes are widely separated
\citep[cf.][]{Abraham1999}. We found widely spaced modes to be a problem
especially for SEDs showing silicate in emission. This inefficiency is
somewhat mitigated by adaptively tuning the step-size in parameter
space \citep{Haario1999,Haario2001}, but our experience is that the tuning can be
adversely affected by parameter boundaries, causing solutions to hang
up on local $\chi^2$ minima. Furthermore, single-chain adaptive RWM
still has difficulty with widely spaced modes. 

The DREAMZS algorithm \citep{Terbraak2008} involves a more robust,
self-adapting stepper with multiple chains, and we employed a modified
version of DREAMZS in clumpyDREAM.  The DREAMZS algorithm uses
differential evolution \citep[DE,][]{Storn1997} to determine steps in parameter
space. Multiple parameter chains are run in parallel, with the current
state of all chains stored in $\mathbf{X}$, where $\mathbf{X}_i$ is
the $i$th vector out of an evolving population of parameter
vectors. The history of accepted 
parameter vectors is stored in $\mathbf{Z}$, where $\mathbf{Z}_j$ is
the parameter vector of the $j$th member of the chain. Trial solutions
$\widetilde{\mathbf{X}}$ are generated by,
\begin{equation}
\widetilde{\mathbf{X}}_i = \mathbf{X}_i
+ \gamma(1+\mathbf{e}) \left(\mathbf{Z}_m
- \mathbf{Z}_n\right)+ \mathbf{\epsilon},
\end{equation}
where $\mathbf{Z}_m$ and $\mathbf{Z}_n$ are randomly drawn, without
replacement, from the history of solutions, $\mathbf{e}$ contains
uniform random deviates in the range $-0.05$ to 0.05, and
$\mathbf{\epsilon}$ is a vector of normal random deviates with
standard deviation $10^{-6}$. The spirit of the DE stepper is to use
the sampling history as a measure of the covariance matrix, and the
term $\gamma$ sets the scale of the step size relative to the spread
of parameter values. 
% replacement text per JG comments 7/30/14 - AR
Following the recommendations of
ter Braak and Vrugt, $\gamma = 2.38 / \sqrt{2 p}$, where $p$ is the
number of parameters being adjusted at a given step.
%In DREAMZS, $\gamma$ is held fixed, but
%clumpyDREAM initializes $\gamma = 0.5$ and tunes $\gamma$ every 1000
%iterations to promote acceptance rates in the range 0.1 -- 0.5. 
The tuning schedule is adapted from that used by \citet{Patil2010}
for their implementation of the adaptive RWM
algorithm. Otherwise, new solutions are accepted or rejected using the
usual Metropolis (likelihood) ratio criterion. The algorithm therefore
returns samples from the posterior probability distribution of the model
parameters subject to prior assumptions or constraints. Here, the main
prior constraints are imposed by the domain of the gridded
parameters. 

clumpyDREAM also employs the snooker updater, a modification of the DE
stepper, every 10 iterations, as described by \citet{Terbraak2008}. 
Since both the DE and snooker methods operate on continuous
parameters, it was necessary to interpolate the gridded models. We
used multidimensional linear splines, which provided stable
solutions. In contrast, cubic spline interpolation introduced
artifacts such as ringing and negative values in the interpolated
SEDs, which indicates that over some regions of parameter space the
model grids sample changes in the shape of the SED insufficiently for
the use of higher-order interpolation methods.

For each Monte Carlo trial, the clumpyDREAM procedure ordinarily
involves two steps: (1) selection of model parameters by either DE or
the snooker updater, and (2) determining the linear coefficients for
each model component by a linear, least-squares fit. The data for
IRASF16399$-$0937 present a challenge, however, that the SEDs for each
of the two nuclei are fit simultaneously subject to the constraint
that the sum of the SEDs must fall under the upper limits imposed by
large aperture data: i.e., far-infrared measurements, which do not
resolve the two nuclei and include diffuse emission outside the
extraction apertures used at shorter wavelengths. We therefore had to
modify clumpyDREAM to treat the coefficients as non-linear parameters
that are sampled by MCMC rather than determined by linear fits. Upper
limits are then handled in a method similar to the In-Out technique
\citep{Friedman1981}
%{\bf (Friedman, J. H., \& Stuetzle, W. 1981, The In-Out Method for Linear
%Regression with Censored Data, Technical Report No. 65, Public Health
%Service Grant 2 RO1 GM21215-06, Division of Biostatistics, Stanford
%University)}: 
if the sum of the model SEDs exceeds an upper limit, that 
upper limit is included in the calculation of the likelihood;
otherwise, it is ignored.

clumpyDREAM randomly initializes $\mathbf{Z}$ with $\sim 100$ randomly
sampled parameter vectors. Initial parameters are drawn assuming
uniform prior distributions before engaging the modified DREAMZS
stepper. For the gridded models, parameters were randomly selected
within their range of support (Table~\ref{tab:modelParms}). Values for the
foreground extinction were randomly selected from values ranging from
$\tau_{\rm V}$ = 0 to 100, but any $\tau_{\rm V} > 0$ was permitted
during MCMC sampling. Foreground covering fractions were sampled from
and restricted to $C_f = 0$ to 1. %\sout{For the specific case of
%IRASF16399$-$0937, the IRAS16399N nucleus was an exception. We assumed that
%the foreground extinction to the AGN is comparable to that for the
%nuclear emission line region; in this case, the foreground extinction
%was restricted to $\tau_{\rm V} <$ 6.4}

Multiple, parallel chains are necessary for the DE algorithm, but they
also provide a means of sampling multimodal distributions more
efficiently than RWM or its variants. To test this assertion, the
modified version of clumpyDREAM was initially applied to the observed
SED of IRASF16399$-$0937 in 20 short-run trials to search for possibly
widely separated modes. In each run, five chains were run in parallel,
and solutions were thinned by concatenating the parameter vector
population $\widetilde{\mathbf{X}}$ onto the chains
$\widetilde{\mathbf{Z}}$ every tenth iteration. All of the short-run
trials converged to indistinguishable posterior distributions despite
starting from different, randomly generated parent chains. Based on
the success of these trial runs, clumpyDREAM was reinitialized and
allowed to run a minimum of $10^5$ iterations and until the
distribution of likelihood values reached a 
stationary state as measured by the Geweke statistic \citep{Geweke1992}.
% {\em Bayesian Statistics 4}, p. 169, eds. J. M. Bernardo,
%J. O. Berger, A. P. Dawid, \& A. F. M. Smith)}. 

\clearpage

\clearpage

\clearpage

\clearpage

\end{document}